\documentclass[acmsmall]{acmart}

\AtBeginDocument{%
  \providecommand\BibTeX{{%
    \normalfont B\kern-0.5em{\scshape i\kern-0.25em b}\kern-0.8em\TeX}}}

\setcopyright{acmcopyright}
\copyrightyear{2021}
\acmYear{2021}

\acmPrice{15.00}
 
\usepackage{verbatim}
\usepackage{booktabs} 
\usepackage{algorithm}
\usepackage{algorithmic}
\usepackage{epsfig}
\usepackage{color}
\usepackage{amsmath}
\usepackage{graphicx,subfigure}
\usepackage{multirow}
\usepackage{makecell}

\begin{document}

\title{Personalized News Recommendation: Methods and Challenges}

 \author{Chuhan Wu}
 \affiliation{%
   \institution{Department of Electronic Engineering \& BNRist, Tsinghua University}
     \city{Beijing}
     \postcode{100084}
   \country{China}
 }
 \email{wuchuhan15@gmail.com}

 \author{Fangzhao Wu}
  \authornotemark[0]
 \authornote{Corresponding Author}
 \affiliation{%
   \institution{Microsoft Research Asia}
   \city{Beijing}
   \country{China}
   \postcode{100080}
 }
 \email{wufangzhao@gmail.com}
 \author{Yongfeng Huang}
 \affiliation{%
   \institution{Department of Electronic Engineering \& BNRist, Tsinghua University}
     \city{Beijing}
     \postcode{100084}
   \country{China}
 }
 \email{yfhuang@tsinghua.edu.cn}

 \author{Xing Xie}
 \affiliation{%
   \institution{Microsoft Research Asia}
   \city{Beijing}
   \country{China}
   \postcode{100080}
 }
 \email{xingx@microsoft.com}

\renewcommand{\shortauthors}{Wu et al.}

\begin{abstract}
 
Personalized news recommendation is important for users to find  interested news information and alleviate information overload.
Although it has been extensively studied over decades and has achieved notable success in improving user experience, there are still many problems and challenges that need to be further studied.
To help researchers master the advances in personalized news recommendation, in this paper we present a comprehensive overview of personalized news recommendation.
Instead of following the conventional taxonomy of news recommendation methods, in this paper we propose a novel perspective to understand personalized news recommendation based on its core problems and the associated techniques and challenges.
We first review the techniques for tackling each core problem in a personalized news recommender system and the challenges they face.  
Next, we introduce the public datasets and evaluation methods for personalized news recommendation. 
We then discuss the key points on improving the responsibility of personalized news recommender systems. 
Finally, we raise several research directions that are worth investigating in the future.
This paper can provide up-to-date and comprehensive views on personalized news recommendation. 
We hope this paper can facilitate research on personalized news recommendation as well as related fields in natural language processing and data mining. 

\end{abstract}


\maketitle

\section{Introduction}

In the era of the Internet, online news distributing platforms such as Microsoft News\footnote{https://microsoftnews.msn.com} have attracted hundreds of millions of users~\cite{wu2020mind}.
Due to the convenience and timeliness of online news services, many users have shifted their news reading habits from conventional newspapers to digital news content~\cite{okura2017embedding}.
However, a large number of news articles are created and published every day, and it is impossible for users to browse through all available news to seek  their interested news information~\cite{wu2019npa}.
Thus, personalized news recommendation techniques, which aim to select news according to users' personal interest, are critical for news platforms to help users alleviate their information overload of users and improve news reading experience~\cite{li2019survey}.
Researches on personalized news recommendation have also attracted increasing attention from both academia and industry in recent years~\cite{okura2017embedding,wu2019neuralnaml}.

\begin{figure}[!t]
    \centering
    \includegraphics[width=0.8\linewidth]{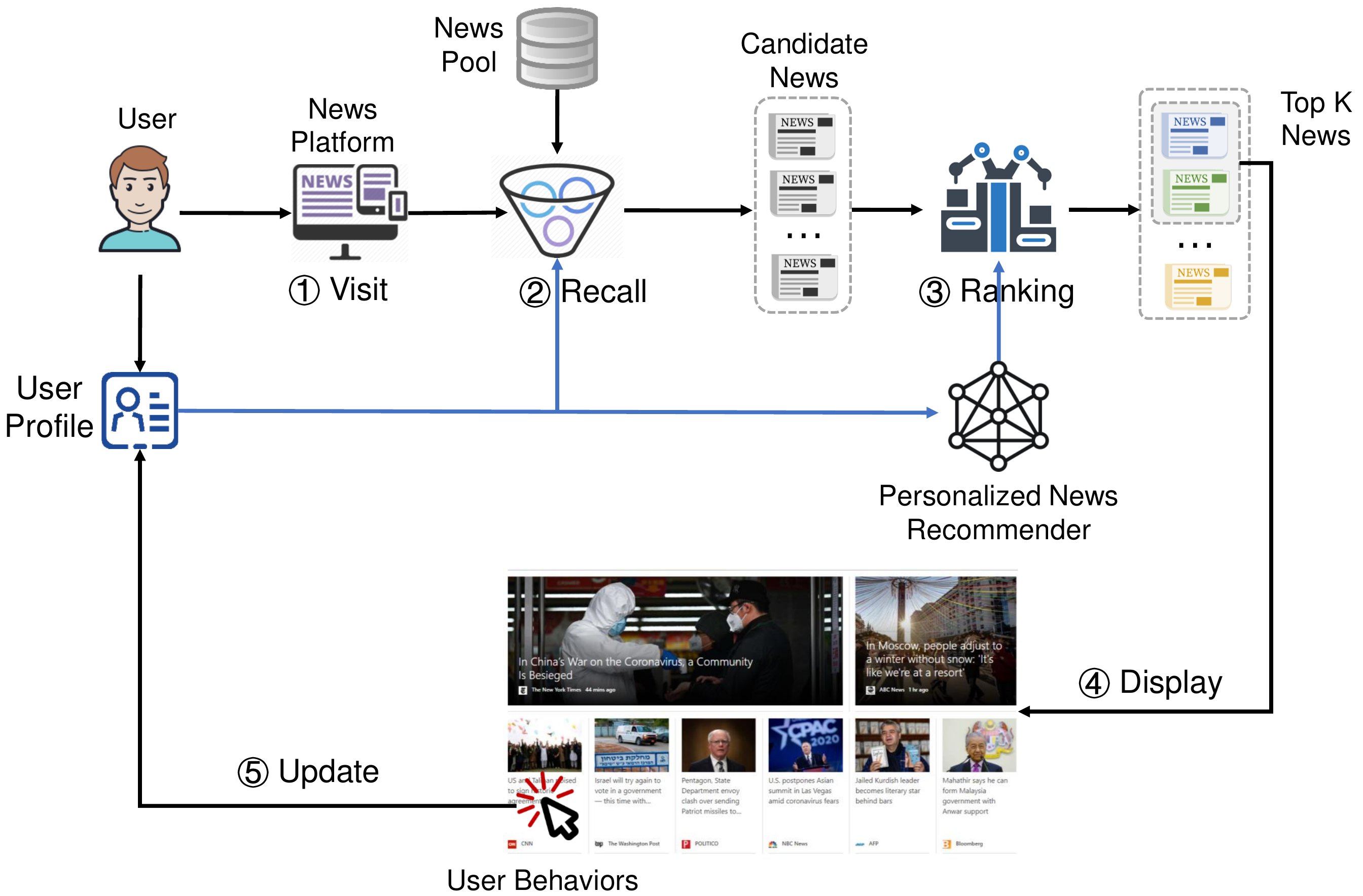}
    \caption{An example workflow of personalized news recommender systems.}
    \label{fig:example}
\end{figure}

An example workflow of personalized news recommender system  is shown in Fig.~\ref{fig:example}.
When a user visits the news platform, the news platform will recall a small set of candidate news from a large-scale news pool, and the personalized news recommender will rank these candidate news articles according to the user interests inferred from user profiles.
Then, the top K ranked news will be displayed to the user, and the user behaviors on these news will be recorded by the platform to update the maintained user profile for providing future services.
Although many prior works have extensively studied these problems in different aspects, personalized news recommendation remains challenging. 
For example, news articles on news websites usually have short life cycles. 
Many new articles emerge every day, and old ones will expire after a short period of time. 
Thus, news recommendation faces a severe cold-start problem.
In addition, news articles usually contain rich textual information such as title and body. 
Thus, it is very important to understand  news content from their texts with advanced natural language processing techniques.
Moreover, there is usually no explicit user feedback such as reviews and ratings on news platforms. 
Thus, we need to infer the personal interests of users from their implicit feedback like clicks. 
However, user interests are usually diverse and dynamic, which poses great challenges to user modeling algorithms. 
The complexity of personalized news recommendation makes it a fascinating research topic with various challenges to be tackled~\cite{feng2020news}.

\begin{table}[t]
\caption{The taxonomy and literature coverage of recent survey papers. Traditional taxonomy means the collaborative, content-based, and hybrid categories. DL stands for deep learning.}\label{survey}
\resizebox{1\textwidth}{!}{
\begin{tabular}{lcll}
\Xhline{1.0pt}
\multicolumn{1}{c}{\textbf{Reference}} & \textbf{Year} & \multicolumn{1}{c}{\textbf{Taxonomy}}                                                                                          & \multicolumn{1}{c}{\textbf{Literature Coverage}} \\ \hline
\cite{ozgobek2014survey}                      & 2014          & Traditional taxonomy                                                                                                           & Feature-based                                    \\
\cite{sood2014survey}                         & 2014          & Traditional taxonomy                                                                                                           & Feature-based                                    \\
\cite{durairaj2014news}                       & 2014          & \begin{tabular}[c]{@{}l@{}}Collaborative, content-based, demographic-based,\\  utility-based, and knowledge-based\end{tabular} & Feature-based                                    \\
\cite{karimi2018news}                         & 2018          & Traditional taxonomy                                                                                                           & Feature-based                                    \\
\cite{li2019survey}                           & 2019          & Traditional taxonomy                                                                                                           & Feature-based                                    \\
\cite{feng2020news}                           & 2020          & Traditional taxonomy and application scenarios                                                                                 & Feature-based \& a few DL-based                  \\
\cite{qin2020research}                        & 2020          & Traditional taxonomy                                                                                                           & Feature-based \& a few DL-based                  \\
\cite{iana2021survey}                         & 2021          & Based on knowledge base structure                                                                                              & Knowledge-based methods                          \\
\cite{raza2021news}                           & 2021          & Traditional taxonomy                                                                                                           & Feature-based \& parts of DL-based               \\ \Xhline{1.0pt}
\end{tabular}
}
\end{table}

A comprehensive overview of existing personalized news recommendation approaches can provide useful guidance for future research in this field.
Over the past years, there are many survey papers that review the techniques of news recommendation~\cite{bogers2007comparing,borges2010survey,li2011personalized,sood2014survey,ozgobek2014survey,durairaj2014news,doychev2014analysis,harandi2015survey,dwivedi2016survey,karimi2018news,li2019survey,feng2020news,qin2020research}.
For example, Li et al.~\cite{li2019survey} reviewed the personalized news recommendation
methods based on handcrafted features to build news and user representations.
They covered many traditional feature-based methods, including collaborative filtering (CF) based ones that use the IDs of users and news, content-based ones that use  features extracted from the content of news and the user behaviors on news, and hybrid ones that rely on content-based collaborative filtering.
They also studied the datasets used by these methods and their techniques for user and news representation construction, data processing and user privacy protection.
Feng et al.~\cite{feng2020news} reviewed news recommendation approaches in many different scenarios including personalized and non-personalized ones.
For personalized news recommendation methods, they also classify them into three categories, i.e., CF-based, content-based, and hybrid.
They mainly studied the techniques adopted by different methods, the challenges they tackled, and the datasets and metrics for evaluation.
We summarize the taxonomy of news recommendation methods and the literature coverage of several recent survey articles in Table~\ref{survey}.
We find that most surveys mainly focus on traditional feature-based methods and only a small part of deep learning-based methods are covered by a few recent surveys, which is not beneficial for researchers to track recent advances in the personalized news recommendation field.
In addition, most surveys follow the canonical taxonomy that categorizes news recommendation methods based on whether they rely on collaborative or content-based filtering techniques.
However, it is difficult for this taxonomy to distinguish between traditional methods and recent deep learning-based approaches.
In addition, the hybrid category contains methods based on quite diverse techniques, e.g., traditional CF-enhanced  content matching and graph neural network, which is not beneficial for researchers to be aware of the evolution of news recommendation technologies.
Moreover, several key techniques in news recommender system design, such as ranking and model training, are rarely discussed in this paradigm. 
Thus, the conventional taxonomy used by most existing surveys cannot meet the development of this field, and a more systematic taxonomy of existing news recommendation methods is needed to help understand their characteristics and inspire further research.

In this paper, we present a comprehensive review of the personalized news recommendation field.
Instead of reviewing existing personalized news recommendation methods based on the conventional taxonomy, in this survey we propose a novel perspective to review them based on the core problems involved in personalized news recommendation and the associated techniques and challenges.
We first introduce the framework of developing a personalized news recommender system in Section~\ref{overview}.
Next, we systematically review the core problems, techniques and challenges in personalized news recommendation, including: news modeling, user modeling, personalized ranking, model training, datasets, benchmarks and evaluation, which are introduced in Sections~\ref{newsmodel}-\ref{evaluation}, respectively.
Through our proposed framework, the characteristics of existing approaches can be more accurately described than using conventional taxonomy, and it is easier for researchers to track the technology evolution in different aspects.
We then present some discussions on developing responsible news recommender systems in Section~\ref{app}, which is an emerging research field in recent years.
Finally, we raise several potential future directions and conclude this paper in Section~\ref{future}.

\section{Framework of Personalized News Recommendation}\label{overview}

Personalized news recommendation techniques have been widely used in many online news websites~\cite{okura2017embedding,wu2020mind}.
Different from non-personalized news recommendation methods that suggest news articles solely based on non-personalized factors~\cite{lavrenko2000language} such as news popularity~\cite{corsini2016clef,yang2016effects,ludmann2017recommending,lommatzsch2018newsreel}, editors' demonstration~\cite{wang2017dynamic} and geographic information~\cite{son2013location,chen2017location}, personalized news recommendation can consider the personal interest of each individual user to provide personalized news services and better satisfy users' need.

\begin{figure}[!t]
    \centering
    \includegraphics[width=0.7\linewidth]{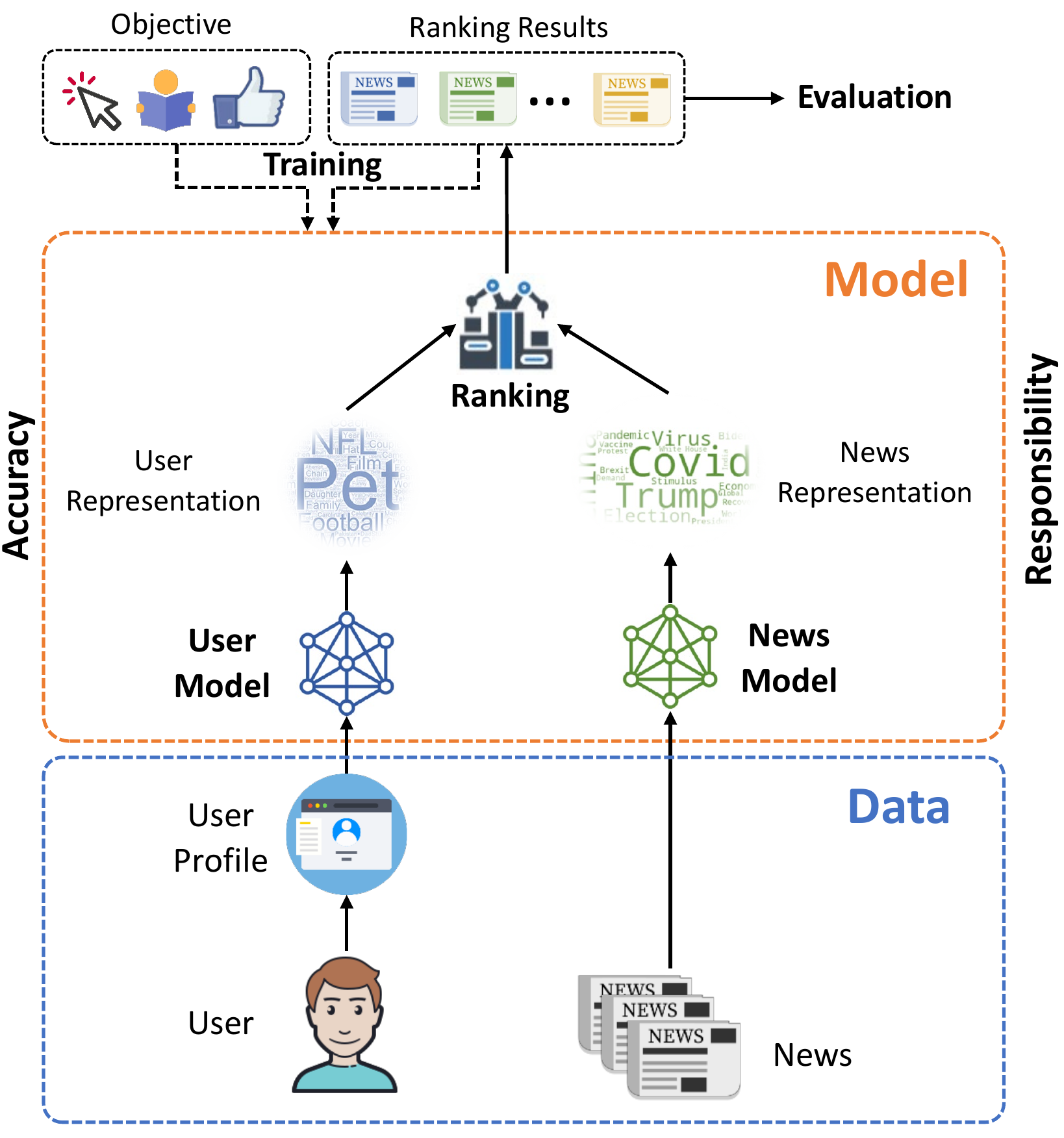}
    \caption{A framework of the key components in developing personalized news recommendation model.}
    \label{fig:framework}
\end{figure}

Existing surveys on personalized news recommendation usually classify methods into three categories, i.e., collaborative filtering-based, content-based and hybrid~\cite{li2019survey}.
However, this taxonomy cannot adapt to the recent advances in news recommendation because many methods with diverse characteristics fall in the same category without distinguishment.
For example, the category of content-based methods includes traditional semantic-based methods, contextual bandit-based methods and recent deep learning-based methods, which is difficult to characterize the paradigm and technical evolution of personalized news recommendation.
Thus, a more systematic overview of existing techniques is required to help understand the development of this field.

Instead of following the conventional taxonomy, in this survey we propose a novel perspective to review existing personalized news recommendation techniques based on the core problems involved in the development of a personalized news recommender system.
A common framework of personalized news recommendation model development is shown in Fig.~\ref{fig:framework}.
We can see that there are several key problems in this framework.
First, news modeling is the backbone of news recommendation and a core problem is how to understand the content and characteristics of news.
In addition, user modeling is required to understand the personal interest of users in news, and it is critical to accurately infer user interest from  user profiles like behaviors.
Based on the news and user representations  built by the news and user models, the next step is ranking candidate news according to certain policies such as the relevance between news and user interest.
Then, it is important to train the recommendation model with proper objectives to make high-quality news recommendations, and evaluating the ranking results given by the recommendation model is also a core problem in the development of personalized recommender systems.
Besides, the datasets and benchmarks for news recommendation are also necessities in designing personalized news recommendation models.
Moreover, beyond developing accurate models, improving the responsibility of intelligent systems has been a spotlight problem in recent years.
How to develop responsible news recommender systems is a less studied but extremely important problem in personalized news recommendation.
Next, we briefly discuss the key problems mentioned above in the following sections.

\subsection{News Modeling}

News modeling aims to understand the characteristics and content of news, which is the backbone of news recommendation.
There are mainly two kinds of techniques for news modeling, i.e., feature-based news modeling and deep learning-based news modeling.
Feature-based news modeling methods usually rely on handcrafted features to represent news articles.
For instance, in many methods based on collaborative filtering (CF), news articles are represented by their IDs~\cite{resnick1994grouplens,das2007google}.
However, on most news websites novel news articles are published continuously and old ones soon vanish.
Thus, representing news articles with their IDs will suffer from severe cold-start problems, and the performance is usually suboptimal.

Considering the drawbacks of ID-based news modeling methods, most approaches incorporate content features to represent news.
Among them, many methods use features extracted from news texts for news modeling.
For instance, Capelle et al.~\cite{capelle2012semantics} proposed to represent news with Synset Frequency-Inverse Document Frequency (SF-IDF), which uses WordNet synonym set to replace the term frequencies in TF-IDF.
Besides the news texts, many methods also explore to incorporate various factors that may have influence on users' news browsing decisions into news modeling, such as news popularity and recency~\cite{li2011scene}.
However, in these methods, the features to represent news are usually manually designed, which usually requires much effort and domain knowledge.
In addition, handcrafted features are usually not optimal in representing the semantic information encoded in news texts.

With the development of natural language processing techniques in recent years, many methods employ neural NLP models to learn deep representations of news.
For example, Okura et al.~\cite{okura2017embedding} proposed to use autoencoders to learn news representations from news content.
Wang et al.~\cite{wang2018dkn} proposed to use a knowledge-aware convolutional neural network (CNN) to learn news representations from news titles and their entities.
Wu et al.~\cite{wu2019neuralnrms} proposed to learn news representations from news titles via a combination of multi-head self-attention and additive attention networks.
Wu et al.~\cite{wu2021empowering} studied to use pre-trained language models to encode news texts.
These deep learning-based news modeling methods can automatically learn informative news representations without heavy effort on manual feature engineering, and they can usually better understand news content than traditional feature-based methods.

\subsection{User Modeling}

User modeling techniques in news recommendation aim to understand users' personal interest in news.
Similar to news modeling, user modeling methods can also be roughly classified into two categories, i.e., feature-based and deep learning-based.
Some feature-based methods like CF represent users with their IDs~\cite{resnick1994grouplens,das2007google}.
However, they usually suffer from the sparsity of user data and cannot model user interest accurately.
Thus, most feature-based methods consider other user information such as click behaviors on news.
For example, Garcin et al.~\cite{garcin2012personalized} proposed to use Latent Dirichlet Allocation (LDA) to extract topics from the concatenation of news title, summary and body.
The topic vectors of all clicked news are further aggregated into a user vector by averaging.
There are also several works that explore to incorporate other user features into user modeling, such as demographics~\cite{lee2007moners}, location~\cite{fortuna2010real} and access patterns~\cite{li2011scene}.
However, feature-based user modeling methods also require an enormous amount of domain knowledge to design informative user features in specific scenarios, and they are usually suboptimal in representing user interests.

There are several methods that use neural networks to learn user representations from users' click behaviors.
For example, Okura et al.~\cite{okura2017embedding} proposed to use a GRU network to learn user representations from clicked news.
Wu et al.~\cite{wu2019npa} proposed a personalized attention network to learn user representations from clicked news in a personalized manner.
Qi et al.~\cite{qi2021hierec} proposed a hierarchical user interest representation method to model the hierarchical structure of user interest.
These methods can automatically learn deep interest representations of users for personalized news recommendation, which are usually more accurate than handcrafted user interest features.

\subsection{Personalized Ranking}

On the basis of news and user interest modeling, the next step is ranking candidate news in a personalized way according to user interest.
Most methods rank news based on their relevance to user interest, and how to accurately measure the relevance between user interest and candidate news is their core problem.
Some methods measure the user-news relevance based on their representations.
For example, Goossen et al.~\cite{goossen2011news} proposed to compute the cosine similarity between the  Concept Frequency-Inverse Document Frequency (CF-IDF) features extracted from candidate news and clicked news, which was further used for personalized candidate news ranking.
Okura et al.~\cite{okura2017embedding} used the inner product between news and user embeddings to compute the click scores,  and ranked candidate news  based on these scores.
Gershman et al.~\cite{gershman2011news} proposed to use an SVM model for each individual user to classify whether this user will click a  candidate news based on news and user interest features.
In several recent methods, the relevance between candidate news and user interest is modeled in a fine-grained way by matching candidate news with clicked news.
For example, Wang et al.~\cite{wang2020fine} proposed to match candidate news and clicked news with a 3-D convolutional neural network to mine the fine-grained relatedness between their content.
However, ranking candidate news and user interest merely based on their relevance may recommend news that are similar to those previously clicked by users~\cite{qi2021hierec}, which may cause the ``filter bubble'' problem.

A few methods use reinforcement learning for personalized ranking.
Li et al.~\cite{li2010contextual} first explore to model the personalized news recommendation task as a contextual bandit problem.
They proposed a LinUCB approach that computes the upper confidence bound (UCB) of each arm  efficiently in closed form based on a linear payoff model, which can match news with users' personal interest and meanwhile explore making diverse recommendations.
DRN~\cite{zheng2018drn} uses a deep reinforcement learning approach to find the interest matching policy that optimizes the long-term reward.
In addition, it uses a Dueling Bandit Gradient Descent (DBGD) method for exploration.
These methods usually optimize the long-term reward rather than the current click probability, which has the potential to alleviate the filter bubble problem by exploring more diverse user interest.

\subsection{Model Training}

Many personalized news recommendation methods employ machine learning models for news modeling, user modeling and interest matching.
How to train these models to make accurate recommendations is a critical problem.
A few methods train their models by predicting the ratings on news given by users.
For example, the Grouplens~\cite{resnick1994grouplens} system is trained by predicting the unknown ratings in the user-news matrix.
However, explicit feedback such as ratings is usually sparse on news platforms.
Thus, most existing methods use implicit feedback like clicks to construct prediction targets for model training.
For example, Wang et al.~\cite{wang2018dkn} formulated the news click prediction problem as a binary classification task, and use crossentropy as the loss function for model training.
Wu et al.~\cite{wu2019npa} proposed to employ negative sampling techniques that combine each positive sample with several negative samples to construct labeled samples for model training.
However, click feedback usually contains heavy noise and may not indicate user interest, which poses great challenges to learning accurate recommendation models.

There are only a few methods that consider user feedback beyond click signals~\cite{wu2020cprs,wu2021feedrec}.
For example, Wu et al.~\cite{wu2020cprs} proposed to model  click preference with click feedback and model reading satisfaction based on the personalized reading speed of users, and train the recommendation model to predict both clicks and user satisfaction.
By optimizing objectives beyond news clicks, these methods are aware of user engagement information and thereby can  better understand user interest.
In addition, these methods have the potential to recommend news articles that are not only clicked by users, but also indeed satisfy their information needs.
Thus, designing engagement-aware training objectives is useful for news recommender systems to provide high-quality news suggestions.

\subsection{Evaluation}

Properly evaluating the performance of personalized news recommendation algorithms is important for developing real-world news recommender systems.
Most existing methods use click-related metrics to measure the accuracy of recommendation results.
Some of them regard the recommendation task as a classification problem~\cite{wang2018dkn,lian2018towards,hu2020graph}, where the performance is evaluated by  classification metrics such as Area Under Curve (AUC) and F1-score.
Many other methods use ranking metrics such as Mean Reciprocal Rank (MRR) and normalized Discounted Cummulative Gain (nDCG).
However, click-based metrics may not indicate user experience.
Thus, a few works explore to use user engagement-based metrics to evaluate the recommendation performance~\cite{wu2021feedrec}, such as dwell time and dislike, which can evaluate the performance of recommendation models more comprehensively.

In most works, the performance of recommendation models is offline evaluated.
However, the data used for offline evaluation is usually influenced by the recommendation results generated by the predecessor recommendation algorithms, and the real user feedback on recommendation results cannot be obtained.
Only a few works reported online evaluation results~\cite{wu2021empowering}, which may better indicate the real performance of the recommender systems. 
To fill the gaps between offline and online experiments, one prior study~\cite{li2011unbiased} proposed an unbiased evaluation method of contextual bandit-based news recommendation methods.
However, there still lacks a general method that can  offline evaluate the potentials of various news recommender algorithms in online environments.

\subsection{Dataset and Benchmark}

Publicly available datasets are important for facilitating researches in the corresponding fields as well as benchmarking their results and findings.
However, in the personalized news recommendation field most researches are conducted on proprietary datasets collected from different news platforms, such as Google News, Microsoft News, Yahoo News, Bing News, etc.
There are only a few datasets that are publicly available for news recommendation research.
Several representative datasets such as plista~\cite{Plista}, Adressa~\cite{gulla2017adressa} and MIND~\cite{wu2020mind} are widely used by recent studies.
The plista dataset is a German news dataset.
A newer version of this dataset is published by the CLEF 2017 NewsREEL~\cite{lommatzsch2017clef} task, and a competition is held based on this data to train and evaluate news recommender systems.
Adressa is a Norwegian dataset that contains not only click information, but also the dwell time of users and rich context information of users and news.
MIND is a large-scale English news recommendation dataset with raw textual information of news.
In addition, MIND is associated with a public leaderboard and an open competition, which can fairly compare the performance of different algorithms.
Thus, many recent researches are conducted on the MIND dataset~\cite{wu2021empowering,wu2021two,wu2021uag}.

\subsection{Responsible News Recommendation}

Most endeavors on personalized news recommendation focus on improving the accuracy of recommendation results.
In recent years, research on the responsibility of machine intelligent systems has gained high attention to help AI techniques better serve humans and avoid their risky and even harmful behaviors that can lead to negative societal impacts and unethical consequences~\cite{dignum2019responsible}.
There are many aspects to improve the responsibility of personalized news recommender systems~\cite{bastian2021safeguarding}.
For example, since many news recommendation methods are learned on private user data, it is important to protect user privacy in recommendation model training and online serving~\cite{qi2021uni}.
Federated learning~\cite{mcmahan2017communication} is a privacy-aware machine learning paradigm, which can empower the construction of privacy-preserving news recommender systems.
Besides optimizing news recommendation accuracy, it is also important to promote the diversity of news recommendation results, which can satisfy users' needs on information variety and alleviate the filter bubble problem~\cite{rao2013taxonomy,wu2020sentirec,raza2021news,hendrickx2021news}.
Moreover, fairness is a critical aspect of responsible news recommendation, since the recommendation models learned on biased user data may inherit unwanted biases, which may lead to the prejudice of algorithms and further unfair recommendation results.
To mitigate the unfairness issue of news recommendation methods, fairness-aware machine learning techniques~\cite{wu2021fairness} can help build inclusive and fair algorithms to provide high-quality news recommendation services to different groups of users.
These research fields on responsible news recommendation emerging in recent years have the potential to improve the quality of news recommender systems to serve users in a more responsible way.
However, there lacks a systematic review on responsible news recommendation in existing survey papers.
In this survey, we first give a comprehensive review on the frontiers of responsible news recommendation research.

Given the overview above, we then present in-depth discussions on each mentioned core problem in the  following sections.
\section{News Modeling}\label{newsmodel}

News modeling is a critical step in personalized news recommendation methods to capture the characteristics of news articles and understand their content.
The techniques for news modeling can be roughly divided into two categories, i.e., feature-based and deep learning-based.
For feature-based methods, news articles are mainly represented by handcrafted features, while deep learning-based methods mainly aim to learn hidden news representations from the raw inputs.
Note that although a few methods may employ some deep learning methods like multi-layer perceptrons to model interactions between sophisticated handcrafted features, we still categorize them into feature-based ones because their news representations are not learned from scratch. 
In addition, some deep learning-based methods may involve some efforts in feature engineering.
Since their news representation methods mainly focus on incorporating additional features to enhance deep representations learned from scratch, we still put them into the deep learning-based category.
The details of the two types of news modeling methods are as introduced follows.

\subsection{Feature-based News Modeling}

Designing informative features to represent news articles is the key problem in feature-based news modeling methods.
As summarized in Fig.~\ref{newsfeat}, there are mainly four types of features used in news modeling, which are introduced as follows.

In many CF-based methods,  news articles are represented by collaborative filtering signals such as news IDs~\cite{resnick1994grouplens,das2007google,7924826,xiao2015time,ji2016regularized,mookiah2018personalized,han2020personalized}.
However, on most news websites novel news are published quickly and old ones will soon vanish.
These methods model news in a content-agnostic manner, which may suffer from the serious cold start problem due to the difficulty in processing newly generated news.
Thus, it is not suitable to simply represent news articles with their IDs~\cite{domann2017highly}.

\begin{figure}[!t]
    \centering
    \includegraphics[width=1.0\linewidth]{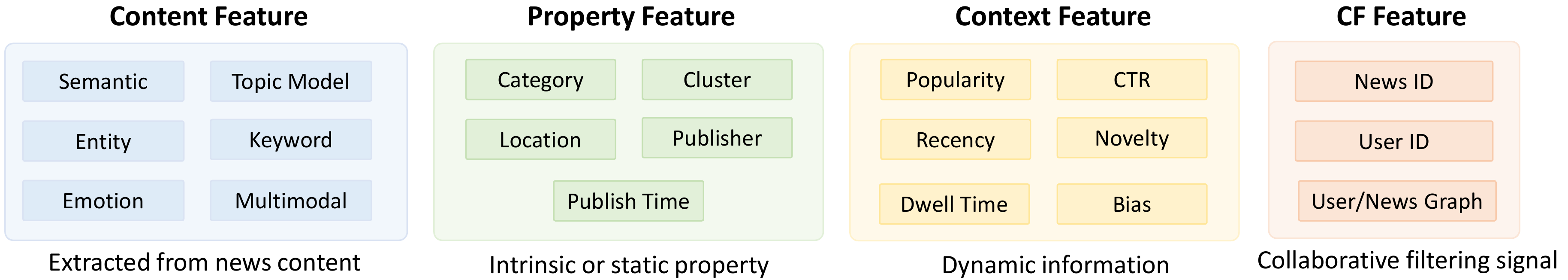}
    \caption{An overview of different types of news features.}
    \label{fig:newsfea}
\end{figure}

Due to the drawbacks of ID-based news modeling, many methods incorporate news content into news modeling.
For instance, 
Gershman et al.~\cite{gershman2011news} considered Term Frequency-Inverse Document Frequency (TF-IDF) features extracted from news texts.
In news articles, entities/concepts are usually more important than other words in understanding news content.
Thus, many methods use the entities/concepts in news texts to represent their  content.
For example, Goossen et al.~\cite{goossen2011news} proposed to use  Concept Frequency-Inverse Document Frequency (CF-IDF) to model news content, which is a variant of TF-IDF that uses the frequency of concepts extracted from WordNet rather than term frequency.
Capelle et al.~\cite{capelle2012semantics} proposed to use Synset Frequency–Inverse Document Frequency (SF-IDF) to model news, which is based on the frequency of synonym sets in WordNet.
SF-IDF is extended by Moerland et al.~\cite{moerland2013semantics} into  SF-IDF+ by additionally considering the relationships of concepts.
They extend the synonym sets of concepts in news by adding other concepts in WordNet that have relationships with the included concepts.
Based on aforementioned approaches, the family of CF-IDF is expanded by a set of later works~\cite{hogenboom2013news,hogenboom2014bing,capelle2015bing,de2018news,brocken2019bing}.

Besides semantic features, some works explore to extract other kinds of content features to enhance  modeling~\cite{lee2007moners,li2011scene,parizi2015emotional}.
For example, Garcin et al.~\cite{garcin2012personalized} proposed to use Latent Dirichlet Allocation (LDA) to extract topics from the concatenation of news title, summary and main content.
Parizi et al.~\cite{parizi2015emotional} proposed to extract emotion features of sentences in news as complementary information of TF-IDF features.
In their method, the emotion is represented by the Ekman model that contains 6 emotion categories.
A variant of this method that uses the sentiment orientation (i.e., positive, neutral and negative) is also developed by Parizi et al.~\cite{parizi2016emonews}.
Beyond news texts, the exploitation of vision-related information such as the videos of news is also studied in~\cite{videorec}. 
These features can provide complementary information to better understand news content.

In addition to content features, many other genres of features are used for news modeling.
They can be roughly divided into two categories, i.e., property features and context features.
Property features such as categories, locations and publishers usually reflect intrinsic properties of news.
The most widely used news property feature is category, since it is an important clue for modeling news content and targeting user interest.
For example, Liu et al.~\cite{liu2010personalized} proposed to represent news using their topic categories.
However, since the category labels of news often need to be manually annotated by editors, in some scenarios news may not have off-the-shelf category labels, 
Thus, several methods explore to cluster news into categories based on their content.
For instance, in the SCENE~\cite{li2011scene} recommender system, news articles are clustered in a hierarchical manner based on their topic features extracted by LDA.
By incorporating the categories or clusters of news into news modeling, the news recommender can be aware of news topics and provide more targeted recommendation services.
Another representative property feature is news location, which is also widely used to provide users with the news related to the locations that they are interested in.
For example, Tavakolifard et al.~\cite{tavakolifard2013tailored} incorporated the geographic information of news to filter news based on their locations.
In addition, since news from different publishers may have differences in their content and topics, the information of news publisher is also considered by several methods to enrich the information for news modeling~\cite{ilievski2013personalized,liang2017clef}.

\begin{table*}[!t]
\begin{center}

\caption{Main features used for news representation. *XF-IDF means TF-IDF and its variants such as CF-IDF and SF-IDF.}  \label{newsfeat}
\resizebox{1\textwidth}{!}{
\begin{tabular}{ll}
\hline
\multicolumn{1}{c}{\textbf{Features for News Modeling}} & \multicolumn{1}{c}{\textbf{References}} \\ \hline
 
BOW/XF-IDF*           &  \begin{tabular}[c]{@{}l@{}}\cite{goossen2011news}\cite{capelle2012semantics}\cite{moerland2013semantics}\cite{capelle2013semantic}\cite{hogenboom2013news}\cite{hogenboom2014bing}\cite{capelle2015bing}\cite{de2018news}\cite{brocken2019bing}\cite{gershman2011news}\cite{chu2009personalized}\cite{lang1995newsweeder}\cite{wen2012hybrid}\cite{nguyen2015semantically}\cite{rao2013personalized}\\
\cite{zhang2017fine}\cite{billsus2000user}\cite{gabrilovich2004newsjunkie}\cite{cantador2009semantic}\cite{gu2014effective}\cite{kirshenbaum2012live}\cite{kompan2010content}\cite{luostarinen2013using}\cite{phelan2009using}\cite{parizi2015emotional}\cite{parizi2016emonews}\cite{liu2021hybrid}\cite{wei2021news}\cite{lu2016hier}\end{tabular}                                       \\ \hline
Entity/Keyword        &  \begin{tabular}[c]{@{}l@{}}\cite{goossen2011news}\cite{capelle2012semantics}\cite{moerland2013semantics}\cite{capelle2013semantic}\cite{hogenboom2013news}\cite{hogenboom2014bing}\cite{capelle2015bing}\cite{de2018news}\cite{brocken2019bing}\cite{li2011scene}\cite{gershman2011news}\cite{tavakolifard2013tailored}\cite{ilievski2013personalized}\cite{lin2014personalized}\cite{mookiah2018personalized}\cite{li2013news}\cite{joseph2019content}
\cite{claypool1999combing}\cite{darvishy2020hypner}\\\cite{rao2013personalized}\cite{zhang2017fine}\cite{billsus2000user}\cite{gabrilovich2004newsjunkie}\cite{zheng2018drn}\cite{caldarelli2016signal}\cite{cantador2009semantic}\cite{cantador2011enhanced}\cite{frasincar2009semantic}\cite{Streamingrec}\cite{kompan2010content}\cite{tran2010user}\cite{Khattar2017leveraging}\cite{wang2020discovery}  \end{tabular}                                         \\ \hline
                              
Cluster/Category         & \begin{tabular}[c]{@{}l@{}}\cite{lee2007moners}\cite{li2011scene}\cite{jonnalagedda2013personalized}\cite{darvishy2020hypner}\cite{epure2017recommending}\cite{chu2009personalized}\cite{liu2010personalized}\cite{saranya2012personalized}\cite{darvishy2020hypner}\cite{sottocornola2018session}\cite{gao2009infoslim}\cite{li2014modeling}\cite{li2011logo}\cite{viana2017hybrid}\cite{li2010contextual}\cite{zheng2018drn}\cite{caldarelli2016signal}\\
\cite{gu2014effective}\cite{jonnalagedda2016incorporating}\cite{kompan2010content}\cite{sood2014preference}\cite{zelenik2011news}\cite{liu2021hybrid}\cite{gharahighehi2020multi}\cite{yang2020double}\cite{symeonidis2021session}\cite{tiwari2022pntrs}\end{tabular}                                            \\ \hline
Topic Distribution     &    \cite{garcin2012personalized}\cite{li2011scene}\cite{noh2014location}\cite{zihayat2019utility}\cite{li2013news}\cite{garcin2013pen}\cite{garcin2013personalized}\cite{saranya2012personalized}\cite{li2014modeling}\cite{li2011logo}\cite{hsieh2016immersive}\cite{li2010user}\cite{luostarinen2013using}\cite{patankar2019bias}\cite{tran2010user}\cite{hsieh2016immersive}   \\ 
Location         & \cite{tavakolifard2013tailored}\cite{noh2014location}\cite{yeung2010proactive}\cite{ilievski2013personalized}\cite{viana2017hybrid}\cite{kazai2016personalised}\cite{wei2021news}                                       \\
Publisher        & \cite{ilievski2013personalized}\cite{liang2017clef}\cite{zheng2018drn}\cite{yang2020double}                                        \\
Popularity       & \cite{7924826}\cite{li2011scene}\cite{gershman2011news}\cite{jonnalagedda2013personalized}\cite{tavakolifard2013tailored}\cite{darvishy2020hypner}\cite{ilievski2013personalized}\cite{zihayat2019utility}\cite{chu2009personalized}\cite{darvishy2020hypner}\cite{li2011logo}\cite{jonnalagedda2016incorporating}\cite{kazai2016personalised}\cite{kirshenbaum2012live}\cite{garcin2013pen}\cite{ilievski2013personalized}                                          \\
CTR              & \cite{chu2009personalized} \\
Recency          & \cite{lee2007moners}\cite{li2011scene}\cite{gershman2011news}\cite{tavakolifard2013tailored}\cite{darvishy2020hypner}\cite{ilievski2013personalized}\cite{zihayat2019utility}\cite{saranya2012personalized}\cite{Khattar2017leveraging}\cite{darvishy2020hypner}\cite{li2011logo}\cite{zheng2018drn}\cite{zelenik2011news}\cite{wei2021news}                                           \\
Novelty          &  \cite{garcin2013personalized}\cite{gabrilovich2004newsjunkie}                                       \\
Dwell Time       & \cite{chen2009hybrid}\cite{gershman2011news}\cite{yi2014beyond}\cite{ilievski2013personalized}\cite{zihayat2019utility}\cite{zheng2018drn}\cite{ilievski2013personalized}                 \\
Time Stamp       & \cite{ilievski2013personalized}\cite{epure2017recommending}\cite{chu2009personalized}\cite{fortuna2010real}\cite{xiao2015time}\cite{yang2020double}              \\
Emotion/Sentiment & \cite{parizi2015emotional}\cite{parizi2016emonews} \\
Bias & \cite{patankar2019bias}\\
Knowledge Graph  &  \cite{joseph2019content}\cite{zhang2017fine}                                        \\
News/User Graph  &  \cite{lin2014personalized}\cite{mookiah2018personalized}\cite{li2013news}\cite{garcin2013personalized}\cite{trevisiol2014cold}\cite{li2014modeling}\cite{li2010user}\cite{phelan2011using}\cite{gharahighehi2020multi}\cite{symeonidis2021session}                                        \\
Ontology         &   \cite{goossen2011news}\cite{capelle2012semantics}\cite{moerland2013semantics}\cite{capelle2013semantic}\cite{hogenboom2013news}\cite{hogenboom2014bing}\cite{capelle2015bing}\cite{de2018news}\cite{brocken2019bing}\cite{wen2012hybrid}\cite{nguyen2015semantically}\cite{rao2013personalized}\cite{shapira2009epaper}\cite{gao2009infoslim}\cite{cantador2009semantic}\cite{cantador2011enhanced}\cite{frasincar2009semantic}                                       \\
Visual Information      & \cite{videorec}\\
\hline
\end{tabular}
}
\end{center}

\end{table*}

Different from property features that are usually static after news publishing, context features of news are  dynamic.
Popularity and recency, which reflect the attractiveness and freshness of news, are two representative context features used by existing methods.
For instance, MONERS~\cite{lee2007moners} is a news recommender system that   represents news articles by news categories, news importance suggested by  providers and the recency of news articles. 
Gershman et al.~\cite{gershman2011news} proposed to use four kinds of features to represent news, i.e., news popularity, news age (recency), TF-IDF features of words and named entities. 
Jonnalagedda et al.~\cite{jonnalagedda2013personalized} proposed to use the timeline on Twitter to enhance news modeling. 
They use the popularity and categories of news on Twitter for news representation.
News recency only considers the time interval between the publishing and display of news, while time stamp of news display can provide finer-grained information, such as seasons, months, days and the time in a day.
Thus, several approaches incorporate the time stamp of news impression~\cite{chu2009personalized,fortuna2010real,ilievski2013personalized,xiao2015time,epure2017recommending}.
For example, Ilievski et al.~\cite{ilievski2013personalized} proposed to incorporate the weekday and the hour of a news impression in news modeling.
In addition to the context features mentioned above, several methods also explore to use weather~\cite{yeung2010proactive}, click-through rate (CTR)~\cite{chu2009personalized}, and fact/opinion bias~\cite{patankar2019bias} to enrich the representations of news. 

Some hybrid methods consider both news IDs and additional features in news modeling~\cite{lommatzsch2014real}.
For example, NewsWeeder~\cite{lang1995newsweeder} represents news articles by their IDs and bag-of-word features.
Claypool et al.~\cite{claypool1999combing} proposed to use news IDs and keywords to model news. 
Liu et al.~\cite{liu2010personalized} proposed to represent news using their IDs and topic categories.
Saranya et al.~\cite{saranya2012personalized} proposed to represent news by their IDs, topics, click frequency and the weights of a news belonging to different categories.
Using the combination of ID-based and content-based news modeling techniques can mitigate the cold-start problem of news to some extent, and have been widely explored by integrating other information like news property features~\cite{wen2012hybrid,darvishy2020hypner}, news sessions~\cite{sottocornola2018session}, ontology~\cite{cantador2011enhanced,nguyen2015semantically,rao2013personalized,shapira2009epaper,gao2009infoslim} and  knowledge graphs~\cite{zhang2017fine}.

To draw a big picture of feature-based news modeling methods, we summarize the major features they used  in Table~\ref{newsfeat}.

\subsection{Deep learning-based News Modeling}

With the development of deep learning techniques, in recent years many methods employ neural networks to automatically learn news representations.
Instead of using handcrafted features like TF-IDF to represent news content, most of them use neural NLP techniques to learn news representations from news texts.
For example, Okura et al.~\cite{okura2017embedding} proposed an embedding-based news recommendation (EBNR) method that uses a variant of denoising autoencoders to learn news representations from news texts.
RA-DSSM~\cite{kumar2017deep} is a neural news recommendation approach which incorporates a similar architecture as DSSM~\cite{huang2013learning}. 
It first builds the representations of news using the doc2vec~\cite{le2014distributed} tool, then uses a two-layer neural network to learn hidden news representations.
This method is also adopted by~\cite{Khattar2017user}.
3-D-CNN~\cite{kumar2017word} represents news by the word2vec~\cite{mikolov2013distributed} embeddings of their words, which is further considered by~\cite{zhang2021research}.
However, it is difficult for these methods to mine the semantic information in news texts with traditional neural NLP models.

Many later approaches use more advanced neural NLP models for text modeling, such as CNN~\cite{wang2020fine,zhang2021learning} and self-attention~\cite{wu2019neuralnrms}.
For instance, WE3CN~\cite{khattar2018weave} uses 2D CNN models to learn representations of news.
NPA~\cite{wu2019npa} uses CNN to generate contextual representations of words in news titles, and use a personalized attention network to form news representations by selecting important words in a personalized manner.
NRMS~\cite{wu2019neuralnrms} learns word representations with a multi-head self-attention network, and uses an additive attention network to form news representations.
Similar news modeling method is also used by many later works~\cite{wu2020cprs,wu2020sentirec,wu2021fairness,wu2021feedrec,wu2021two}.
NRNF~\cite{wu2020ccf} uses self-attention to model the contexts of words in news title and body, and it uses an interactive attention network to model the relatedness between title and body.
A similar co-attention mechanism is used by~\cite{mao2021neuralnews} to model the interactions between news title and abstract.
FedRec~\cite{qi2020privacy} learns news representations from news titles via a combination of CNN and multi-head self-attention networks.
These methods usually learn news representations based on shallow text models and non-contextualized word embeddings such as GloVe~\cite{pennington2014glove}, which may be insufficient to capture the deep semantic information in news.
WG4Rec~\cite{shi2021wg4rec} introduces a word-graph based news text modeling method.
It constructs a word graph based on semantic similarity, co-occurrence and news co-click, and learns word embeddings through a GNN model.
These methods can enhance news text understanding with various neural architectures.
However, these models are still rather shallow and may not be strong enough in capturing the deep semantic information in news texts.

In recent years, big and powerful pre-trained language models (PLMs) such as BERT~\cite{devlin2019bert} have been greatly successful in NLP, and a few recent works explore to empower news modeling with PLMs~\cite{wu2021empowering,xiao2021training,jia2021rmbert,zhang2021amm,zhang2021unbert,raza2021deep,wu2021newsbert,yi2021efficient}.
For example, PLM-NR~\cite{wu2021empowering} uses different PLMs to empower English and multilingual news recommendation, and the  online flight results in Microsoft News showed notable performance improvement.
UNBERT~\cite{zhang2021unbert} incorporates the concatenation of news texts as the input of a BERT model.
The findings in these works imply the effectiveness of large PLMs in empowering  text understanding in news recommendation.

Instead of merely modeling semantic information in news texts, several methods study to use entities or keywords in news texts to enhance news modeling by introducing complementary knowledge and commonsense information.
A direct way is regarding entities as texts and combining them with news text modeling~\cite{ji2021attention}.
For instance, Gao et al.~\cite{gao2018fine} proposed a knowledge-aware news recommendation approach with hierarchical attention networks.
In their method, a word attention network is used to learn word-based  news representations by using the embeddings of keywords as attention queries, and these representations are concatenated with both entity embeddings and the average embeddings of the entities in their contexts. 
An item attention network is used to aggregate these three kinds of news representations by modeling their informativeness.
DAN~\cite{zhu2019dan} learns news representations from news titles and entities via two parallel CNN networks with max pooling operations.
Saskr~\cite{chu2019next} builds news representations from news titles and bodies based on the average word embeddings of their entities.
DNA~\cite{zhang2019dynamic} learns news representations from the news body, news ID and the elements  (entities and keywords).
More specifically, the sentences in a news body are transformed into their embeddings via doc2vec~\cite{le2014distributed}, and then are aggregated into a unified one via a sentence-level candidate-aware attention network.
Each news element is represented by averaging the embeddings of its words, and elements representations are synthesized together via an element-level candidate-aware attention network.
The embeddings of the ID, texts, and elements of each piece of news are concatenated together into a unified news representation.
HieRec~\cite{qi2021hierec} uses text self-attention and entity self-attention to model the contexts in news titles and the relations between entities in news texts, respectively.
These methods can easily unify the use of texts and knowledge entities, but they cannot effectively exploit the relatedness between entities.

Another way to exploit entity information is incorporating knowledge graph embeddings~\cite{wang2018dkn,liu2019news,sheu2020context,tran2021deep,sun2021hybrid}.
For example, DKN~\cite{wang2018dkn} learns news representations from the titles of news and the entities within titles via a knowledge-aware CNN.
The representations of entities are learned from a knowledge graph using the TransD~\cite{ji2015knowledge} knowledge graph embedding algorithm.
Liu et al.~\cite{liu2019news} proposed to construct a news-relevant knowledge graph on the basis of the Microsoft Satori knowledge graph by extracting additional knowledge entities and topic entities from news and connecting entities in the same news, entities clicked by the same user and entities appearing in the same browsing session to enrich the relations between entities in the knowledge graph.
They combine the entity embeddings learned by TransE~\cite{bordes2013translating} with the news text embeddings learned by LDA and DSSM.
CAGE~\cite{sheu2020context,sheu2021knowledge} constructs subgraphs of KG by using one-hop neighbors of entities, and uses the TransE embeddings of entities as complements to text embeddings learned by CNN.
However, these knowledge graph embeddings mainly condense low-level interactions between entities.
To enhance the modeling of rich entity relatedness, TEKGR~\cite{lee2020news} enriches the knowledge graph with topical relations between entities.
It predicts the topic of news based on texts and concepts, and uses the predicted topic to enrich the knowledge graph and learn topic enriched knowledge representations of news with graph neural networks.
KRED~\cite{liu2020kred} first learns entity embeddings from knowledge graph with graph attention networks, then incorporates additional entity features  such as frequency, category and position, and finally selects entities according to the texts representations of news. 
KIM~\cite{qi2021kim} incorporates a knowledge-aware interactive news modeling method that can model the relations between the entities and their neighbors of clicked news and candidate news through graph co-attention networks.
KOPRA~\cite{tian2021joint} only uses the TransE embeddings of knowledge entities to represent news, and it uses a recurrent graph convolution network to learn hidden entity representations.
These methods can encode richer knowledge information of news than pure text-based methods to empower news recommendation.

To better model the characteristics of news articles,  several methods explore to incorporate other types of news information beyond texts into news modeling.
Among them, topic categories and tags are widely considered by existing methods~\cite{park2017deep,zhang2018deep,wu2019neural,wu2019neuralnaml,han2021neural}.
For example, DeepJoNN~\cite{zhang2018deep} learns news representations from news IDs, categories, keywords and entities via a character-level CNN.
Park et al.~\cite{park2017deep} proposed a neural news recommendation method based on LSTM.
They use a proprietary corpus to train a doc2vec~\cite{le2014distributed} model to encode news articles into their vector representations, and use an LSTM network to generate user representations from the representations of news.
In addition, they incorporate the categories of news into news representations, which are predicted by a CNN~\cite{kim2014convolutional} model.
TANR~\cite{wu2019neural} learns news representation from news titles via a combination of CNN and attention network, which is also used in~\cite{wu2019neuralnrhub,yi2021debiasedrec}.
Moreover, TANR incorporates an auxiliary news topic prediction task to learn topic-aware news representations. 
NAML~\cite{wu2019neuralnaml} is a news recommendation method with attentive multi-view learning, which incorporates different kinds of news information as different views of news.
In this method, news titles, bodies, categories and subcategories are processed by different models, and their embeddings are further aggregated together into a unified one via a view-level attention network.
A similar method is also used by~\cite{zhang2021combining,wu2021uag} to model candidate news.
LSTUR~\cite{an2019neural} uses a combination of CNN and attention network to process news titles, and incorporates categories and subcategories by applying a non-linear transformation to their embeddings.
CHAMELEON~\cite{gabriel2019contextual,de2018newssession} learns news representations from news bodies by using CNN with different kernel sizes, and these textual representations are fused with news metadata features such as topics, categories and tags using a fully connected layer.
It also predicts the metadata features of news via auxiliary tasks.

In addition to topical information, several methods consider other types of content information of news.
For example, SentiRec~\cite{wu2020sentirec} considers the sentiment orientation of news to  learn sentiment-aware news representations.
It uses the VADER~\cite{hutto2014vader} algorithm to compute real-valued sentiment scores of news.
MM-Rec~\cite{wu2021mm} uses a visiolinguistic model ViLBERT~\cite{lu2019vilbert} to learn news multi-modal representations from both news texts and images.
IMRec~\cite{xun2021we} models the rich visual impression information of news such as texts, image regions, the arrangement of different fields, and spatial positions of different words on the impression.
These methods can usually learn more accurate news representations by characterizing their content in multiple aspects.

Another major group of additional information is context features, such as popularity and positions.
For example, PP-Rec~\cite{qi2021pprec} uses both news title, entities and news popularity information in news modeling. 
It uses gating mechanisms to synthesize the near-real-time CTR, recency and popularity predicted from news title into a unified news popularity score.
TSHGNN~\cite{ji2021temporal} incorporates the active time of users on a news page into the modeling of news texts.
DCAN~\cite{meng2021dcan} uses the time from news publishing and near-real-time CTR to model the current positions of news articles in their lifecycles.  
CTX~\cite{cho2021overlooked} studies the exploitation of CTR, Popularity, and Freshness features.
The results show that these context features may even have a stronger impact than personalized interest signals on click prediction.
DebiasRec~\cite{yi2021debiasedrec} uses CNN and attention network to learn news content representations from news titles, and learns news bias representations from the  size and positions of news displayed on websites with a bias model.
These methods can usually better understand users' interaction patterns with news by incorporating additional context information.
However, some news features (e.g., near-real-time CTR) may not be available in some real-world news recommender systems, which hinders the exploitation of these features.

\begin{table*}[!t]
\begin{center}
 
\caption{Comparison of different methods on news modeling. }\label{deeplearning}
\resizebox{0.98\textwidth}{!}{
\begin{tabular}{l|c|l|l}
\Xhline{1.0pt}
\multicolumn{1}{c|}{\textbf{Method}} & \multicolumn{1}{c|}{\textbf{Year}} & \multicolumn{1}{c|}{\textbf{Information Used}} & \multicolumn{1}{c}{\textbf{Model}} \\ \hline
EBNR~\cite{okura2017embedding}      &  2017    & Body                                           & Autoencoder                        \\
RA-DSSM~\cite{kumar2017deep}       &  2017    & Title+Body                                     & Doc2vec+NN                         \\
Khattar et al.~\cite{Khattar2017user} &  2017 & Title+Body                                     & Doc2vec+NN                        \\
3-D-CNN~\cite{kumar2017word} &    2017        & Title+Body                                     & Word2vec                           \\
WE3CN~\cite{khattar2018weave} &  2018         & Title+Body                                     & 2-D CNN                            \\
NPA~\cite{wu2019npa} &         2019           & Title                                          & CNN+Personalized Attention               \\
NRMS~\cite{wu2019neuralnrms} &    2019        & Title                                          & Self-Attention+Attention                    \\
NRHUB~\cite{wu2019neuralnrhub} &   2019       & Title                                          & CNN+Attention                            \\
DAINN~\cite{zhang2019dynamicattention} & 2019 & Body                                           & CNN+Dynamic Topic Model            \\
FIM~\cite{wang2020fine} &  2020  & Title                                          &  Dilated CNN                 \\
NRNF~\cite{wu2020ccf} & 2020  & Title                                          & Transformer+Attention                    \\
FedRec~\cite{qi2020privacy} &  2020       & Title                                          & CNN+Self-Attention+Attention                \\ 
CPRS~\cite{wu2020cprs} & 2020  &   Title+Body   &    Self-Attention+Attention  \\
UniRec~\cite{wu2021two} &    2021        & Title                                          & Self-Attention+Attention                    \\
FeedRec~\cite{wu2021feedrec} &  2021          & Title                                          & Transformer+Attention                    \\
FairRec~\cite{wu2021fairness} & 2021  & Title                                          & Transformer+Attention                    \\   
EEG~\cite{zhang2021combining} &  2021          & Title+Abstract+Body                & CNN+Attention                            \\
AMM~\cite{zhang2021amm} & 2021 & Title+Abstract+Body                                          & PLM                \\ 
RMBERT~\cite{jia2021rmbert} & 2021 & Title                                          & PLM                \\ 
UNBERT~\cite{zhang2021unbert} & 2021 & Title                                          & PLM+Attention               \\ 
PLM-NR~\cite{wu2021empowering} & 2021 & Title                                          & PLM+Attention               \\ 
SFI~\cite{zhang2021learning} &2021   & Title                                          & CNN+Attention               \\ 
TempRec~\cite{wu2021can} & 2021 & Title                                          & Transformer               \\ 
WG4Rec~\cite{shi2021wg4rec} & 2021 & Title+Word Graph                                          &   GNN+Attention             \\ 
CNE-SUE~\cite{mao2021neuralnews} & 2021 & Title+Abstract                                         &   LSTM+Self-Attention+Co-Attention             \\ 
\hline
DKN~\cite{wang2018dkn} &    2018                & Title+Entity                                   & KCNN                               \\
Gao et al.~\cite{gao2018fine} &2018& Body+Entity                                    & Attention   \\                        DAN~\cite{zhu2019dan} &       2019              & Title+Entity                                   & CNN                                \\
DNA~\cite{zhang2019dynamic} & 2019              & Body+Element+ID                                & Doc2vec+Candidate-Aware Attention+ID Embedding          \\
Saskr~\cite{chu2019next} &    2019              & Entity                                         & Entity Embedding                   \\ 
Liu et al.~\cite{liu2019news} & 2019  & Title+Entity                                    & Entity Embedding+Attention                               \\
TEKGR~\cite{lee2020news} &  2020 & Title+Entity                                    & Entity Embedding+Candidate-aware Attention                               \\
CAGE~\cite{sheu2020context} &   2020 & Title+Entity                                         & CNN+Entity Embedding    \\
KRED~\cite{liu2020kred} &    2020& Title+Entity+Entity Context Feature  &  Attention \\
HieRec~\cite{qi2021hierec} & 2021  & Title+Entity & Transformer+Attention \\  
KIM~\cite{qi2021kim} & 2021  & Title+Entity & CNN+Transformer+Co-Attention+Graph Co-Attention \\
KOPRA~\cite{tian2021joint} &  2021 & Entity & TransE+Recurrent Graph Convolution \\

\hline
Park et al.~\cite{park2017deep} &  2017         & Title+Body+Query+Category                      & Doc2vec                            \\
DeepJoNN~\cite{zhang2018deep} &    2018         & Keywords/Entities+Category+ID                  & Char CNN                           \\
TANR~\cite{wu2019neural} &       2019           & Title+Category                                 & CNN+Attention+Topic Prediction           \\
LSTUR~\cite{an2019neural}  &      2019          & Title+Category+Subcategory                     & CNN+Attention                            \\
NAML~\cite{wu2019neuralnaml}  &    2019         & Title+Body+Category+Subcategory                & CNN+Attention                            \\
CHAMELEON~\cite{gabriel2019contextual} &  2019  & Body+Metadata+Context Features                 & CNN+Attribute Prediction           \\ 
SentiRec~\cite{wu2020sentirec} & 2020  & Title+Sentiment & Self-Attention   \\
PP-Rec~\cite{qi2021pprec} & 2021  & Title+Entity+CTR+Recency  &  Self-Attention+Co-Attention+Gating \\
CTX~\cite{cho2021overlooked} &  2021  & Title+CTR+Popularity+Freshness  &  Add on Existing Methods \\
MM-Rec~\cite{wu2021mm} &  2021 & Title+Image & 
ViLBERT  \\ 
IMRec~\cite{xun2021we}  & 2021 & Title+category+Image+Spatial Position & 
Pre-trained CNN+Memory Network+Self-Attention  \\ 
DebiasRec~\cite{yi2021debiasedrec} &  2021 & Title+Position+Size & CNN+Attention+Bias Embedding   \\
User-as-Graph~\cite{wu2021uag} &  2021                & Title+Category+Subcategory+Entity                                & Transformer+Attention           \\
AGNN~\cite{ji2021attention} & 2021  & Title+Entity &  CNN \\
TSHGNN~\cite{ji2021temporal} & 2021  & Title+Entity+Active Time &  CNN \\
CUPMAR~\cite{tran2021deep} & 2021   & Title+Entity+Body+Category+Subcategory & Self-Attention+Attention   \\          
KG-LSTUP~\cite{sun2021hybrid} & 2021  & Title+Entity+Abstract+Category+Subcategory & LSTM+CNN+Attention   \\  
DCAN~\cite{meng2021dcan} &  2021 & Title+Body+Entity+Time from Publishing+CTR & CNN   \\  
D2NN~\cite{raza2021deep} & 2021  & Title+Abstract+Category+Subcategory & BERT+CNN+Attention   \\  
EENR~\cite{han2021neural} & 2021 & Title+Event+Category+Event Type Graph & LSTM+Attention+Node2Vector  \\  
\hline
IGNN~\cite{qian2019interaction} &  2019   & Title+Entity+User-News Graph                   & KCNN+GNN                           \\
INNR~\cite{REN2019113115} &        2019         & Heterogeneous Graph                            & Node2vec                           \\
GNewsRec~\cite{hu2020graph} &    2020           & Title+Entity+Heterogeneous Graph               & CNN+GNN                            \\
GERL~\cite{ge2020graph} &      2020             & Title+Category+User-News Graph                 & Transformer+GAT                    \\
MVL~\cite{santosh2020mvl} &  2020 & Title+Body+Category+User-News Graph & CNN+Attention+GAT \\
GNUD~\cite{hu2020graph2} & 2020  & Title+Entity+User-News Graph &  CNN+Disentangled GCN \\ 
GBAN~\cite{ma2021graph} &  2021 & Keywords+Tag+Category+Heterogeneous Graph & CNN+DeepWalk \\
\Xhline{1.0pt}
\end{tabular}
}

\end{center}

\centering

\end{table*}

There are a few methods that learn news representations from graphs.
For example, IGNN~\cite{qian2019interaction} uses KCNN~\cite{wang2018dkn} to learn text-based news representations from news titles, and learn graph-based news representations from the user-news graph.
GERL~\cite{ge2020graph} learns news title representations with a combination of multi-head self-attention and additive attention networks, and combines title representations with the embeddings of news categories.
MVL~\cite{santosh2020mvl} uses a content view to incorporate news title, body and category, and uses a graph view to enhance news representations with their neighbors on the user-news graph. 
In addition, it uses a graph attention network to enhance representations of news by incorporating the information of their first- and second-order neighbors on the user-news graph.
GNUD~\cite{hu2020graph2} uses the same news encoder as DAN to learn text-based news representations, and uses a graph convolution network (GCN) with a preference disentanglement regularization to learn disentangled news representations on user-news graphs.
In addition to the user-news graphs used by the above methods, a few methods incorporate heterogeneous graphs that condense richer collaborative information~\cite{REN2019113115,hu2020graph,ma2021graph}. 
For example, GNewsRec~\cite{hu2020graph} is a hybrid approach which considers graph information of users and news as well as news topic categories.
It also uses the same architecture with DAN to learn text-based news representations, and uses a two-layer graph neural network (GNN) to learn graph-based news representations from a heterogeneous user-news-topic graph.
These methods can exploit the high-order information on  graphs to enhance news modeling. 
However, it is difficult for these methods to handle newly generated news with few connections to existing nodes on the old graph used for training.

To help better understand the relatedness and differences between the methods reviewed above, we summarize the information and models they used for learning news representations in Table~\ref{deeplearning}.
Next, we provide several discussions on the aforementioned methods for news modeling.

\subsection{Discussions on News Modeling}

\subsubsection{Feature-based News Modeling}

In feature-based news modeling methods, mining textual information of news is critical for representing news content.
Many methods incorporate BOW/TF-IDF features or their variants to represent news texts, which are also popular in the NLP field.
In addition, topic models like LDA are employed by various methods to extract topics from texts.
This is probably because topic models are capable of mining the topic distributions of news articles and can also provide useful clues for inferring user interest on different topics.
Moreover, since users may focus more on the entities or keywords in news, they are considered by many methods to summarize the content and topic of news, and can also be useful links to find similar news or map news on knowledge graphs.
Especially, some methods also use ontology such as Wikipedia to extract entity features to represent them more accurately.

Besides the texts of news, many methods utilize other information of news.
For instance, the categories or clusters of news are popular news features to help model news content. 
In addition, several dynamic features of news are also widely employed in feature-based news modeling methods, such as popularity and recency.
Since many users may pay more attention to popular events and news usually vanish quickly, incorporating news popularity and recency can help build more informative news representations.
Besides, several environmental factors, such as locations and time are also utilized by several methods.
This is because considering locations of news can provide news related to  users' neighbors, and using the timestamps of news may be useful for providing time-aware news services.

A few methods also study incorporating other interesting features.
For example, the sentiment information of news is useful for news understanding, because users may have different tastes on the sentiment of news.
The bias of news may also need to be taken into consideration, because recommending news with biased opinions and facts may hurt user experience and the reputation of news platforms.
Finally, although several non-personalized news recommendation methods have used news images to build news representations~\cite{lommatzsch2018newsreel}, few personalized ones consider the visual information of news, which is very useful for news modeling.

Although feature-based news modeling methods have  comprehensive coverage of various news information, they usually require a large amount of domain knowledge for feature design.
In addition, handcrafted features are usually not optimal in representing the textual content of news due to the absence of the contexts and orders of words.

\subsubsection{Deep Learning-based News Modeling}

Among all the reviewed methods, only two methods, i.e., DNA~\cite{zhang2019dynamic} and DeepJoNN~\cite{zhang2018deep}, directly incorporate the embeddings of news IDs.
This is probably because of the short lifecycle of news articles and the quick generation of novel news, which make the coverage of news IDs in the training set very limited.
Thus, it is very important to understand news from their content.

News text modeling is critical for news understanding.
Most methods use news titles to model news  since news titles, because news titles usually have decisive influence on users' click behaviors.
Several methods such as EBNR~\cite{okura2017embedding}, NAML~\cite{wu2019neuralnaml} and CPRS~\cite{wu2020cprs} use news bodies to enhance news representations, since news bodies are contain more detailed information of news.
In existing methods, CNN is the most frequently used architecture for text modeling.
This is because local contexts in news articles are important for modeling news content, and CNN is effective and efficient in capturing local contexts.
In addition, since different news information may have different informativeness in modeling news content and user interest, attention mechanisms are also widely used to build news representations by selecting important features.
With the success of Transformer in NLP, many methods also use Transformer-like architectures for news modeling, such as NRMS~\cite{wu2019neuralnrms} and CPRS~\cite{wu2020cprs}.
In addition, a few methods use pre-trained language or and visiolinguistic models to empower news modeling~\cite{wu2021empowering,wu2021mm}.
These advanced NLP techniques can greatly improve news content understanding, which is very important for personalized news recommendation.
However, these methods mainly aim to capture the semantic information of news and may not be aware of the knowledge and commonsense information encoded in news.

To address this issue, many methods incorporate news entities into news modeling to learn knowledge-aware news representations~\cite{liu2021reinforced}.
Some methods such as DAN~\cite{zhu2019dan} directly use entity texts to represent entities, while several other methods like DKN~\cite{wang2018dkn} use knowledge graph embeddings to represent entities.
These entity representations are usually combined with representations learned from news texts to better model news content.
However, there are many new entities and concepts emerging in news and it may be difficult to accurately represent them with off-the-shelf knowledge bases.

Several methods incorporate the topic categories of news into news modeling, because news topics are very useful  for understanding news content and inferring user interest.
Considering the scenarios that some news articles are not labeled with topic categories, some methods  such as TANR~\cite{wu2019neural} and CHAMELEON~\cite{gabriel2019contextual} also adopt auxiliary tasks  by predicting news topic categories to encode topic information into news representations.
In addition, a few methods study using other kinds of news features such as sentiment~\cite{wu2020sentirec}, popularity~\cite{cho2021overlooked}, recency~\cite{qi2021pprec}, which can help better understand the characteristics of news.
However, some additional news features (e.g., category and CTR) may be unavailable in certain scenarios, which limits the application of these methods.

There are also a few methods that explore to enhance news modeling with graph information~\cite{ge2020graph,hu2020graph}.
These methods can incorporate the high-order information on user-news bipartite graphs~\cite{qian2019interaction,ge2020graph,hu2020graph2,santosh2020mvl} or more complicated heterogeneous graphs~\cite{REN2019113115,hu2020graph}, which can provide useful contexts on understanding the characteristics of news for news recommendation.
However, since the graphs used in these methods are static, they may have some difficulties in accurately representing newly published news.

In summary, by reviewing news modeling techniques used in existing news recommendation methods, we can see that news modeling is still a quite challenging problem in news recommendation due to the variety, dynamic, and timeliness of online news information.
\section{User Modeling}\label{usermodel}

User modeling is also a critical step in personalized news recommender systems to infer users' personal interests in news.
It is usually important for user modeling algorithms to understand users from their behaviors~\cite{wu2019neural}.
An example user modeling framework in personalized news recommendation is shown in Fig.~\ref{usermodel}.
We can see that user modeling is based on the modeling of news that users have interactions with, and it introduces additional user features to achieve better personalized user understanding.
The techniques for user modeling in existing news recommendation methods can also be classified into feature-based ones and deep learning-based ones.
Feature-based user modeling techniques mainly rely on manually designed user modeling rules or heuristic patterns to represent user interest.
By contrast, deep learning-based methods usually focus on automatically finding useful patterns from user behaviors to infer user interest.
The details of the two kinds of user modeling methods are introduced in the following sections.

\begin{figure}[!t]
    \centering
    \includegraphics[width=0.6\linewidth]{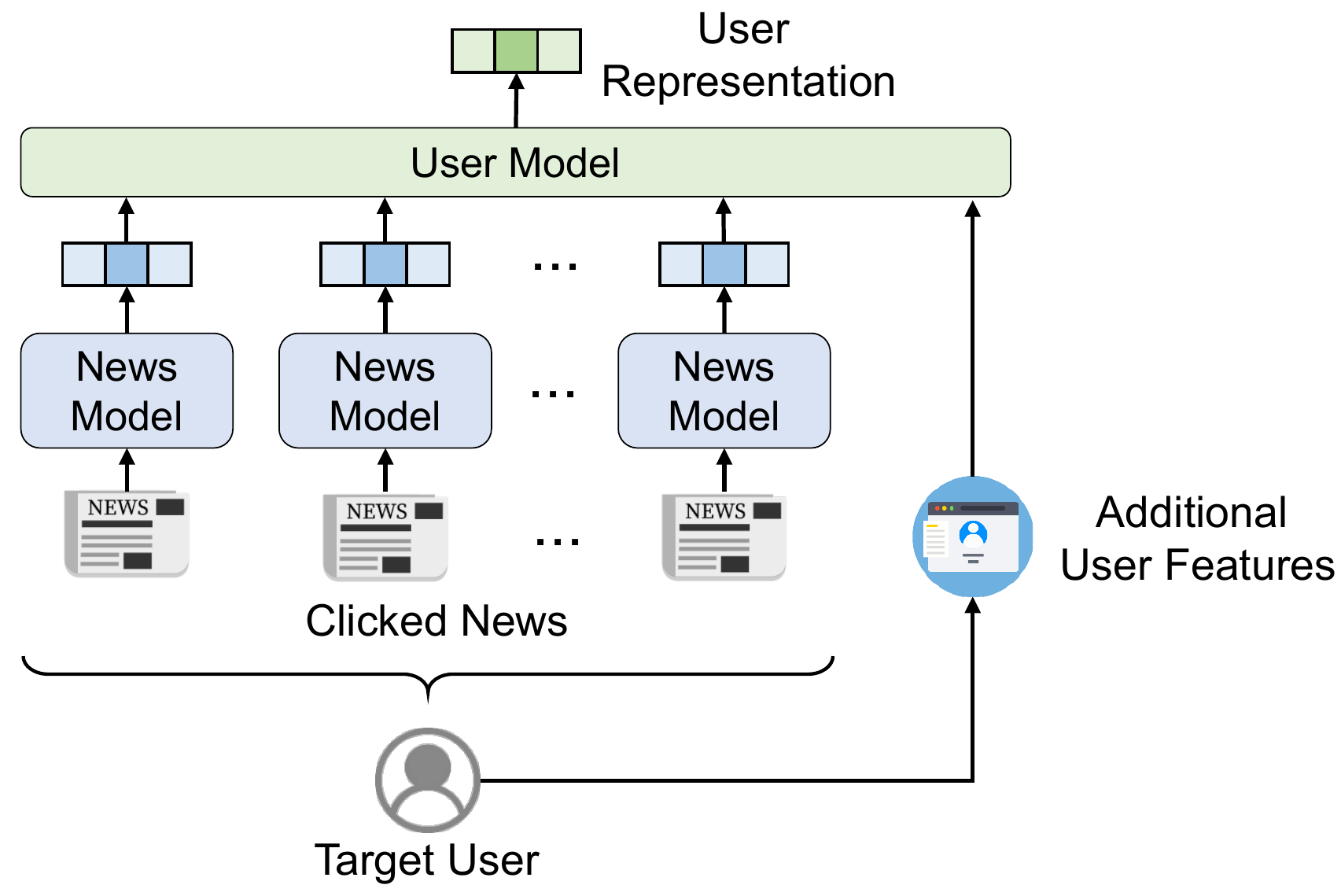}
    \caption{An example framework of user modeling.}
    \label{fig:usermodel}
\end{figure}

\subsection{Feature-based User Modeling}

Feature-based user modeling methods use  handcrafted features to represent users.
Similar to news modeling, in CF-based methods  users are also represented by their IDs~\cite{resnick1994grouplens,das2007google}.
However, ID-based user modeling methods usually suffer from the data sparsity.
Thus, most methods consider the behaviors of users such as news clicks to model their interest.
An intuitive way is to use the features of clicked news to build user features.
For example, Goossen et al.~\cite{goossen2011news} used the CF-IDF features of clicked news to represent user interest.
Capelle et al.~\cite{capelle2012semantics} proposed to use the SF-IDF features of clicked news for user modeling.
Garcin et al.~\cite{garcin2012personalized} proposed to model users by aggregating the LDA features of all clicked news into a user vector by averaging.
However, it is difficult for these methods to model users accurately when their news click behaviors are sparse.

Besides news features, many methods consider other supplementary information of users in user modeling.
For instance, in the MONERS~\cite{lee2007moners} recommender system, users are clustered into segments, and the preferences of user segments on news categories and news articles are used to represent users.
In addition, the demographics of users, such as age, gender and profession, are also useful information for user modeling because users in different demographic groups usually have different preferences on news.
Thus, user demographic features are incorporated by several methods~\cite{lee2007moners,yeung2010proactive,ilievski2013personalized}.
For instance, Yeung et al.~\cite{yeung2010proactive} proposed to use the age, gender, occupation status and social economic grade of users to help identify their different preferences on news in different categories.
Chu et al.~\cite{chu2009personalized} used the age and gender categories of users to model their characteristics.
Besides, the location information of users is also very useful for accurate user modeling, and it has been used by several location-aware news recommendation methods~\cite{fortuna2010real,noh2014location}.
However, some kinds of user features such as locations and demographics are privacy-sensitive, and many users may not provide their accurate personal information.

Since news clicks may not necessarily indicate user interests, several methods also consider other kinds of user behaviors or feedback.
For example, Gershman et al.~\cite{gershman2011news} proposed to represent users by the news they carefully read (regarded as positive news), rejected, and scrolled (both are regarded as negative news).
In addition, users' dwell time on clicked news is also an important indication of user interest, and Yi et al.~\cite{yi2014beyond} studied to use  dwell time as the weights of clicked news for user modeling.
Besides these user behaviors, several other kinds of user behavior information such as access patterns, are utilized by a few methods~\cite{li2011scene,saranya2012personalized} to capture the users' habits on news reading.

Several methods also consider graph information (e.g., news-user graphs) in user modeling~\cite{gharahighehi2020multi}.
For example, Li et al.~\cite{li2013news}  proposed a news personalization method by using hypergraph to model various high-order interactions between different news information, where users are represented by subgraphs of the hypergraph.
Garcin et al.~\cite{garcin2013personalized} proposed to use context trees for user modeling.
They constructed context trees based on the sequence of articles, the sequence of topics and the distribution of topics. 
Trevisiol et al.~\cite{trevisiol2014cold} proposed to build a browsing graph from the news browsing histories of users on Yahoo News.
Joseph et al.~\cite{joseph2019content} proposed to represent users by regarding the clicked news as subgraphs of a knowledge graph, which are constructed via entity linking.
These methods can consider the high-order information on graphs to help understand user behaviors, which can improve user modeling.

A few methods combine user IDs with other user features in user modeling~\cite{lommatzsch2014real}.
For example, NewsWeeder~\cite{lang1995newsweeder} used user IDs and the bag-of-words features of clicked news to represent users.
Claypool et al.~\cite{claypool1999combing} used  user IDs and keywords of clicked news for user modeling
Liu et al.~\cite{liu2010personalized} proposed to represent users using their IDs and user interest features predicted by a Bayesian model.
These methods can mitigate the drawbacks of ID-based user modeling and meanwhile incorporate useful personal information encoded by user IDs.

Considering the evolutionary characteristics of user interest, some methods model both long-term and short-term user interests~\cite{cantador2011enhanced,li2014modeling}. 
NewsDude~\cite{billsus2000user} may be one of the earliest methods that consider long short-term user interests.
In this approach, users are represented by a hybrid model, which models short-term interest of users based on recently browsed news, and models long-term user interest by sorting words of news in each category with respect to their TF-IDF values and selecting the top ranked words.
Li et al.~\cite{li2011logo} proposed LOGO, which is a news recommendation method that models both long-term and short-term user interests.
LOGO uses a weighted summation of the topic distributions of news clicked by users to indicate long-term user interest, and it uses
the topic distribution of the latest clicked news as the short-term user interest.
Viana et al.~\cite{viana2017hybrid} proposed another news recommendation method based on long short-term user interest. 
In their method, the long-term interest of users is represented by  the frequency of a specific tag being read by this user, and short-term interest is represented by several recently clicked news.
Different from other methods that only consider short-term or long-term user interests, these methods can better model the evolution of user interests by capturing long short-term user interests.

To help readers better understand  feature-based user modeling methods in personalized news recommender systems, we summarize the additional user features (ID and news features are excluded) used in these methods in Table~\ref{userfeat}.

\begin{table*}[!t]

\begin{center}

  \caption{Additional features used for user representation. *ID/textual features of clicked news are excluded because they are incorporated by most methods.}\label{userfeat}  
 \resizebox{0.8\linewidth}{!}{
\begin{tabular}{ll}
\hline
\multicolumn{1}{c}{\textbf{Features for User Representation*}} & \multicolumn{1}{c}{\textbf{References}} \\ \hline
 
Demographic           & \cite{lee2007moners}\cite{yeung2010proactive}\cite{ilievski2013personalized}\cite{chu2009personalized}\cite{li2010contextual}\cite{wei2021news}                                         \\
Cluster/Segment    &   \cite{lee2007moners}\cite{yeung2010proactive}\cite{zheng2018drn}\cite{darvishy2020hypner}\cite{manoharan2020intelligent}                            \\
Tag/Keyword     & \cite{jonnalagedda2013personalized}\cite{yeung2010proactive}\cite{darvishy2020hypner}\cite{gao2009infoslim}\cite{billsus2000user}\cite{cantador2011enhanced}\cite{jonnalagedda2016incorporating}\cite{sood2014preference}\cite{tiwari2022pntrs}                        \\
Location     & \cite{fortuna2010real}\cite{yeung2010proactive}\cite{tavakolifard2013tailored}\cite{noh2014location}\cite{ilievski2013personalized}\cite{viana2017hybrid}\cite{li2010contextual}\cite{kazai2016personalised}\cite{chu2009personalized}\cite{wei2021news}                                         \\
Access Pattern     &  \cite{li2011scene}\cite{saranya2012personalized}\cite{wei2021news}                                      \\
Behaviors on Other Platforms     &  \cite{li2010contextual}\cite{gu2014effective}\cite{bai2017exploiting}\cite{hsieh2016immersive}\cite{kazai2016personalised}\cite{li2010user}\cite{phelan2011using}\cite{phelan2009using}\cite{hsieh2016immersive} \cite{lian2018towards}                                   \\
 \hline
\end{tabular}
}
\end{center}
\end{table*}

\subsection{Deep Learning-based User Modeling}

In recent years, many personalized news recommendation methods use deep learning techniques for user modeling to remove the need of manual feature engineering.
Most existing methods infer user interests from historical news click behaviors.
Several methods focus on aggregating the representations of historical clicked news~\cite{han2021neural}.
For example, Khattar et al.~\cite{Khattar2017user} used the summation of clicked news representations weighted by an exponential discounting function, where more recent clicks gain higher weights.
NAML~\cite{wu2019neuralnaml} and KRED~\cite{liu2020kred} learn user representations from the representations of clicked news using a news-level attention network, and AMM~\cite{zhang2021amm} also uses attention network to aggregate different information of clicked news and candidate news for user modeling.
DKN~\cite{wang2018dkn} learns user representations from the representations of clicked news via a candidate-aware attention network, i.e., computing the attention weight of each clicked news according to its relevance to candidate news.
The candidate-aware attention mechanism is also used by TEKGR~\cite{lee2020news} for user modeling.
Liu et al.~\cite{liu2019news} use a simple time-decayed averaging of the embeddings of clicked news to build the user embedding.
MM-Rec~\cite{wu2021mm} uses a crossmodal candidate-aware attention network that selects clicked news based on their  crossmodal relatedness with candidate news for user modeling.
HieRec~\cite{qi2021hierec} uses a hierarchical user interest representation method that first models subtopic-level user interest from the news within the same subtopic, then aggregates subtopic-level interest representations into coarse-grained topic-level user interest representations, and finally synthesizes topic-level interest representations into an over interest representation.
DebiasRec~\cite{yi2021debiasedrec} uses a bias-aware user modeling module to learn debiased user interest representations by incorporating the influence of presentation bias information on click behaviors into attentive behavior aggregation.
These methods can select important click behaviors for user modeling.
However, the relations among different clicked news, which provide rich contexts of behaviors that are useful to user modeling, cannot be modeled by these methods.

Therefore, many methods consider the contexts of news click behaviors.
Recurrent neural network (RNN) is a popular choice to model the sequential dependency between different clicked news~\cite{okura2017embedding,kumar2017deep,zhu2019dan,raza2021deep}.
For example, EBNR~\cite{okura2017embedding} learns representations of users from the representations of their browsed news via a GRU network.
RA-DSSM~\cite{kumar2017deep} uses a bi-directional long short-term memory (Bi-LSTM) network to process the historical news click sequence, and then use a news-level attention network to form a user representation.
DAN~\cite{zhu2019dan} learns user representations from clicked news using a combination of attentive LSTM and candidate-aware attention, which generate user historical sequential embedding and  user interest embedding, respectively.
The advantage of RNN-based user models is their strong ability in modeling user interest dynamics.
However, they are somewhat weak in capturing the global interest information of users.
In addition, as pointed by~\cite{wu2021can}, news recommendation may not be suitable to be modeled as a conventional sequential recommendation problem because users have a strong preference on the diversity between past and future clicked news.

Many other common deep models, such as CNN~\cite{zhang2021learning}, self-attention~\cite{wu2019neuralnrms} and co-attention~\cite{meng2021dcan}, have also been applied in user modeling.
For example, WE3CN~\cite{khattar2018weave} learns representations of users from the 3D representation tensors of their clicked news using a 3D CNN model.
SFI~\cite{meng2021dcan} further introduces a hard selection mechanism to reduce the computational cost in 3D CNN-based user modeling.
NRMS~\cite{wu2019neuralnrms} learns contextual news representations by using a news-level multi-head self-attention network, and uses an additive attention network to form the user representation.
This method is also adopted by many methods like FairRec~\cite{wu2021fairness}, IMRec~\cite{xun2021we} and SentiRec~\cite{wu2020sentirec}, and the variant that uses ``CLS'' token representation of Transformer is used by~\cite{zhang2021unbert}.
FIM~\cite{wang2020fine} uses a fine-grained interest modeling method that can capture the word-level relatedness between news with a 3D CNN model.
UniRec~\cite{wu2021two}  learns  the user embedding for news ranking with the NRMS~\cite{wu2019neuralnrms} model, and then uses this embedding as the attention query to select a set of basis interest embeddings to aggregate them into a user embedding for news recall.
KIM~\cite{qi2021kim} uses a user-news co-encoder that models the interactions between candidate news and clicked news to collaboratively learn a candidate-aware user interest representation and a user-aware candidate news representation.
PP-Rec~\cite{qi2021pprec} uses a popularity-aware user modeling method that first uses self-attention to model the contexts of user behaviors and then uses a content-popularity joint attention network that selects clicked news according to their content and popularity for user interest modeling.
RMBERT~\cite{jia2021rmbert} uses a reasoning memory network~\cite{fu2020recurrent} to capture the sophisticated interactions between user behaviors and candidate news in user interest modeling.
These methods can effectively capture the relations of different user behaviors to enhance user modeling.

In recent years, graph neural networks have also been applied to model the contexts of user behaviors by capturing their high-order relations~\cite{mao2021neuralnews}.
For example, CAGE~\cite{sheu2020context} first uses a GCN model to capture the relations between different behaviors within a news session to refine the behavior representations, and then uses a GRU network to build user representations.
User-as-Graph~\cite{wu2021uag} is probably the first work in news recommendation that represents each user with a personalized heterogeneous graph constructed from click behaviors, where the user modeling task is modeled as a graph pooling problem.
It uses a heterogeneous graph pooling method named HG-Pool to iteratively summarize the personalized heterogeneous graph for learning user interest representations.
EEG~\cite{zhang2021combining} models each user as an entity graph.
It first uses a graph neural network to learn hidden entity representations, and then uses an attention network to aggregate them into an entity-based user representation. 
KOPRA~\cite{tian2021joint} also models users as entity graphs and uses recurrent graph convolution to process the entity graphs.
It models both long-term and short-term user interests with the entire graph and the subgraph inferred from recently clicked news, respectively.
In addition, it introduces an entity neighbor pruning technique to select entity neighbors according to user interests.
CNE-SUE~\cite{mao2021neuralnews} applies a GCN to different subgraphs of an entire user behavior graph to learn different interest representations for different behavior clusters.
It further employs an intra-cluster attention mechanism to pool node representations and uses an intra-cluster attention mechanism to aggregate cluster representations.
These methods can usually capture the rich high-order relatedness between users' click behaviors to discover latent user interest.

In addition to click behaviors, a few methods also consider the ID information of users~\cite{zhang2018deep,zhang2019dynamic}.
For example, NPA~\cite{wu2019npa} uses a news-level personalized attention network to select important news  according to user characteristics, where the embeddings of user IDs are used to generate the attention queries.
LSTUR~\cite{an2019neural} learns short-term user interest embeddings by a GRU network, and models long-term user interests by the embeddings of user IDs.
To fuse the two kinds of user representations, LSTUR explores two methods, i.e., concatenating two vectors together, or using the long-term user interest embedding to initialize the hidden state of the GRU network.
This framework is further used by CUPMAR~\cite{tran2021deep} and KG-LSTUP~\cite{sun2021hybrid}.
These methods can usually better serve active users with rich behaviors to tune their ID embeddings.
However, they have some difficulties in handling cold-start users without well-tuned user embeddings.

All the aforementioned methods mainly rely on the information of users' click behaviors. 
However, click behaviors are very noisy and may not necessarily indicate user interest, and it is difficult to comprehensively and accurately infer user interest from click feedback only.
Thus, a few methods study incorporating other kinds of user information to enhance user interest modeling~\cite{zhang2019dynamicattention}.
One major direction is adding context features to enhance user modeling.
For example, CHAMELEON~\cite{gabriel2019contextual,de2018newssession} uses several user context features like time, device, location and referrer.
It uses a UGRNN network to learn representations of users in a session, and the click score is evaluated by the cosine similarity between user and candidate news representations.
The context features used in these methods can provide rich information for inferring users' current preferences to improve subsequent news recommendation.

Another main direction is incorporating various kinds of user behaviors~\cite{ma2021graph}.
For example, NRHUB~\cite{wu2019neuralnrhub} considers heterogeneous user behaviors, including news clicks, search queries, and browsed webpages.
It incorporates different kinds of user behaviors as different views of users by learning a user embedding from each kind of user behaviors separately, where  a combination of CNN and attention network is used to learn behavior representations and a behavior attention network is used to learn a user embedding by selecting important user behaviors.
The user embeddings from different views are aggregated into a unified one via a view attention network.
The effectiveness of webpage browsing behaviors in user modeling for news recommendation is also studied by WG4Rec~\cite{shi2021wg4rec}.
CPRS~\cite{wu2020cprs} considers users' click and reading behaviors in user modeling.
It models the click preference of users from the titles of clicked news, and models their reading satisfaction from the body of clicked news as well as the personalized reading speed metric derived from dwell time and body length. 
NRNF~\cite{wu2020ccf} uses a dwell time threshold to divide click news into positive ones and negative ones.
It uses separate Transformers and attention networks to learn positive and negative user interest representations.
FeedRec~\cite{wu2021feedrec} uses various kinds of user feedback including click, nonclick, finish, quick close, share and dislike to model user interest.
It uses a heterogeneous Transformer to model the relatedness between all kinds of feedback and uses different homogeneous Transformers to model the interactions between the same kind of feedback.
In addition, it uses a strong-to-weak attention network that uses the representations of strong feedback to distill real positive and negative user interest information from weak feedback.
These methods can usually infer user interests more accurately by mining complementary information encoded in multiple kinds of user behaviors.

There are also several methods that learn user representations on graphs that involve the collaborative information of users and news.
For example, IGNN~\cite{qian2019interaction} learns content-based user representations using the average embedding of clicked news, and learns graph-based user representations from the user-news graph via a graph neural network.
The content-based user representation is concatenated with graph-based user representation to form a unified one.
GNewsRec~\cite{hu2020graph} uses the same architecture with DAN to learn  short-term user representations, and uses a two-layer graph neural network (GNN) to learn long-term user representations from a heterogeneous user-news-topic graph.
Both short-term and long-term user representations are concatenated to build a unified user representation.
GERL~\cite{ge2020graph} uses multi-head self-attention and additive attention networks to form content-based user representations from the click history.
In addition, it uses a graph attention network to learn graph-based representations of  users by capturing  high-order information on the user-news graph, which are further combined with the content-based user representations.
MVL~\cite{santosh2020mvl} uses attention networks to learn user interest representations in a content view, and uses a graph attention network to model user interest from the user-news graph in a graph view.
GNUD~\cite{hu2020graph2} uses a disentangled graph convolution network to learn user representations from the user-news graph.
These methods can exploit the high-order information on graphs to enhance user modeling.
GBAN~\cite{ma2021graph} combines user embeddings learned by an LSTM and heterogeneous graph embeddings.
It further introduces subgraph core and coritivity scores that measure the importance of a target user-news pair in the subgraph to enhance user representations.
These methods can take the advantage of high-order interaction information between user and news as well as the associated meta features.
However, it is challenging for them to accurately represent new users that do not participate in the model training.

\begin{table*}[t]
\centering
\caption{Comparison of different methods on user modeling.}\label{deeplearningu}
\resizebox{1\linewidth}{!}{
\begin{tabular}{l|c|l|l}
\Xhline{1.0pt}
\multicolumn{1}{c|}{\textbf{Method}} &
\multicolumn{1}{c|}{\textbf{Year}} &\multicolumn{1}{c|}{\textbf{Information Used}} & \multicolumn{1}{c}{\textbf{Model}} \\ \hline
EBNR~\cite{okura2017embedding}           &  2017 & News Click                                     & GRU                                \\
RA-DSSM~\cite{kumar2017deep}            & 2017 &  News Click                                     & Bi-LSTM+Attention                        \\
Khattar et al.~\cite{Khattar2017user}  &  2017 & News Click                                     & Exponential-decayed Average                        \\
3-D-CNN~\cite{kumar2017word}            & 2017  & News Click                                     & Word2vec                           \\
Park et al.~\cite{park2017deep}         & 2017 &  News Click                                     & LSTM                               \\
WE3CN~\cite{khattar2018weave}           & 2018 &  News Click                                     & 3-D CNN                            \\
DKN~\cite{wang2018dkn}                  & 2018 &  News Click                                     & Candidate-Aware Attention                             \\
Gao et al.~\cite{gao2018fine}           &  2018 & News Click                                     & Candidate-Aware+Multi-Head Attention                        \\
Saskr~\cite{chu2019next}                & 2019 & News Click                                     & Self-Attention+Candidate-Aware Attention                   \\ 
NAML~\cite{wu2019neuralnaml}            &  2019&  News Click                                     & Attention                                \\
TANR~\cite{wu2019neural}                &  2019 & News Click                                     & Attention                                \\
NRMS~\cite{wu2019neuralnrms}            & 2019 &  News Click                                     &  Self-Attention+Attention                    \\
Liu et al.~\cite{liu2019news}  & 2019  & News Click                                     & Time-decayed Average \\
DAN~\cite{zhu2019dan}                   & 2019 &  News Click                                     & LSTM+Self-Attention+Candidate-Aware Attention                      \\
KRED~\cite{liu2020kred}            & 2020  & News Click                                     & Attention                                \\
TEKGR~\cite{lee2020news}                  & 2020 &  News Click                                     & Candidate-Aware Attention                             \\
FIM~\cite{wang2020fine}    &  2020 &  News Click                                          &  3-D CNN    \\
FedRec~\cite{qi2020privacy}         &  2020 & News Click                                     &  Self-Attention+Attention+GRU                   \\ 
SentiRec~\cite{wu2020sentirec}            & 2020  & News Click                                     & Transformer+Attention                    \\
CAGE~\cite{sheu2020context} & 2020  & News Click   &   GCN+GRU  \\
PLM-NR~\cite{wu2021empowering}            & 2021 &  News Click                                     & Attention                                \\
UniRec~\cite{wu2021two}            &  2021 & News Click                                     &  Self-Attention+Attention+Basis User Embedding                    \\
FairRec~\cite{wu2021fairness}            &  2021 & News Click                                     & Transformer+Attention                    \\
MM-Rec~\cite{wu2021mm} &  2021 & News Click   &   Crossmodal Candidate-Aware Attention   \\
IMRec~\cite{xun2021we} &  2021 & News Click   &    Self-Attention+Attention  \\ 
HieRec~\cite{qi2021hierec} &   2021& News Click   &   Hierarchical Attention  \\
KIM~\cite{qi2021kim} &  2021 & News Click   &   User-News  Co-Encoder  \\
PP-Rec~\cite{qi2021pprec} &  2021 & News Click   &   Content-Popularity Joint Attention Network  \\
CTX~\cite{cho2021overlooked}  & 2021  & News Click   &  Add on existing methods \\
DCAN~\cite{meng2021dcan} & 2021 & News Click+Click Time+Context Features & LSTM+Co-Attention   \\  
DebiasRec~\cite{yi2021debiasedrec} & 2021  & News Click   &   Content Attention+Bias Attention  \\
User-as-Graph~\cite{wu2021uag} & 2021  & News Click   &   Heterogeneous Graph Pooling \\ 
AMM~\cite{zhang2021amm} &  2021 & News Click                                          & Attention                \\ 
RMBERT~\cite{jia2021rmbert}& 2021  & News Click                                           & Reasoning Memory Network                 \\ 
UNBERT~\cite{zhang2021unbert}&  2021 & News Click                                          & Transformer               \\ 
EEG~\cite{zhang2021combining} &  2021 & News Click   &   GNN+Attention \\ 
KOPRA~\cite{tian2021joint} &  2021 & News Click  & Recurrent Graph Convolution \\
SFI~\cite{zhang2021learning} &  2021 & News Click                                          & Hard Similarity Selection+3-D CNN               \\ 
TempRec~\cite{wu2021can}& 2021  & News Click                                                & Temporal Diversity-aware Transformer               \\ 
D2NN~\cite{raza2021deep} &2021   & News Click     & Embedding Sum+LSTM+Attention   \\  
EENR~\cite{han2021neural}& 2021   & News Click & Attention+Event Type/Category Distribution  \\  
CNE-SUE~\cite{mao2021neuralnews}& 2021  & News Click                                        &   GCN+Attention             \\ 
\hline
DeepJoNN~\cite{zhang2018deep}           & 2018  & News Click+User ID                             & LSTM                               \\
NPA~\cite{wu2019npa}                    & 2019  & News Click+User ID                             & Personalized Attention                   \\
LSTUR~\cite{an2019neural}               & 2019  & News Click+User ID                             & GRU+ID Embedding                   \\
DNA~\cite{zhang2019dynamic}             & 2019  & News Click+User ID+Time Interval               & Attention+CNN                            \\ 
CUPMAR~\cite{tran2021deep}             &  2021 & News Click+User ID               & GRU+ID Embedding                            \\ 
KG-LSTUP~\cite{sun2021hybrid} & 2021& News Click+User ID & GRU+ID Embedding    \\  
\hline
CHAMELEON~\cite{gabriel2019contextual}  &  2019 & News Click+Context Features                    & UGRNN                              \\ 
DAINN~\cite{zhang2019dynamicattention}  &2019& News Click+Time+Location                       & Attention                                \\
NRHUB~\cite{wu2019neuralnrhub}          &  2019 & News Click+Query+Webpage                       & Attention                                \\
CPRS~\cite{wu2020cprs} & 2020  & News Click+News Reading                       & Attention+Content-Satisfaction Attention  \\
NRNF~\cite{wu2020ccf} & 2020  & Positive News Click+Negative News Click                      & Transformer+Attention  \\
FeedRec~\cite{wu2021feedrec} &  2021 & News Click+Nonclick+Finish+Quick Close+Share+Dislike                       & Transformer+Strong-to-Weak Attention  \\
WG4Rec~\cite{shi2021wg4rec}& 2021  & News Click+Webpage                                          &   GRU+Attention             \\ 
\hline

IGNN~\cite{qian2019interaction}   & 2019  & News Click+User-News Graph                     & GNN                                \\
INNR~\cite{REN2019113115}               &  2019 & Heterogeneous Graph                            & Node2vec                           \\
GNewsRec~\cite{hu2020graph}             & 2020  & News Click+Heterogeneous Graph                 & LSTM+Attention+GNN                       \\
GERL~\cite{ge2020graph}                 & 2020  & News Click+User-News Graph                     &  Self-Attention+GAT                    \\ 
MVL~\cite{santosh2020mvl} & 2020  & News Click+User-News Graph                     &  Self-Attention+GAT     \\
GNUD~\cite{hu2020graph2} &  2020 & User-News Graph &  Disentangled GCN \\
AGNN~\cite{ji2021attention} &2021   & News Click+Heterogeneous Graph &   LSTM Variant~\cite{gers2000learning}+GNN+Self-Attention+Attention  \\
TSHGNN~\cite{ji2021temporal} & 2021  & News Click+Heterogeneous Graph &  LSTM Variant+GNN+Self-Attention+Attention \\
GBAN~\cite{ma2021graph} & 2021 &  Click+Nonclick+Like+Follow+Comment+Share+Heterogeneous Graph & Core and DeepWalk+Coritivity Scores+LSTM+Attention \\
\Xhline{1.0pt}
\end{tabular}
}

\end{table*}

We summarize the user information and user modeling techniques used in these deep learning-based methods in Table~\ref{deeplearningu}.
We then provide several discussions on the user modeling methods introduced in this section.

\subsection{Discussions on User Modeling}

\subsubsection{Feature-based User Modeling}

Most feature-based methods construct user profiles based on the collections of features extracted from the clicked news.
Besides the news information, some methods leverage additional  user features to facilitate user modeling.
For example, the demographics of users (e.g., age, gender and profession) are used in several methods, since users with different demographics usually have different preferences on news.
The location of users can be used to identify the news related to the user's neighborhood, and the access patterns of users can also help understand the news click behaviors of users.
In addition, many methods use the tags or keywords of users to indicate user interest, and cluster users based on their characteristics.
In this way, the recommender system can more effectively recommend news according to users' interest in different topics.
Moreover, several methods incorporate user behaviors on other platforms, such as social media, search engines and e-commerce platforms.
These behaviors can not only facilitate user interest modeling, but also has the potential to mitigate the problem of cold-start on the news platform if user data can be successfully aligned.
However, feature-based user modeling methods usually require massive expertise for feature design and validation, and may not be optimal for representing user interests.

\subsubsection{Deep Learning-based User Modeling}

Deep learning-based user modeling methods usually aim to learn user representations from user behaviors without feature engineering.
Many of them infer user interests merely from click behaviors, because click behaviors are implicit indications of users interest in news.
However, click behaviors are usually noisy and they do not necessarily indicate real user interests.
Thus, many methods consider other kinds of information in user modeling.
For example, some methods such as NPA and LSTUR incorporate the IDs of users to better capture users' personal interest.
CHAMELEON and DAINN consider the context features of users such as devices and user locations.
CPRS, FeedRec and GBAN incorporate multiple kinds of user feedback on the news platform to consider user engagement information in user interest modeling.
GERL and GNewsRec can exploit the high-order information on  graphs to encode user representations.
However, it is still difficult for these methods to accurately infer user interests when user behaviors on the news platforms are sparse.
There are only two methods, i.e., NRHUB and WG4Rec, that consider users' behaviors on multiple platforms, which can still model users accurately even user behaviors on the news platform are sparse.
However, there may exist some difficulties in linking user data on different platforms due to privacy reasons.

According to the summarization in Table~\ref{deeplearning}, we can see that the model architectures used for user representation learning are diverse.
Some methods utilize recurrent neural networks to capture the relatedness of news clicked by users, such as EBNR, DAN and CHAMELEON.
With the great success of Transformer models, many methods also use self-attention or Transformer networks to model the  global contexts of user behaviors.
However, these sequential models cannot effectively model the high-order relations between user behaviors, which can provide useful contexts for user interest understanding.
Instead of  modeling user behaviors as a sequence, several methods like User-as-Graph model each user as a personalized graph, where the high-order relations between behaviors can be fully modeled.
In addition, several works such as GERL and GNUD use graph neural networks to capture the high-order interactions between users and news on the global user-news graphs, which can also help better understand user interest by incorporating collaborative information.
However, the computational cost of these graph-based architectures is usually much heavier than sequential models, and collaborative signals are usually not available for cold-start users and news.

To select clicked news that is informative for inferring user interest, attention mechanisms are widely used by many methods.
In some works such as NAML and KRED, the attention query is a global parameter vector, which is invariant with respect to different users.
In the NPA method, the attention query is generated by the embedding of user ID, which can achieve personalized news selection.
Both kinds of attention mechanisms are efficient in the online test phase because user representations can be prepared in advance~\cite{wu2019npa}.
However, the relatedness between candidate news and clicked news cannot be fully modeled, which may not be optimal in modeling user interests in a specific candidate news.
Another kind of attention mechanism, i.e., candidate-aware attention, is also widely used by many methods such as DKN,  DAN and KIM.
In candidate-aware attention networks, the representation of candidate news is used as the attention query, and user representations can be dynamically constructed based on candidate news.
However, they need to memorize the representations of all clicked news in the test phase, which may lead to some sacrifice in efficiency.

Some methods study modeling multiple types of user interests.
For example, LSTUR, GNewsRec and FedRec consider both long-term and short-term interests of users to better capture their interest dynamics.
HieRec models the hierarchical structure of user interests, which can capture the user interests in different granularities.
These methods can improve user interest understanding of user interests by taking different kinds of user interest into consideration.
However, user interests are diverse and evolutional, which are still difficult to be comprehensively and accurately modeled by these methods.

\subsubsection{Differences to User Modeling in General Recommendation}

The user modeling techniques used for personalized news recommendation have close relations to the user modeling methods in general recommendation scenarios such as e-commerce~\cite{zhou2018deep} and movie recommendation~\cite{diao2014jointly}.
For example, the core neural architectures such as RNN, CNN, self-attention and graph neural networks, are also widely used for sequential recommendation.
In addition, several useful user modeling paradigms such as long short-term user interest modeling are also popular in other recommendation fields.
However, by scrutinizing recent literature, we find there are several unique characteristics of user modeling in personalized news recommendation:

(1) Short news lifecycles. 
Different from the common e-commerce recommendation scenarios where items can be actively interacted with for months or even years~\cite{zhou2020s3}, most news articles have very short lifecycles (i.e., a few days).
Thus, in news recommendation it is important to take this unique characteristic into consideration when designing user modeling algorithms.
For example, in GNN-based methods for general recommendation, item nodes can be simply represented by ID embeddings.
However, in GNN-based news recommender, it is better to learn embeddings of news nodes from news content to handle uncovered news in user modeling~\cite{hu2020graph}.
In addition, the quick vanishment of old news de facto limits the exploitation of collaborative signals in user modeling due to the large fraction of cold news in the inference stage.

(2) Fine-grained candidate-aware user modeling. 
In many recommendation tasks, items are mainly represented by their overall embeddings~\cite{sun2019bert4rec}, and modeling feature interactions is important for candidate-aware user modeling~\cite{zhou2018deep}.
By contrast, news articles have rich content and context information, and the interactions between user behaviors and candidate news can be modeled in a more fine-grained way (e.g., word-level interactions).
Capturing the fine-grained relevance between user behaviors and candidate news is very important for understanding user interest in a specific candidate news article.
Thus, fine-grained candidate-aware user modeling is a core technique used in many recent news recommendation methods.

(3) User modeling as document modeling.
Different from many recommendation scenarios where user behaviors are not associated with sufficient textual information~\cite{zhou2019deep}, in news recommendation user behaviors are usually clicked news that contain rich texts.
Thus, the user modeling problem in news recommendation can be formulated as a document modeling problem, where the texts of clicked news are embedded in user ``documents''.
Several recent methods follow this setting and employ strong pre-trained language models to empower user modeling~\cite{zhang2021unbert,jia2021rmbert}.
The success of these PLM-based user modeling techniques indicates that there may not be a huge barrier between NLP and user modeling for news recommendation, which is a unique characteristic of the news recommendation field.

(4) Potential strong temporal diversity preference.
Different from general recommendation scenarios where users may prefer to click very similar items, in news recommendation users tend to click news that are somewhat different from the previously clicked ones~\cite{abdollahpouri2021toward} (i.e., preference on serendipity).
As pointed by a recent study~\cite{wu2021can}, it may not be very suitable to model news recommendation as a standard sequential recommendation task, and it is important to consider such temporal diversity preference in user modeling.
The results show that many sequential models such as RNN~\cite{hidasi2016session} and casual self-attention~\cite{kang2018self} are inferior to standard self-attention that focus more on global context rather than sequential dependency.
Further study on this unique phenomenon is needed to better understand user modeling mechanism in news recommendation.

In summary, by reviewing user modeling techniques used in existing news recommendation methods, we argue that user modeling is also remained challenging due to many reasons, such as the noise and sparsity of user behaviors, the diverse and dynamic characteristics of user interests, and the difficulties in modeling user interests in a specific candidate news effectively and efficiently.
\section{Personalized Ranking}\label{ranking}

On the basis of news and user modeling, news ranking aims to rank candidate news for personalized  display according to users' personal interest.
Common news ranking techniques can be divided into two categories, i.e., relevance-based and reinforcement learning-based.
We introduce them in the following sections.

\subsection{Relevance-based Personalized Ranking}

Relevance-based news ranking methods usually rank candidate news with user interests based on their personalized relevance.
In these methods, how to accurately measure the relevance between candidate news and user interest is a core problem.
Many methods directly evaluate the user-news relevance  based on the similarities of their final representations.
For instance, Goossen et al.~\cite{goossen2011news} computed the cosine similarities between the CF-IDF feature vectors of user and news to measure their relevance.
Garcin et al.~\cite{garcin2012personalized} used the similarities between the news topic vectors and the user topic vector to evaluate their relevance.
Okura et al.~\cite{okura2017embedding} used the inner product between news and user representations to predict the relevance scores.
DFM~\cite{lian2018towards} uses an inception module that combines neural networks with different depths to compute the relevance scores from news and user features.
These methods usually employ two-tower architectures, which enable efficient inference by computing news and user features in advance.
However, user interests are usually diverse, and candidate news may only match the user interests indicated by a part of the clicked news.
These methods cannot fully consider the relatedness between candidate news and clicked news, and the matching between candidate news and user interest may not be very accurate.

A few methods use fine-grained interest matching techniques to better model the relevance between users' interest and candidate news. 
For example, FIM~\cite{wang2020fine} first multiplies together the word representations of candidate news and clicked news, and then uses a  matching module with 3-D CNN networks to compute relevance scores by  capturing the fine-grained relatedness between  candidate news and clicked news.
KIM~\cite{qi2021kim} first uses a knowledge-aware news co-encoder to model the relatedness between words and entities in candidate news and clicked news, and further uses a user-news co-encoder to further help model the interactions between clicked news and candidate news for better relevance modeling.
HieRec~\cite{qi2021hierec} has a hierarchical interest matching mechanism that matches candidate news with the fine-grained subtopic-level user interest, the coarse-grained topic-level user interest and the overall user interest.
AMM~\cite{zhang2021amm} uses a multi-field matching scheme to model the interactions between each pair of views of a clicked news and a candidate news.
These single-tower methods can more accurately evaluate the relevance between candidate news and user interest  by modeling their fine-grained and multi-grained relatedness, which can help generate  news ranking results that better target user interest. 
However, these methods usually have much larger computational costs in the inference stage than coarse-grained interest matching, which may hinder their application in some low-latency or low-resource scenarios.

In most methods, candidate news with higher relevance to user interest will gain higher ranks.
However, these methods may tend to recommend news that are similar to those previously clicked by users, which is also called the ``filter bubble'' problem.
Thus, some news ranking methods explore to recommend news that are somewhat different from previously clicked ones to introduce diversity and serendipity~\cite{abdollahpouri2021toward}.
For example, Newsjunkie~\cite{gabrilovich2004newsjunkie} is a system that ranks news articles based on their novelty in the context of the news that users previously clicked. 
SCENE~\cite{li2011scene} first ranks news articles based on their relevance to user interests, and then refines the ranking list based on news popularity and recency to form the final recommendation list.
Different from the methods that are solely based on the relevance between candidate news and user interests, these methods have the potential to provide more diverse recommendations.

\subsection{Reinforcement Learning-based Personalized Ranking}

Different from relevance-based ranking methods that mainly aim to optimize the objectives (e.g., clicks) on current candidate news articles, reinforcement learning-based ranking methods usually aim to optimize the total reward in a long term~\cite{zheng2018drn,song2021dql}. 
A representative reinforcement learning-based approach to personalized news recommendation is LinUCB~\cite{li2010contextual}, which models the problem of personalized news recommendation as a contextual bandit problem.
In this method, LinUCB computes the payoff by a hybrid linear model, which means that some parameters are shared by all arms, while the others are not.
LinUCB can outperform context-free bandit methods such as $\epsilon$-greedy and Upper Confidence Bound (UCB), and it is computationally efficient because the  block parameters in LinUCB have fixed dimensions and can be incrementally updated~\cite{li2010contextual}. 
It is also latterly evaluated by~\cite{li2011unbiased} in an unbiased manner by estimating the  per-trial payoff with log data directly rather than a simulator.
In the CLEF NewsREEL 2017 challenge, Liang et al.~\cite{liang2017clef} also developed a system based on LinUCB.
The LinUCB model is used to help choose the appropriate recommender from a pool of recommendation algorithms based on user and news features.
Deep reinforcement learning is also explored in news recommendation~\cite{zheng2018drn,islambouli2021user,song2021dql}.
For example, DRN~\cite{zheng2018drn} uses a Deep Q-Network (DQN) to estimate the policy reward, which is a weighted summation of click labels and the activeness of users that is computed based on their return time after recommendations.
In addition, DRN applies the Dueling Bandit Gradient
Descent~\cite{yue2009interactively} algorithm to eliminate the recommendation performance decline brought by classical exploration methods such as $\epsilon$-greedy and UCB.
Different from relevance-based ranking methods, reinforcement learning-based ranking methods have the ability of exploration, which can increase the diversity of recommendation results and further discover potential user interests.

\subsection{Discussions on Personalized Ranking}

In this section we provide some discussions on the news ranking methods in existing personalized news recommender systems.
Relevance-based news ranking methods mainly need to  accurately evaluate the relevance between candidate news and user interest for subsequent news ranking.
Many methods model their overall relevance by evaluating the relevance between the unified representations of user interest and candidate news.
However, candidate news usually can only match part of user interests, and directly match the overall user interest with candidate news may be suboptimal.
A few methods explore to evaluate the relevance between user interest and candidate news in a fine-grained way by modeling the relatedness between candidate news and clicked news, which can improve the accuracy of relevance modeling for news ranking.
However, these methods are much more time-consuming because the representations of users are dependent on candidate news and cannot be computed in advance.
Moreover, pure relevance-based interest matching methods may tend to recommend news that are similar to previously clicked news, which is not beneficial for users to receive diverse news information.
Thus, a few works explore to adjust the news ranking strategy by incorporating other factors such as news novelty, popularity and recency, which have the potential to make more diverse news recommendations and mitigate the filter bubble problem in news recommender systems.

In relevance-based news ranking methods, candidate news is usually greedily matched with users, i.e., choosing the news in each impression that mostly satisfy the ranking policy on the current candidate news list.
However, it may not be optimal in improving long-term user experience.
In reinforcement learning-based methods, the ranking algorithm aims to find the optimal ranking policy to maximize the long-term reward.
Thus, RL-based news ranking methods may be more suitable for exploring potential user interest and improving long-term user experience and engagement, while it may have some sacrifice in short-term news CTRs.

In summary, news ranking in news recommendation also faces many challenges, including how to accurately and efficiently evaluate the relevance between candidate news and user interest indicated by user behaviors, how to mitigate the ``filter bubble'' problem in news recommender systems, and how to explore potential user interests without hurting user experience.

\section{Model Training}\label{training}

Many personalized news recommendation methods exploit machine learning models for news modeling, user modeling and interest matching.
Training these models is a necessary step in building an accurate news recommender system.
In this section, we review the techniques used for model training in news recommendation.

\subsection{Training Methods}

In a few methods based on collaborative filtering, the news recommendation task is formulated as a rating prediction problem, i.e., predicting the ratings that users give to news~\cite{ji2016regularized}.
To learn their models, they usually use loss functions such as the mean squared error (MSE) computed between the predicted ratings and the gold ratings, which are further used to optimize the model~\cite{claypool1999combing}.
However, explicit user feedback like rating is usually very sparse, which may be insufficient to train an accurate recommendation model.

Since implicit feedback such as click is abundant, most methods use the click feedback of users as the prediction target.
They formulate the news recommendation task as a click prediction task.
Some methods simply classify whether a candidate news will be clicked by a target user~\cite{fortuna2010real,gershman2011news,wang2018dkn}.
However, these methods cannot exploit the relatedness between clicked and nonclicked samples.
Thus, a few methods use contrastive training techniques to maximize the margin between the predicted click scores of clicked and nonclicked news.
For example, PP-Rec~\cite{qi2021pprec} uses the Bayesian Personalized Ranking (BPR) loss for model training by comparing each clicked sample with an nonclicked one.
However, the BPR loss can only exploit a small part of nonclicked samples.
NPA~\cite{wu2019npa} uses the InfoNCE~\cite{oord2018representation} loss for model training.
For each clicked sample (regarded as a positive sample), it randomly samples a certain number of nonclicked ones (regarded as negative samples) and jointly predicts their click scores.
These click scores are further normalized by the softmax function to compute the posterior click probabilities, and the model aims to maximize the negative log-likelihood of the posterior click probability of positive samples.
In this way, the model can exploit the information of more negative samples.

Besides click feedback, a few methods also consider other kinds of feedback to construct training tasks.
For example, CPRS~\cite{wu2020cprs} trains the recommendation model collaboratively in the click prediction task and an additional reading satisfaction prediction task, which  aims to infer the personalized reading speed based on user interest and news body.
FeedRec~\cite{wu2021feedrec} trains the model in three tasks, including click prediction, dwell time prediction and finish prediction.
GBAN~\cite{ma2021graph} models the recommendation task as a future behavior classification problem to predict the behavior type of a user on a specific candidate (i.e., click, nonclick, like, follow, comment, and share).\footnote{We argue that this setting is somewhat difficult to be implemented in real-world news recommender systems because it is not clear how to rank news accordingly and users may also have multiple types of behaviors on the same news.}
These methods can encourage the model to optimize not only CTR but also user engagement, which can help learn engagement-aware news recommendation models.

There are several methods that use additional  news information to design auxiliary training tasks.
For example, EBNR~\cite{okura2017embedding} uses autoencoder to learn news representations and it uses another weak supervision task by encouraging the embeddings of news in the same topic to be similar than the embeddings of news in different topics.
TANR~\cite{wu2019neural} uses an auxiliary news topic prediction task to help learn topic-aware news representations.
SentiRec~\cite{wu2020sentirec} uses a news sentiment orientation score prediction task to learn sentiment-bearing news representations.
KRED~\cite{liu2020kred} trains the model in various tasks including item recommendation, item-to-item recommendation, category classification, popularity prediction and local news detection.
These methods can also effectively encode additional information into the recommendation model without taking it as the input.
However, it is usually non-trivial to balance the main recommendation task and the auxiliary tasks.

\subsection{Training Environment}

Existing researches mainly focus on the model training methods while ignoring the implementation environment of model training, which is in fact important in developing real-world news recommender systems.
In many existing methods, the news recommendation models are offline trained on centrally stored data with centralized computing resources~\cite{wu2020mind}.
This model training paradigm can help quick development of news recommender systems, but it also has several  main drawbacks.
First,  user behavior data for model training is usually abundant and many recent news recommendation models are in large size~\cite{wu2021empowering}, which require a large amount of computing resource to train accurate models. 
Although some recent works like~\cite{wu2021empowering} explore to train models in parallel on multiple GPUs, it is still  insufficient to train huge models. 
Thus, distributed model learning with proper acceleration methods like data rearrangement and cache mechanisms may be required in industrial practice~\cite{xiao2021training}.
Second, the model learned on offline data only may also have some mismatches with the characteristics of recommendation scenarios~\cite{zheng2018drn}.
Moreover, the distribution of user interest and news topics may also evolve, and it is shown in previous research that the performance of offline trained models may decline with time~\cite{wu2019npa}.
Thus, instead of re-training models periodically, online model training on streaming data is needed.
Third, most existing news recommendation methods are trained on centrally stored user data, which may have some privacy risks because user data usually contains private user information.
Several recent works like~\cite{qi2020privacy,qi2021uni,yi2021efficient} explore to train news recommendation models based on decentralized data with federated learning techniques, which can better protect user privacy in model training.

\subsection{Discussions on Model Training}

Next, we provide some discussions on the model training techniques used in news recommendation methods.
In some CF-based methods, news recommendation is modeled as a regression task where the ratings given by users are regarded as prediction targets.
However, on news platforms explicit user feedback such as rating is usually scarce, which poses great challenges to model training.
Therefore, most methods adopt implicit feedback to construct training tasks.
Click feedback is one of the most widely used signals for model training because it can implicitly indicate user interests in news and help the model optimize the CTR of recommendation results.
However, click signals also have some gaps with the real user interests~\cite{yi2014beyond}, and increasing CTR only may lead to recommending  clickbait news to users, which is actually harmful to user experience.
Thus, a few methods incorporating other user engagement signals such as dwell time and finish into model training, which can help learn user engagement-aware recommendation model to improve user experience.
Besides user feedback, some methods also consider using additional news information as auxiliary prediction objectives.
By jointly training the model in both recommendation task and auxiliary tasks, the model can be aware of the additional news information.
Since these methods do not take the additional features as the input, they can handle the scenarios where the additional features  are unavailable.
However, in  multi-task learning based methods, it is difficult to choose the proper coefficients for weighting the loss functions of different tasks, and these coefficients may also be sensitive to the dataset characteristics.

Another important problem in model training is designing effective strategies for constructing labeled training samples.
In most methods the negative samples  are randomly  drawn from the entire news set or the impression list~\cite{wu2021two}, which are further packed with the positive samples.
However, researchers have found that randomly selected negative samples may be too easy for the model to distinguish, which is not beneficial for learning discriminative recommendation models~\cite{li2019sampling}. 
It is also an interesting problem to study the influence of the number of negative samples on model training~\cite{wu2021rethinking}.

Besides, the environment for news recommendation model training is a less studied but important problem.
Most researches are offline conducted by learning models on centralized data with centralized computing resources.
As discussed in the  previous section, this model training environment may pose many potential challenges like the limitation of centralized computing resources, the gaps between offline data and online applications, and the privacy concerns and risks of centralized model training, which need to be extensively studied in the future.

In summary, model training is critical for news recommendation while it still has much room for improvement, such as designing more effective training tasks, choosing more representative training samples, adaptively tuning the loss coefficients for multi-task learning, and building more effective, efficient and privacy-preserving  environment for news recommendation model training.

\subsubsection{A Bird's-eye View on Recent Approaches}

To help readers better understand the details of recent news recommendation methods in terms of their news modeling, user modeling, ranking, and model training techniques, we illustrate a joint table that summarizes their details in these aspects.
Due to the limitation of page sizes, we do not include it in the main content, and readers can refer to it in a public repository (https://github.com/wuch15/News-Rec-Survey).

\section{Evaluation Metrics}\label{evaluation}

There are many metrics to quantitatively evaluate the performance of news recommender systems.
Most metrics aim to measure the recommendation performance in terms of the ranking relevance. 
For methods that regard the task of news recommendation as a classification problem, the Area Under Curve (AUC) score is a widely used metric, which is formulated as follows:
\begin{equation}
    \rm{AUC}=\frac{|\{(i,j)|Rank(p_i)<Rank(n_j))\}|}{N_pN_n},
\end{equation}
where $N_p$ and $N_n$ are  the numbers of positive and negative samples, respectively.
$p_i$ is the predicted score of the $i$-th positive sample and $n_j$ is  the score of the $j$-th negative sample.
Another set of popular metrics are precision, recall and F1 scores, which are computed as:
\begin{align}
\rm{Precision}&=\frac{TP}{TP+FP},\\
\rm{Recall}&=\frac{TP}{TP+FN},\\
\rm{F1}&=\frac{2*\rm{Precision}*\rm{Recall}}{\rm{Precision}+\rm{Recall}}, \end{align}
where TP, FP and FN respectively denote true positive, false positive and false negative.

For methods that model news recommendation as a regression task (e.g., predict the ratings of news), several common metrics for regression such as mean absolute error (MAE), mean squared error (MSE), rooted  mean squared error (RMSE) and Pearson correlation coefficient (PCC) are used to indicate the recommendation performance, which are respectively formulated as follows:
\begin{align}
\rm{MAE}&=\frac{1}{N}\sum_{i=N}|r_i-p_i|,\\
\rm{MSE}&=\frac{1}{N}\sum_{i=N}(r_i-p_i)^2,\\
\rm{RMSE}&=\sqrt{\frac{1}{N}\sum_{i=N}(r_i-p_i)^2},\\ 
\rm{PCC}&=\frac{1}{N-1}\sum_{i=1}^N(\frac{r_i-\bar{r}}{\sigma_r})(\frac{p_i-\bar{p}}{\sigma_p}),
\end{align}
where $r_i$ and $p_i$ are the real and predicted ratings of the $i$-th sample, $\bar{r}$  and $\bar{p}$  
respectively denote the arithmetic mean of the real and predicted ratings, and $\sigma$ is the standard deviation.

For methods that regard news recommendation as a ranking task,  besides the AUC metric there are also several other metrics such as Average Precision (AP), Hit Ratio (HR), Mean Reciprocal Rank (MRR) and normalized Discounted Cummulative Gain (nDCG).
Note that these metrics may be applied to the top K recommendation lists, e.g., HR@K and nDCG@K.
These metrics are respectively formulated as follows:
\begin{align}
\rm{AP}&=\frac{1}{N_p}\sum_{i=1}^{N_p}\frac{|\{k|Rank(p_k)\leq Rank(p_i)\}|}{Rank(p_i)},\\
\rm{HR@K}&=\frac{|\{k|Rank(p_k)\leq K\}|}{K},\\
\rm{MRR}&=\frac{1}{N_p}\sum_{i=1}^{N_p}\frac{1}{Rank(p_i)},\\
\rm{nDCG@K}&=\frac{\sum_{i=1}^{K}(2^{r_i}-1)/\log_2(1+i)}{\sum_{i=1}^{N_p}1/\log_2(1+i)},
\end{align}
where $r_i$ is a relevance score of news with the $i$-th rank, which is 1 for clicked news and 0 for non-clicked news.
There are several other metrics such as Click-Through Rate (CTR) and dwell time, which can be only used to measure the performance of online news recommenders.

Besides the metrics for measuring ranking accuracy, there are several other objective or subjective metrics to evaluate news recommender systems in other aspects.
In~\cite{gabrilovich2004newsjunkie} the recommendation results are evaluated by novelty, which is subjectively judged by a group of human subjects by rating the news sets from most novel to least novel.
In FeedRec~\cite{wu2021feedrec}, the recommendation results are further evaluated by a set of user engagement-related metrics, such as the average dwell time, finish ratio, dislike ratio and share ratio of the top ranked news.
These metrics can help comprehensively evaluate the performance of news recommender systems and further improve user experience.
A few methods also measure the diversity of recommendation results in different aspects.
For example, in~\cite{zheng2018drn}, an Intra-List Similarity (ILS) function is used to measure the diversity of recommendation results.
More specifically, given a ranking list $L$, its ILS score is calculated as follows:
\begin{equation}
    ILS(L) = \frac{\sum_{b_j\in L}\sum_{b_j\in L, b_j\neq b_i} S(b_i, b_j)}{\sum_{b_j\in L}\sum_{b_j\in L, b_j\neq b_i} 1},
\end{equation}
where $S(b_i, b_j)$ represents the cosine similarity between the item $b_i$ and $b_j$.
A similar diversity metric  ILAD is also used in~\cite{qi2021pprec,qi2021hierec}.
SentiRec~\cite{wu2020sentirec} uses a set of sentiment diversity metrics to measure the sentiment difference between historical clicked news and candidate news, which are formulated as follows:
\begin{equation}
    \begin{aligned}
    Senti_{MRR}&=max(0,\bar{s}\sum_{i=1}^N\frac{s^c_i}{i}),\\
    Senti@5&=max(0,\bar{s}\sum_{i=1}^5 s^c_i),\\
    Senti@10&=max(0,\bar{s}\sum_{i=1}^{10} s^c_i),\\
    \end{aligned}
\end{equation}
where $N$ is the number of candidate news in an impression, $s^c_i$ denotes the sentiment score of the $i$-th ranked candidate news.

Besides diversity, several fairness metrics are used to measure whether a news recommender system is fair to different groups of users or different news publishers.
For example, FairRec~\cite{wu2021fairness} uses the accuracy of sensitive attribute (e.g., gender) prediction based on top recommendation results as the fairness metric, where a higher accuracy means more serious unfairness because the recommendation results are more heavily influenced by sensitive attributes.
\cite{gharahighehi2021fair} studies the news recommendation fairness to different groups of authors.
It uses an Equity Attention for group fairness (EAGF) measurement and a Supplier Popularity Deviation (SPD) measurement for evaluating such kind of fairness, which if formulated as follows:
\begin{equation}
    \begin{aligned}
    EAGF&=\sum_{i=1}^{|g|}\sqrt{|L(i)|},\\
    SPD&=\frac{\sum_{i=1}^{|g|}|\frac{|L(i)|}{|L|}-\frac{|A(i)|}{|A|}|}{|g|},\\
    \end{aligned}
\end{equation}
where $g$ is the set of author groups and $L(i)$ is the set of recommended news belonging to the $i$-th group, $L$ is the set of all recommended items, $A(i)$ is the set of items in the training set belonging to the $i$-th group, and $A$ is the whole set of items.
A higher EAGF and a lower SPD score indicate better fairness.
These metrics used in the two works can be used to measure user-side fairness and provider-side fairness, respectively.

With the development of privacy-preserving news recommendation methods based on federated learning, a few measurements can be used to evaluate the degree of privacy protection in news recommendation.
For example, in FedRec~\cite{qi2020privacy} the privacy protection ability of the model can be directly indicated by the privacy budget of model gradients.
In addition, privacy protection can also be measured by conducting membership inference attack on user behavior histories to guess whether a behavior belongs to a target user~\cite{qi2021uni}.
These metrics can indicate whether private user information encoded in exchanged models results is well-protected.

\section{Dataset, Competition and Benchmark}\label{dataset}

Many works in the news recommendation field are based on proprietary datasets, such as those collected from Google News~\cite{das2007google}, Yahoo's news~\cite{okura2017embedding}, Bing news~\cite{lian2018towards} and MSN news~\cite{wu2019npa}.
There are only a few publicly available datasets for the research on personalized news recommendation, which are respectively introduced as follows.

The first one is the plista~\cite{Plista} dataset.
It is constructed by collecting the 70,353 news articles from  13 German news portals as well as 1,095,323 news click logs of users.
In the CLEF 2017 NewsREEL task, the organizers publish a new version of the plista dataset, which records users' interactions with news from eight publishers in February 2016. 
This dataset contains 2 million notifications, 58 thousand news updates, and 168 million recommendation requests.
The language used in the plista datasets is German since it is mainly based on the news websites and users in German speaking world.
Note that the number of users is not provided.

The second one is the Adressa~\cite{gulla2017adressa} dataset, which was constructed by collecting  the news logs of the Adresseavisen website in three months. 
It has a full version with logs in 10 weeks and a small version with logs in one week.
The small version contains 561,733 users, 11,207 articles and 2,286,835 clicks, and the full version contains 3,083,438 users, 48,486 articles and  27,223,576 clicks.
The news articles in Adressa are written in Norwegian.

The third one is the Globo~\cite{de2018newssession} dataset, which is retrieved from the Globo news portal in Brazil.
This dataset contains about 314,000 users, 46,000 news articles and 3 million news clicks.
This dataset is in Portuguese, and there is no original news text in this dataset, and it only provides the embeddings of words generated by a neural model that is pre-trained in a news metadata classification task.

The fourth one is a Yahoo!\footnote{https://webscope.sandbox.yahoo.com/catalog.php?datatype=l} dataset for session-based news recommendation.
It contains 14,180 news articles and 34,022 click events.
In this dataset, no news text is provided and the number of users is also unknown because there is no information about user ID.

The fifth one is the MIND~\cite{wu2020mind}\footnote{https://msnews.github.io/} dataset, which is a large-scale English dataset for news recommendation.
This dataset is recently released by MSN News, which contains the real news logs of  1 million users in 6 weeks (from October 12 to November 22, 2019).
It involves 161,013 news articles, 15,777,377 impressions and 24,155,470 news clicks.

We present a comparison of the volume, textual information and leaderboard information of these datasets in Table~\ref{compare}.
We can see that only the MIND dataset is associated with a public leaderboard.
In fact, many researches conducted on other datasets such as Adressa use different dataset preprocessing methods~\cite{zhu2019dan,hu2020graph}, making it difficult to make head-to-head comparisons between the results reported in different papers.
On the contrary, on the MIND dataset the training, validation and test samples are given, and the evaluation metrics are consistent.
Thus, MIND can serve as a standard testbed for news recommendation research.

\begin{table*}[!t]
\centering
\caption{Comparisons of the five public datasets for news recommendation.}\label{compare}
\resizebox{1\textwidth}{!}{
\begin{tabular}{ccccclc}
\Xhline{1.0pt}
\textbf{Dataset} & Language & \# Users & \# News & \# Clicks & News Text & Has Leaderboard? \\ \hline
Plista           &   German       &  Unknown        & 70,353        &  1,095,323         &    title, body   & $\times$     \\ 
Adressa          &   Norwegian       &    3,083,438      &     48,486    &    27,223,576       &  title, body, category, entities   & $\times$           \\ 
Globo            &  Portuguese        &   314,000       &  46,000       &  3,000,000         &   word embeddings of texts     & $\times$       \\ 
Yahoo!          &   English       &    Unknown      &    14,180    &   34,022      &    anonymized word IDs & $\times$ \\ 
MIND             &  English        &    1,000,000  &   161,013  &   24,155,470  &  title, abstract, body, category, entities    & $\checkmark$       \\\Xhline{1.0pt}
\end{tabular}
}
\end{table*}

Based on the datasets introduced above, several competitions and benchmarks on personalized news recommendation have been established.
One representative one is the NEWSREEL challenge held from 2013 to 2017 (in 2013 the challenge is named NRS).\footnote{https://www.newsreelchallenge.org/}
There are usually two tasks in the NEWSREEL challenge.
The first one is news recommendation in a living lab, which are conducted on an operating news recommendation service. 
The goal of recommendation algorithms in this task is  achieving  high news CTRs.
The second one is offline evaluation of news recommendation methods in a simulated environment.
This task is performed based on the plista dataset, and the goal is to predict which news articles a visitor would read in the future.
In the 2017 edition of NewsREEL 87 participants are registered~\cite{kille2017clef}, and two systems achieved CTRs higher than 2\% in the online evaluation task.

Another recent competition is the MIND News Recommendation Competition\footnote{https://msnews.github.io/competition.html}, which is conducted on the MIND dataset.
The goal of this challenge is to predict the click scores of candidate news based on user interests and rank candidate news in each impression.
This challenge attracted more than 200 registered participants and the top submission achieved 71.33\% in terms of AUC.
The leaderboard of this challenge opens after the challenge, and researchers can submit their predictions on the test set to obtain the official evaluation scores.
The current top result on this leaderboard is 73.04\% in terms of AUC, which is achieved  by a recommender named ``UniUM-Fastformer-Pretrain'' based on the techniques in~\cite{wu2021empowering} and~\cite{wu2021fastformer}.
The MIND dataset, challenge and the public leaderboard can form a good benchmark to facilitate research and engineering on personalized news recommendation.
\section{Responsible Personalized News Recommendation}\label{app}

Although personalized news recommendation techniques have achieved notable success in targeting user interest, they still have several issues that may affect user experience and even lead to potential negative social impacts.
There are several critical problems in developing more responsible personalized news recommender systems, including privacy protection, debiasing and fairness, diversity, and content quality, which are discussed in the following sections, respectively.

\subsection{Privacy Protection}

Most existing personalized news recommender systems rely on centralized storage of users' behavior data for user modeling and model training.
However, user behaviors are usually privacy sensitive, and centrally storing them may lead to users' privacy concerns and further risks on data leakage~\cite{li2020federated}.
There are only a few works that study the privacy preservation problem  in news recommendation~\cite{desarkar2014diversification,qi2020privacy}.
For example, FedRec~\cite{qi2020privacy} may be the first attempt to learning privacy-preserving news recommendation model.
Instead of collecting and storing user behavior data in a central server, in FedRec users' news click data are locally stored on user devices.
FedRec uses a federated learning based framework to collaboratively learn news recommendation model.
Each client keeps a local copy of the model and locally computes the model updates based on local data.
The local model updates are uploaded to a central server that coordinates a number of user clients for model training.
The server aggregates the local gradients into a global one to update its maintained global model, and distributes the updated global model to user devices for local update.
In addition, to further protect user privacy, FedRec applies local differential privacy (LDP) techniques to perturb the local model gradients.
Since the protected model gradients usually contain much less private information, user privacy can be better protected.
However, FedRec is only a framework for privacy-preserving   news recommendation model training, and privacy-preserving online serving is still a challenging problem.

Uni-FedRec~\cite{qi2021uni} is an improved version of FedRec that considers both privacy-preserving training and serving.
It has a recall stage to generate dozens of candidates from the news pool in the server and a ranking stage to locally rank candidate news.
In the recall stage, Uni-FedRec generates multiple user embeddings to better cover user interest.
Instead of sending the original user embedding learned by the user model, it decomposes each user embedding into a linear combination of several basis user embeddings, and the combination weights are protected by LDP before sending to the server.
The server reconstructs user embeddings to retrieve candidate news and send them to clients for local ranking.
This framework can be used for both model training and serving in a privacy preserving way.
However, there are still considerable communication costs in this framework.
Efficient-FedRec~\cite{yi2021efficient} further studies how to reduce the communication costs of federated news recommendation model learning.
It decomposes the whole model into a heavy news model and a light-weight user model, where the news model is placed on the server while the user model is kept by clients.
The hidden news representations inferred by the news model on the server are distributed to clients in the model training.
This method provides the potential of incorporating big models such as BERT in federated news recommendation.

Although existing works on privacy-preserving news recommendation have made notable progresses, there are still many challenges in this field, such as the huge performance sacrifice of differential privacy mechanism, the difficulty of involving some context features (e.g., CTR) and collaborative information in GNN, and the difficulty of real-world deployment of federated news recommender systems.

\subsection{Debiasing}

User behavior data usually encodes various kinds of biases.
Some kinds of biases are related to news.
For example, click behaviors are influenced by the positions and sizes of news displayed on the webpages (i.e., presentation bias)~\cite{yi2021debiasedrec}.
In addition, popular news may have higher chances to be clicked than unpopular news (i.e., popularity bias)~\cite{qi2021pprec}.
These types of bias information may affect the accuracy of user interest modeling and model training.
A few works explore to eliminate the influence of certain kinds of bias information to improve personalized news recommendation.
For instance, DebiasRec~\cite{yi2021debiasedrec} aims to reduce the influence of position and size biases on news recommendation.
It uses a bias-aware user modeling method to learn debiased user interest representations, and uses a bias-aware click prediction method that decomposes the overall click score into a bias score and a bias-independent user preference score.
PP-Rec~\cite{qi2021pprec} uses a popularity-aware user modeling method to learn calibrated user interest representations, and it separately models the popularity of news and users' personal preference on news, which can help better model personalized user interest.
These methods mainly aim to infer debiased user interest from biased user data.
However, without any prior knowledge about unbiased data distribution, the bias information usually cannot be fully eliminated.
In addition, many kinds of bias such as exposure and selection biases are rarely studied in the news recommendation field.
Thus, it is important for future research to understand how different biases affect user behaviors and the recommendation model as well as how to eliminate their effect in model training and evaluation.

\subsection{Fairness}

Making fair recommendations is an important problem in responsible news recommendation.
Researchers have studied various kinds of fairness problems in recommendation, such as provider-side fairness and consumer-side fairness~\cite{burke2017multisided}.
In personalized news recommendation, a representative kind of unfairness is brought by the biases related to sensitive user attributes, such as genders and professions.
Users with the same sensitive attributes may have similar patterns in news click behaviors, e.g., fashion news are more preferred by female users.
The model may capture these biases and produce biased recommendation results, e.g., tend to only recommend fashion news to female users.
This will lead to the unfairness problem that some users cannot obtain their interested news information, which is harmful to user experience.
To address this problem, FairRec~\cite{wu2021fairness} uses a decomposed adversarial learning framework with independent user models to learn a bias-aware user embedding and a bias-free user embedding.
The bias-aware user embedding mainly aims to capture bias information related to sensitive user attributes, and the bias-free user embedding aims to model bias-independent user interest.
Both embeddings are regularized to be orthogonal thereby the bias-free user embedding can contain less bias information.
The bias-free user embedding is further used for making fair news recommendations.
By learning user embeddings that are agnostic to the sensitive user attributes, the unfairness brought by the bias information related to sensitive user attributes can be effectively mitigated.
However, adversarial learning based methods are usually brittle and it is difficult to tune their hyperparameters to  fully  remove the bias information.
In addition, many other genres of fairness  (e.g., provider-side fairness) are less studied in news recommendation.
In summary, there are many types of fairness to be improved in news recommendation and it is non-trivial to make both fair and accurate news recommendations.

\subsection{Diversity}

Diversity is critical for personalized news recommendation~\cite{reuver2021no,liu2021interaction,prawesh2021complex}.
Users may not prefer to click news with homogeneous information and improving the information variety is important for improving user experience and engagement~\cite{bernstein2020diversity}.
However, most existing news recommendation methods focus on optimizing recommendation accuracy while ignoring recommendation diversity, and it is shown in~\cite{wu2020sentirec,qi2021pprec,qi2021hierec} that many existing news recommendation methods cannot make sufficiently diverse recommendations.
There are only a few methods that consider the diversity of news recommendation.
Some methods aim to recommend news that are diverse from previously clicked news~\cite{gabrilovich2004newsjunkie,wu2020sentirec}, and several other works explore to diversify the top  news recommendation list~\cite{li2011scene,gabriel2019contextual}.
However, there is still no work on promoting both kinds of diversity in news recommendation.
In addition, many diversity-aware news recommendation methods rely on reranking strategies to improve recommendation diversity, which may not be optimal for achieving a good tradeoff between recommendation accuracy and diversity.
Thus, further research on learning unified diversity-aware news recommendation models is important for improving the quality of online news services.

\subsection{Content Moderation}

The moderation of news content in news recommendation is a rarely studied problem.
In fact, some news articles published online are clickbaits, fake news or containing misinformation.
In addition, some news may encode adversarial clues~\cite{descampe2021automated} or contain low-quality or even harmful content (e.g., racialism and hate speech).
Recommending these news will damage user experience and the reputation of news platforms, and may even lead to negative societal impact~\cite{lazer2018science}.
Although online news platforms can perform  manual moderation on news content quality, the huge amount of online news information makes it too difficult or even impossible to filter all news articles with harmful and useless content. 
Thus, it is important to design news recommendation algorithms that can avoid recommending news with low-quality content.
Researchers have found that news with high ratios of short reading dwell time (e.g., less than 10 seconds) are probably clickbaits~\cite{wu2020ccf}.
In addition, user behaviors such as comments and sharing on social media may also provide rich clues for detecting news that contain misinformation and harmful content~\cite{shu2017fake,banko2020unified}.
Thus, incorporating the various user feedback has the potential to help recommend news with high-quality content, which can improve the responsibility of news recommendation algorithms.

\section{Future Direction and Conclusion}\label{future}

By comprehensively reviewing existing news recommendation techniques in different aspects, we can see that personalized news recommendation techniques have achieved substantial progress over the past years.
However, there remain many challenges and unresolved problems.
Thus, in this section we raise several potential directions that worth exploring in the future.

\subsection{Deep News Understanding}

News modeling is at the heart of personalized news recommendation.
It can be improved in the following aspects.
First, text understanding is a core problem in news modeling, and existing methods may not be capable of  understanding the textual content of news deeply.
Thus, using more advanced NLP techniques (e.g., knowledge-aware PLMs) may help better understand news texts and improve news modeling.
Second, besides textual information, news also contain rich multimodal information such as images, videos and slides.
The multimodal news content can provide complementary information on news understanding.
Thus, using multimodal content modeling techniques has the potential to improve the comprehensiveness of news understanding.
Third, there are many useful factors for news modeling that are not covered by news content, such as publisher, popularity and recency.
A unified framework is required to incorporate various kinds of news information (e.g., property features and context features) and meanwhile effectively model the relatedness between different features.
Further research on these directions can help understand news more accurately and deeply to  empower subsequent user modeling and news ranking.

\subsection{Universal User Modeling}

User modeling is critical for understanding users' interest in news.
However, it is difficult to model the dynamic and diverse user interest accurately and comprehensively for news recommendation.
To tackle this problem, a universal user modeling framework that can model various kinds of user interest is needed.
We argue that this framework should satisfy the following requirements.
First, the user modeling framework needs to comprehensively infer user interest from multiple kinds of user behaviors and feedback.
This is because click behaviors are very noisy and may be sparse for some users, and it is insufficient to model user interests solely from click behaviors.
Fortunately, different kinds of user behaviors and feedback (e.g., read and dislike) can provide rich complementary information like user engagement, and incorporating them in a unified framework can better support user modeling.
Second, the framework needs to model the diverse and multi-grained user interest.
Since a single user embedding may be insufficient to comprehensively model user interests, it may be a promising way to represent user interest with more sophisticated structures such as embedding sets and graphs to improve the understanding of user interest. 
Third, the framework needs to capture the dynamics of user interests.
Since user interest usually evolves with time, it is important to understand user interest in different periods and further model their inherent relations.
To meet this end, using more advanced sequence modeling techniques may help improve user interest modeling in personalized news recommendation.

\subsection{Effective and Efficient Personalized Ranking}

News ranking is an essential step to make personalized news recommendations.
There are mainly three research directions to improve news ranking.
First, most existing personalized ranking methods are mainly based the coarse-grained relevance between candidate news and user interest, which may not be optimal for accurately targeting user interest.
Although a few methods can model the fine-grained relatedness between user and news, they are inefficient and may not be suitable for scenarios with limited computation resources and latency tolerance.
Thus, developing both effective and efficient personalized ranking methods is important for improving online news recommendation.
Second, ranking news solely based on relevance may lead to the filter bubble problem.
It is important to design more sophisticated news ranking strategies to achieve a good tradeoff between accuracy and diversity.
Third, most existing news ranking methods are greedy, i.e., only consider the current ranking list in the ranking policy.
However, they may not be optimal for achieving  good  user engagement in the long-term.
Thus, designing proper news ranking strategies to optimize long-term rewards may be beneficial for user experience.

\subsection{Hyperbolic Representation Learning for News Recommendation}

In most existing news recommendation methods, news and users representations are learned in Euclidean space.
Matching functions such as inner product and cosine similarity are widely used for computing relevance scores for news ranking.
However, representation learning in Euclidean space is ineffective in capturing the hierarchical structure of data, while hyperbolic representation learning is much better at it.
There are many inherent hierarchical data structures in personalized news recommendation, such as different levels of user interests, news topics, and commonsense knowledge encoded by knowledge graphs.
Thus, news recommendation with hyperbolic representation learning may be a promising solution.
There are several existing neural architectures in hyperbolic space, such as hyperbolic attention~\cite{gulcehre2018hyperbolic} and hyperbolic GCN~\cite{chami2019hyperbolic}, which can serve as the core model components in news recommendation.
In addition, there have been several successful applications of hyperbolic representation learning to CF-based recommendation~\cite{vinh2018hyperbolic,sun2021hgcf} and knowledge graph embedding~\cite{chami2020low,wang2020h2kgat}, which can provide useful guidance of collaborative signal modeling and knowledge exploitation in news recommendation.
Future research on hyperbolic representation learning may create a new direction to overcome several drawbacks of current user/news modeling and personalized ranking techniques conducted in Euclidean space.

\subsection{Unified Model Training}

Model training techniques are also important for learning effective and robust personalized news recommendation models.
There are four potential directions for future works to improve model training.
First, most methods only use click signals for model training, which may be inaccurate because click signals are  usually noisy and biased.
In addition, the supervision signals in specific tasks may also be insufficient~\cite{wu2020ptum}.
Thus, a unified framework to incorporate various kinds of supervised and self-supervised training signals and objectives for collaborative model learning can effectively improve the model quality. 
Second, although several methods explore to use multi-task learning frameworks to incorporate multiple objectives into model training, they need to manually tune the loss coefficients of different tasks in model training, which usually require much human effort and may be sensitive to the characteristics of datasets.
Thus, a self-adaptive  multi-task learning framework to automatically tune hyperparameters like loss coefficients can reduce the  developing effort and improve the model generality.
Third, many methods use randomly selected negative samples for model training, which may be noisy and  less informative.
Thus, using more effective negative sampling can help train more robust and accurate news recommendation models.
Fourth, offline trained models may have gaps with the online scenarios and may suffer from the performance decline with time.
Thus, it is important to incorporate both offline and online learning techniques to help the model better adapt to the latest online serving requirements.

\subsection{News Recommendation in Social Context}

On some news platforms, users may have social interactions with other users in many ways, such as leaving comments, replies, and sharing to their social media blogs like Twitter.
The social interactions among users concerning certain news can usually reflect their opinions, preferences, and satisfaction on the recommended news~\cite{wei2011effective}, which can provide rich complementary information to user modeling.
In addition, users' discussions and dissemination behaviors can also help understand the content, quality and authenticity of news~\cite{shu2017fake}.
Besides, they can help recognize breaking news and adjust recommendation results accordingly~\cite{phelan2011terms}.
Therefore, the social contexts of news recommendation play an important role.
However, they are usually neglected by news recommendation researches in recent years.
In future researches, it is an interesting topic to study the impacts of users' online social interactions on the accuracy, timeliness and quality of news personalization.

\subsection{Privacy-preserving News Recommendation}

In recent years, the ethical issues of intelligent systems have attracted much attention from both the academia and public.
Developing more responsible news recommender systems can help better serve users of online news services with smaller risks.
One important direction for improving the responsibility of personalized news recommendation is user privacy protection.
Although a few works like~\cite{qi2020privacy} explore to use federated learning techniques to train news recommendation models in a privacy-preserving way, there are still many challenges in developing a privacy-preserving news recommender system.
First, given a model learned in a federated way, it is still challenging to deploy it online to serve users efficiently.
Second, there may also be potential privacy risks during the training and serving of news recommendation models, and canonical differential privacy techniques usually lead to a heavy sacrifice on model utility.
Third, the data isolation problem in federated learning framework settings makes it difficult to exploit some context features like CTR and collaborative information in GNN.
Thus, further researches on developing more effective, efficient and privacy-preserving news recommendation methods are needed.

\subsection{Secure and Robust News Recommendation}

Existing researches on news recommendation focus on building algorithms in a trusted environment.
However, in real-world scenarios there may be various kinds of threats brought by malicious users and platforms.
For example, existing news recommendation methods are vulnerable to poisoning attacks, which aim to promote certain items, trigger certain backdoors, or degrade the recommender system performance.
In addition, news recommender systems may be sensitive to adversarial samples.
When news recommender systems are trained in the federated learning framework, the threats from the untrusted outside environment become even more serious.
Unfortunately, although the security and robustness of personalized news recommender systems are critical, researches on this problem are rather limited.
Future studies on secure and robust news recommendation are important for the stability and reliability of online news platforms.

\subsection{Diversity-aware News Recommendation}

Besides accuracy, diversity in news recommendation also has decisive influence on user experience.
There are three main research directions to improve the diversity of news recommendation.
The first one is temporal-spatial diversity-aware  news recommendation, which aims to recommend news that are diverse from each other and meanwhile diverse from historical clicked news.
This can help the recommendation results better satisfy users' preference on information variety.
The second one is personalizing the diversity in news recommendation.
Different users may have different preferences on the tradeoff between accuracy and diversity, and it may be better to consider their personalized preference to improve user experience.
The third one is fine-grained diversity, which aims to not only diversify the content and topic of news, but also many other factors like publishers, locations, opinions and emotions.
It has the potential to make higher-quality diversity-aware news recommendations.

\subsection{Bias-free News Recommendation}

Debiasing is another important problem in improving the  responsibility of news recommendation.
The biases encoded by user behavior data will propagate to the recommendation model and may further be amplified in the loops of recommendation.
Thus, designing effective methods to eliminate the influence of the various kinds of biases on recommendation results is important for making high-quality news recommendations.
There are several potential research directions in this field.
First, it is important to understand the influence of different kinds of biases on user behaviors and the recommendation model, which can help the subsequent debiasing.
Second, different users may be influenced by the same bias information in different ways, and considering the personalized preference of users on bias information can help better eliminate the effects of biases.
Third, there are various kinds of biases in news recommendation.
A unified debiasing framework that can simultaneously reduce the effects of different biases can greatly improve the accuracy and robustness of news recommendation algorithms.

\subsection{Fairness-aware News Recommendation}

Fairness is an essential  but often ignored factor in personalized news recommendation.
A fair news recommender system is required to provide fair recommendation services to different groups of users and  give fair chances to news from different providers to be recommended.
Future research on fair news recommendation can be conducted in the following three directions.
First, it is important to reduce the consumer-side unfairness related to  sensitive user attributes. 
Although adversarial learning techniques are  mature solutions to this problem, they are usually brittle and difficult to tune.
Thus, more robust and effective methods are required to remove the biases introduced by sensitive user attributes.
Second, different news providers and publishers are diverse in their characteristics, such as topic preference and reputation.
Thus, it is non-trivial to properly balance the recommendation chances of  news from different providers and publishers to achieve better provider-side fairness.
Third, there are different types of fairness in the personalized news recommendation scenario, and it is very challenging to simultaneously achieve multi-side fairness without a heavy sacrifice of recommendation accuracy.

\subsection{Content Moderation in News Recommendation}

The moderation of news content is important for online news platforms to avoid recommending news with low quality or harmful content to users  and mitigate their impact on users and society.
However, this issue is rarely studied and cannot be resolved by most existing news recommendation methods.
There are three key research directions on this problem.
First, it is essential to understand the generation and spreading mechanism of harmful news as well as their impact on users, which can help news platforms better defend toxic content.
Second, it may be useful to incorporate content moderation techniques like fake news detection~\cite{shu2017fake} and clickbait detection~\cite{wu2020clickbait} into news recommendation to adjust the recommendation results according to the quality of news content.
Third, without the assistance of additional tasks and resources, we can learn content quality-aware news recommendation models with the guidance of certain kinds of user feedback such as comments and dislikes, which is expected to help recommend high-quality news to users.

\subsection{Societal Impact of News Recommendation}

News recommender systems can generate societal impact when they serve a certain number of users.
They may imperceptibly influence the opinions and views of users when displaying personalized news content~\cite{makhortykh2021can}. 
Thus, it is valuable for further research  to identify and analyze the societal impact of personalized news recommendation algorithms, such as their influence on political events, economic activities and psychological health.
In addition, research on how to reduce the potential negative societal impact of personalized news recommendation methods can help avoid their risky behaviors and better serve online users. 

\subsection*{Conclusion}

Finally, we present a conclusion to this survey paper.
In this survey, we conduct a comprehensive overview of the personalized news recommendation field, including the technologies involved in different core modules of a personalized news recommender, the dataset and metrics for performance evaluation, the key points for developing responsible personalized news recommender systems, and potential directions to be explored in the future.
Different from existing survey papers that follow the conventional taxonomy of news recommendation methods, in this paper we provide a novel perspective to understand personalized news recommendation from its key problems and the associated techniques and challenges.
In addition, this is the first survey paper that comprehensively covers both traditional and up-to-date deep learning techniques for personalized news recommendation, which can provide rich insights for extending the frontier of this field.
We hope this paper can facilitate future research on personalized news recommendation as well as related fields in NLP and data mining. 

\section*{Acknowledgments}
This work was supported by the National Natural Science Foundation of China under Grant numbers U1936208, U1936216, 61862002, and U1705261.

\bibliographystyle{ACM-Reference-Format}
\bibliography{main}


\begin{thebibliography}{249}


\ifx \showCODEN    \undefined \def \showCODEN     #1{\unskip}     \fi
\ifx \showDOI      \undefined \def \showDOI       #1{#1}\fi
\ifx \showISBNx    \undefined \def \showISBNx     #1{\unskip}     \fi
\ifx \showISBNxiii \undefined \def \showISBNxiii  #1{\unskip}     \fi
\ifx \showISSN     \undefined \def \showISSN      #1{\unskip}     \fi
\ifx \showLCCN     \undefined \def \showLCCN      #1{\unskip}     \fi
\ifx \shownote     \undefined \def \shownote      #1{#1}          \fi
\ifx \showarticletitle \undefined \def \showarticletitle #1{#1}   \fi
\ifx \showURL      \undefined \def \showURL       {\relax}        \fi
\providecommand\bibfield[2]{#2}
\providecommand\bibinfo[2]{#2}
\providecommand\natexlab[1]{#1}
\providecommand\showeprint[2][]{arXiv:#2}

\bibitem[\protect\citeauthoryear{Abdollahpouri, Malthouse, Konstan, Mobasher,
  and Gilbert}{Abdollahpouri et~al\mbox{.}}{2021}]%
        {abdollahpouri2021toward}
\bibfield{author}{\bibinfo{person}{Himan Abdollahpouri},
  \bibinfo{person}{Edward~C Malthouse}, \bibinfo{person}{Joseph~A Konstan},
  \bibinfo{person}{Bamshad Mobasher}, {and} \bibinfo{person}{Jeremy Gilbert}.}
  \bibinfo{year}{2021}\natexlab{}.
\newblock \showarticletitle{Toward the next generation of news recommender
  systems}. In \bibinfo{booktitle}{\emph{Companion Proceedings of WWW}}.
  \bibinfo{pages}{402--406}.
\newblock


\bibitem[\protect\citeauthoryear{An, Wu, Wu, Zhang, Liu, and Xie}{An
  et~al\mbox{.}}{2019}]%
        {an2019neural}
\bibfield{author}{\bibinfo{person}{Mingxiao An}, \bibinfo{person}{Fangzhao Wu},
  \bibinfo{person}{Chuhan Wu}, \bibinfo{person}{Kun Zhang},
  \bibinfo{person}{Zheng Liu}, {and} \bibinfo{person}{Xing Xie}.}
  \bibinfo{year}{2019}\natexlab{}.
\newblock \showarticletitle{Neural news recommendation with long-and short-term
  user representations}. In \bibinfo{booktitle}{\emph{ACL}}.
  \bibinfo{pages}{336--345}.
\newblock


\bibitem[\protect\citeauthoryear{Bai, Cambazoglu, Gullo, Mantrach, and
  Silvestri}{Bai et~al\mbox{.}}{2017}]%
        {bai2017exploiting}
\bibfield{author}{\bibinfo{person}{Xiao Bai}, \bibinfo{person}{B~Barla
  Cambazoglu}, \bibinfo{person}{Francesco Gullo}, \bibinfo{person}{Amin
  Mantrach}, {and} \bibinfo{person}{Fabrizio Silvestri}.}
  \bibinfo{year}{2017}\natexlab{}.
\newblock \showarticletitle{Exploiting search history of users for news
  personalization}.
\newblock \bibinfo{journal}{\emph{Information Science}}  \bibinfo{volume}{385}
  (\bibinfo{year}{2017}), \bibinfo{pages}{125--137}.
\newblock


\bibitem[\protect\citeauthoryear{Banko, MacKeen, and Ray}{Banko
  et~al\mbox{.}}{2020}]%
        {banko2020unified}
\bibfield{author}{\bibinfo{person}{Michele Banko}, \bibinfo{person}{Brendon
  MacKeen}, {and} \bibinfo{person}{Laurie Ray}.}
  \bibinfo{year}{2020}\natexlab{}.
\newblock \showarticletitle{A unified taxonomy of harmful content}. In
  \bibinfo{booktitle}{\emph{WOAH@EMNLP}}. \bibinfo{pages}{125--137}.
\newblock


\bibitem[\protect\citeauthoryear{Bastian, Helberger, and Makhortykh}{Bastian
  et~al\mbox{.}}{2021}]%
        {bastian2021safeguarding}
\bibfield{author}{\bibinfo{person}{Mariella Bastian}, \bibinfo{person}{Natali
  Helberger}, {and} \bibinfo{person}{Mykola Makhortykh}.}
  \bibinfo{year}{2021}\natexlab{}.
\newblock \showarticletitle{Safeguarding the Journalistic DNA: Attitudes
  towards the Role of Professional Values in Algorithmic News Recommender
  Designs}.
\newblock \bibinfo{journal}{\emph{Digital Journalism}} \bibinfo{volume}{9},
  \bibinfo{number}{6} (\bibinfo{year}{2021}), \bibinfo{pages}{835--863}.
\newblock


\bibitem[\protect\citeauthoryear{Bernstein, de~Vreese, Helberger, Schulz,
  Zweig, Baden, Beam, Hauer, Heitz, J{\"u}rgens, et~al\mbox{.}}{Bernstein
  et~al\mbox{.}}{2020}]%
        {bernstein2020diversity}
\bibfield{author}{\bibinfo{person}{Abraham Bernstein}, \bibinfo{person}{Claes
  de Vreese}, \bibinfo{person}{Natali Helberger}, \bibinfo{person}{Wolfgang
  Schulz}, \bibinfo{person}{Katharina Zweig}, \bibinfo{person}{Christian
  Baden}, \bibinfo{person}{Michael~A Beam}, \bibinfo{person}{Marc~P Hauer},
  \bibinfo{person}{Lucien Heitz}, \bibinfo{person}{Pascal J{\"u}rgens},
  {et~al\mbox{.}}} \bibinfo{year}{2020}\natexlab{}.
\newblock \showarticletitle{Diversity in news recommendations}.
\newblock \bibinfo{journal}{\emph{arXiv preprint arXiv:2005.09495}}
  (\bibinfo{year}{2020}).
\newblock


\bibitem[\protect\citeauthoryear{Billsus and Pazzani}{Billsus and
  Pazzani}{2000}]%
        {billsus2000user}
\bibfield{author}{\bibinfo{person}{Daniel Billsus} {and}
  \bibinfo{person}{Michael~J Pazzani}.} \bibinfo{year}{2000}\natexlab{}.
\newblock \showarticletitle{User modeling for adaptive news access}.
\newblock \bibinfo{journal}{\emph{User Modeling and User-adapted Interaction}}
  \bibinfo{volume}{10}, \bibinfo{number}{2-3} (\bibinfo{year}{2000}),
  \bibinfo{pages}{147--180}.
\newblock


\bibitem[\protect\citeauthoryear{Bogers and Van~den Bosch}{Bogers and Van~den
  Bosch}{2007}]%
        {bogers2007comparing}
\bibfield{author}{\bibinfo{person}{Toine Bogers} {and} \bibinfo{person}{Antal
  Van~den Bosch}.} \bibinfo{year}{2007}\natexlab{}.
\newblock \showarticletitle{Comparing and evaluating information retrieval
  algorithms for news recommendation}. In \bibinfo{booktitle}{\emph{Recsys}}.
  \bibinfo{pages}{141--144}.
\newblock


\bibitem[\protect\citeauthoryear{Bordes, Usunier, Garcia-Duran, Weston, and
  Yakhnenko}{Bordes et~al\mbox{.}}{2013}]%
        {bordes2013translating}
\bibfield{author}{\bibinfo{person}{Antoine Bordes}, \bibinfo{person}{Nicolas
  Usunier}, \bibinfo{person}{Alberto Garcia-Duran}, \bibinfo{person}{Jason
  Weston}, {and} \bibinfo{person}{Oksana Yakhnenko}.}
  \bibinfo{year}{2013}\natexlab{}.
\newblock \showarticletitle{Translating embeddings for modeling
  multi-relational data}. In \bibinfo{booktitle}{\emph{NIPS}}.
  \bibinfo{pages}{1--9}.
\newblock


\bibitem[\protect\citeauthoryear{Borges and Lorena}{Borges and Lorena}{2010}]%
        {borges2010survey}
\bibfield{author}{\bibinfo{person}{Hugo~L Borges} {and} \bibinfo{person}{Ana~C
  Lorena}.} \bibinfo{year}{2010}\natexlab{}.
\newblock \showarticletitle{A survey on recommender systems for news data}.
\newblock In \bibinfo{booktitle}{\emph{Smart Information and Knowledge
  Management}}. \bibinfo{publisher}{Springer}, \bibinfo{pages}{129--151}.
\newblock


\bibitem[\protect\citeauthoryear{Brocken, Hartveld, de~Koning, van Noort,
  Hogenboom, Frasincar, and Robal}{Brocken et~al\mbox{.}}{2019}]%
        {brocken2019bing}
\bibfield{author}{\bibinfo{person}{Emma Brocken}, \bibinfo{person}{Aron
  Hartveld}, \bibinfo{person}{Emma de Koning}, \bibinfo{person}{Thomas van
  Noort}, \bibinfo{person}{Frederik Hogenboom}, \bibinfo{person}{Flavius
  Frasincar}, {and} \bibinfo{person}{Tarmo Robal}.}
  \bibinfo{year}{2019}\natexlab{}.
\newblock \showarticletitle{Bing-CF-IDF+: A Semantics-Driven News Recommender
  System}. In \bibinfo{booktitle}{\emph{CAiSE}}. Springer,
  \bibinfo{pages}{32--47}.
\newblock


\bibitem[\protect\citeauthoryear{Burke}{Burke}{2017}]%
        {burke2017multisided}
\bibfield{author}{\bibinfo{person}{Robin Burke}.}
  \bibinfo{year}{2017}\natexlab{}.
\newblock \showarticletitle{Multisided fairness for recommendation}.
\newblock \bibinfo{journal}{\emph{arXiv preprint arXiv:1707.00093}}
  (\bibinfo{year}{2017}).
\newblock


\bibitem[\protect\citeauthoryear{Caldarelli, Gurini, Micarelli, and
  Sansonetti}{Caldarelli et~al\mbox{.}}{2016}]%
        {caldarelli2016signal}
\bibfield{author}{\bibinfo{person}{Sirian Caldarelli},
  \bibinfo{person}{Davide~Feltoni Gurini}, \bibinfo{person}{Alessandro
  Micarelli}, {and} \bibinfo{person}{Giuseppe Sansonetti}.}
  \bibinfo{year}{2016}\natexlab{}.
\newblock \showarticletitle{A Signal-Based Approach to News Recommendation.}.
  In \bibinfo{booktitle}{\emph{UMAP (Extended Proceedings)}}.
\newblock


\bibitem[\protect\citeauthoryear{Cantador and Castells}{Cantador and
  Castells}{2009}]%
        {cantador2009semantic}
\bibfield{author}{\bibinfo{person}{Iv{\'a}n Cantador} {and}
  \bibinfo{person}{Pablo Castells}.} \bibinfo{year}{2009}\natexlab{}.
\newblock \showarticletitle{Semantic contextualisation in a news recommender
  system}.
\newblock  (\bibinfo{year}{2009}).
\newblock


\bibitem[\protect\citeauthoryear{Cantador, Castells, and
  Bellog{\'\i}n}{Cantador et~al\mbox{.}}{2011}]%
        {cantador2011enhanced}
\bibfield{author}{\bibinfo{person}{Iv{\'a}n Cantador}, \bibinfo{person}{Pablo
  Castells}, {and} \bibinfo{person}{Alejandro Bellog{\'\i}n}.}
  \bibinfo{year}{2011}\natexlab{}.
\newblock \showarticletitle{An enhanced semantic layer for hybrid recommender
  systems: Application to news recommendation}.
\newblock \bibinfo{journal}{\emph{Int. Journal on Semantic Web and Inf. Syst.}}
  \bibinfo{volume}{7}, \bibinfo{number}{1} (\bibinfo{year}{2011}),
  \bibinfo{pages}{44--78}.
\newblock


\bibitem[\protect\citeauthoryear{Capelle, Frasincar, Moerland, and
  Hogenboom}{Capelle et~al\mbox{.}}{2012}]%
        {capelle2012semantics}
\bibfield{author}{\bibinfo{person}{Michel Capelle}, \bibinfo{person}{Flavius
  Frasincar}, \bibinfo{person}{Marnix Moerland}, {and}
  \bibinfo{person}{Frederik Hogenboom}.} \bibinfo{year}{2012}\natexlab{}.
\newblock \showarticletitle{Semantics-based news recommendation}. In
  \bibinfo{booktitle}{\emph{WIMS}}. \bibinfo{pages}{1--9}.
\newblock


\bibitem[\protect\citeauthoryear{Capelle, Hogenboom, Hogenboom, and
  Frasincar}{Capelle et~al\mbox{.}}{2013}]%
        {capelle2013semantic}
\bibfield{author}{\bibinfo{person}{Michel Capelle}, \bibinfo{person}{Frederik
  Hogenboom}, \bibinfo{person}{Alexander Hogenboom}, {and}
  \bibinfo{person}{Flavius Frasincar}.} \bibinfo{year}{2013}\natexlab{}.
\newblock \showarticletitle{Semantic news recommendation using wordnet and bing
  similarities}. In \bibinfo{booktitle}{\emph{SAC}}. \bibinfo{pages}{296--302}.
\newblock


\bibitem[\protect\citeauthoryear{Capelle, Moerland, Hogenboom, Frasincar, and
  Vandic}{Capelle et~al\mbox{.}}{2015}]%
        {capelle2015bing}
\bibfield{author}{\bibinfo{person}{Michel Capelle}, \bibinfo{person}{Marnix
  Moerland}, \bibinfo{person}{Frederik Hogenboom}, \bibinfo{person}{Flavius
  Frasincar}, {and} \bibinfo{person}{Damir Vandic}.}
  \bibinfo{year}{2015}\natexlab{}.
\newblock \showarticletitle{Bing-SF-IDF+ a hybrid semantics-driven news
  recommender}. In \bibinfo{booktitle}{\emph{SAC}}. \bibinfo{pages}{732--739}.
\newblock


\bibitem[\protect\citeauthoryear{Chami, Wolf, Juan, Sala, Ravi, and
  R{\'e}}{Chami et~al\mbox{.}}{2020}]%
        {chami2020low}
\bibfield{author}{\bibinfo{person}{Ines Chami}, \bibinfo{person}{Adva Wolf},
  \bibinfo{person}{Da-Cheng Juan}, \bibinfo{person}{Frederic Sala},
  \bibinfo{person}{Sujith Ravi}, {and} \bibinfo{person}{Christopher R{\'e}}.}
  \bibinfo{year}{2020}\natexlab{}.
\newblock \showarticletitle{Low-Dimensional Hyperbolic Knowledge Graph
  Embeddings}. In \bibinfo{booktitle}{\emph{ACL}}. \bibinfo{pages}{6901--6914}.
\newblock


\bibitem[\protect\citeauthoryear{Chami, Ying, R{\'e}, and Leskovec}{Chami
  et~al\mbox{.}}{2019}]%
        {chami2019hyperbolic}
\bibfield{author}{\bibinfo{person}{Ines Chami}, \bibinfo{person}{Zhitao Ying},
  \bibinfo{person}{Christopher R{\'e}}, {and} \bibinfo{person}{Jure Leskovec}.}
  \bibinfo{year}{2019}\natexlab{}.
\newblock \showarticletitle{Hyperbolic graph convolutional neural networks}.
\newblock \bibinfo{journal}{\emph{NeurIPS}}  \bibinfo{volume}{32}
  (\bibinfo{year}{2019}), \bibinfo{pages}{4868--4879}.
\newblock


\bibitem[\protect\citeauthoryear{Chen, Meng, Xu, and Lukasiewicz}{Chen
  et~al\mbox{.}}{2017}]%
        {chen2017location}
\bibfield{author}{\bibinfo{person}{Cheng Chen}, \bibinfo{person}{Xiangwu Meng},
  \bibinfo{person}{Zhenghua Xu}, {and} \bibinfo{person}{Thomas Lukasiewicz}.}
  \bibinfo{year}{2017}\natexlab{}.
\newblock \showarticletitle{Location-aware personalized news recommendation
  with deep semantic analysis}.
\newblock \bibinfo{journal}{\emph{IEEE Access}}  \bibinfo{volume}{5}
  (\bibinfo{year}{2017}), \bibinfo{pages}{1624--1638}.
\newblock


\bibitem[\protect\citeauthoryear{Chen, Zhang, Chen, and Bu}{Chen
  et~al\mbox{.}}{2009}]%
        {chen2009hybrid}
\bibfield{author}{\bibinfo{person}{Wei Chen}, \bibinfo{person}{Li-jun Zhang},
  \bibinfo{person}{Chun Chen}, {and} \bibinfo{person}{Jia-jun Bu}.}
  \bibinfo{year}{2009}\natexlab{}.
\newblock \showarticletitle{A hybrid phonic web news recommender system for
  pervasive access}. In \bibinfo{booktitle}{\emph{Proc. of the WRI Int. Conf.
  on Commun. and Mobile Comput.}}, Vol.~\bibinfo{volume}{3}. IEEE,
  \bibinfo{pages}{122--126}.
\newblock


\bibitem[\protect\citeauthoryear{Cho, Lim, Park, Yoo, and Park}{Cho
  et~al\mbox{.}}{2021}]%
        {cho2021overlooked}
\bibfield{author}{\bibinfo{person}{Sungmin Cho}, \bibinfo{person}{Hongjun Lim},
  \bibinfo{person}{Keunchan Park}, \bibinfo{person}{Sungjoo Yoo}, {and}
  \bibinfo{person}{Eunhyeok Park}.} \bibinfo{year}{2021}\natexlab{}.
\newblock \showarticletitle{On the Overlooked Significance of Underutilized
  Contextual Features in Recent News Recommendation Models}.
\newblock \bibinfo{journal}{\emph{arXiv preprint arXiv:2112.14370}}
  (\bibinfo{year}{2021}).
\newblock


\bibitem[\protect\citeauthoryear{Chu, Liu, Sun, and Zhou}{Chu
  et~al\mbox{.}}{2019}]%
        {chu2019next}
\bibfield{author}{\bibinfo{person}{Qianfeng Chu}, \bibinfo{person}{Gongshen
  Liu}, \bibinfo{person}{Huanrong Sun}, {and} \bibinfo{person}{Cheng Zhou}.}
  \bibinfo{year}{2019}\natexlab{}.
\newblock \showarticletitle{Next News Recommendation via Knowledge-Aware
  Sequential Model}. In \bibinfo{booktitle}{\emph{CCL}}. Springer,
  \bibinfo{pages}{221--232}.
\newblock


\bibitem[\protect\citeauthoryear{Chu and Park}{Chu and Park}{2009}]%
        {chu2009personalized}
\bibfield{author}{\bibinfo{person}{Wei Chu} {and} \bibinfo{person}{Seung-Taek
  Park}.} \bibinfo{year}{2009}\natexlab{}.
\newblock \showarticletitle{Personalized recommendation on dynamic content
  using predictive bilinear models}. In \bibinfo{booktitle}{\emph{WWW}}.
  \bibinfo{pages}{691--700}.
\newblock


\bibitem[\protect\citeauthoryear{Claypool, Gokhale, Miranda, Murnikov, Netes,
  and Sartin}{Claypool et~al\mbox{.}}{1999}]%
        {claypool1999combing}
\bibfield{author}{\bibinfo{person}{Mark Claypool}, \bibinfo{person}{Anuja
  Gokhale}, \bibinfo{person}{Tim Miranda}, \bibinfo{person}{Paul Murnikov},
  \bibinfo{person}{Dmitry Netes}, {and} \bibinfo{person}{Matthew Sartin}.}
  \bibinfo{year}{1999}\natexlab{}.
\newblock \showarticletitle{Combing content-based and collaborative filters in
  an online newspaper}. In \bibinfo{booktitle}{\emph{Recommender
  Systems@SIGIR}}.
\newblock


\bibitem[\protect\citeauthoryear{Corsini and Larson}{Corsini and
  Larson}{2016}]%
        {corsini2016clef}
\bibfield{author}{\bibinfo{person}{F Corsini} {and} \bibinfo{person}{MA
  Larson}.} \bibinfo{year}{2016}\natexlab{}.
\newblock \showarticletitle{CLEF NewsREEL 2016: Image based Recommendation}. In
  \bibinfo{booktitle}{\emph{CLEF}}. \bibinfo{pages}{593--605}.
\newblock


\bibitem[\protect\citeauthoryear{Darvishy, Ibrahim, Sidi, and
  Mustapha}{Darvishy et~al\mbox{.}}{2020}]%
        {darvishy2020hypner}
\bibfield{author}{\bibinfo{person}{Asghar Darvishy}, \bibinfo{person}{Hamidah
  Ibrahim}, \bibinfo{person}{Fatimah Sidi}, {and} \bibinfo{person}{Aida
  Mustapha}.} \bibinfo{year}{2020}\natexlab{}.
\newblock \showarticletitle{HYPNER: A Hybrid Approach for Personalized News
  Recommendation}.
\newblock \bibinfo{journal}{\emph{IEEE Access}}  \bibinfo{volume}{8}
  (\bibinfo{year}{2020}), \bibinfo{pages}{46877--46894}.
\newblock


\bibitem[\protect\citeauthoryear{Das, Datar, Garg, and Rajaram}{Das
  et~al\mbox{.}}{2007}]%
        {das2007google}
\bibfield{author}{\bibinfo{person}{Abhinandan~S Das}, \bibinfo{person}{Mayur
  Datar}, \bibinfo{person}{Ashutosh Garg}, {and} \bibinfo{person}{Shyam
  Rajaram}.} \bibinfo{year}{2007}\natexlab{}.
\newblock \showarticletitle{Google news personalization: scalable online
  collaborative filtering}. In \bibinfo{booktitle}{\emph{WWW}}.
  \bibinfo{pages}{271--280}.
\newblock


\bibitem[\protect\citeauthoryear{de~Koning, Hogenboom, and Frasincar}{de~Koning
  et~al\mbox{.}}{2018}]%
        {de2018news}
\bibfield{author}{\bibinfo{person}{Emma de Koning}, \bibinfo{person}{Frederik
  Hogenboom}, {and} \bibinfo{person}{Flavius Frasincar}.}
  \bibinfo{year}{2018}\natexlab{}.
\newblock \showarticletitle{News recommendation with CF-IDF+}. In
  \bibinfo{booktitle}{\emph{Int. Conf. on Adv. Information Syst. Eng.}}
  Springer, \bibinfo{pages}{170--184}.
\newblock


\bibitem[\protect\citeauthoryear{de~Souza Pereira~Moreira, Ferreira, and
  da~Cunha}{de~Souza Pereira~Moreira et~al\mbox{.}}{2018}]%
        {de2018newssession}
\bibfield{author}{\bibinfo{person}{Gabriel de Souza Pereira~Moreira},
  \bibinfo{person}{Felipe Ferreira}, {and} \bibinfo{person}{Adilson~Marques da
  Cunha}.} \bibinfo{year}{2018}\natexlab{}.
\newblock \showarticletitle{News session-based recommendations using deep
  neural networks}. In \bibinfo{booktitle}{\emph{DLRS@RecSys}}.
  \bibinfo{pages}{15--23}.
\newblock


\bibitem[\protect\citeauthoryear{Desarkar and Shinde}{Desarkar and
  Shinde}{2014}]%
        {desarkar2014diversification}
\bibfield{author}{\bibinfo{person}{Maunendra~Sankar Desarkar} {and}
  \bibinfo{person}{Neha Shinde}.} \bibinfo{year}{2014}\natexlab{}.
\newblock \showarticletitle{Diversification in news recommendation for privacy
  concerned users}. In \bibinfo{booktitle}{\emph{DSAA}}. IEEE,
  \bibinfo{pages}{135--141}.
\newblock


\bibitem[\protect\citeauthoryear{Descampe, Massart, Poelman, Standaert, and
  Standaert}{Descampe et~al\mbox{.}}{2021}]%
        {descampe2021automated}
\bibfield{author}{\bibinfo{person}{Antonin Descampe},
  \bibinfo{person}{Cl{\'e}ment Massart}, \bibinfo{person}{Simon Poelman},
  \bibinfo{person}{Fran{\c{c}}ois-Xavier Standaert}, {and}
  \bibinfo{person}{Olivier Standaert}.} \bibinfo{year}{2021}\natexlab{}.
\newblock \showarticletitle{Automated news recommendation in front of
  adversarial examples and the technical limits of transparency in algorithmic
  accountability}.
\newblock \bibinfo{journal}{\emph{AI \& SOCIETY}} (\bibinfo{year}{2021}),
  \bibinfo{pages}{1--14}.
\newblock


\bibitem[\protect\citeauthoryear{Devlin, Chang, Lee, and Toutanova}{Devlin
  et~al\mbox{.}}{2019}]%
        {devlin2019bert}
\bibfield{author}{\bibinfo{person}{Jacob Devlin}, \bibinfo{person}{Ming-Wei
  Chang}, \bibinfo{person}{Kenton Lee}, {and} \bibinfo{person}{Kristina
  Toutanova}.} \bibinfo{year}{2019}\natexlab{}.
\newblock \showarticletitle{BERT: Pre-training of Deep Bidirectional
  Transformers for Language Understanding}. In
  \bibinfo{booktitle}{\emph{NAACL-HLT}}. \bibinfo{pages}{4171--4186}.
\newblock


\bibitem[\protect\citeauthoryear{Diao, Qiu, Wu, Smola, Jiang, and Wang}{Diao
  et~al\mbox{.}}{2014}]%
        {diao2014jointly}
\bibfield{author}{\bibinfo{person}{Qiming Diao}, \bibinfo{person}{Minghui Qiu},
  \bibinfo{person}{Chao-Yuan Wu}, \bibinfo{person}{Alexander~J Smola},
  \bibinfo{person}{Jing Jiang}, {and} \bibinfo{person}{Chong Wang}.}
  \bibinfo{year}{2014}\natexlab{}.
\newblock \showarticletitle{Jointly modeling aspects, ratings and sentiments
  for movie recommendation (JMARS)}. In \bibinfo{booktitle}{\emph{KDD}}.
  \bibinfo{pages}{193--202}.
\newblock


\bibitem[\protect\citeauthoryear{Dignum}{Dignum}{2019}]%
        {dignum2019responsible}
\bibfield{author}{\bibinfo{person}{Virginia Dignum}.}
  \bibinfo{year}{2019}\natexlab{}.
\newblock \bibinfo{booktitle}{\emph{Responsible artificial intelligence: how to
  develop and use AI in a responsible way}}.
\newblock \bibinfo{publisher}{Springer Nature}.
\newblock


\bibitem[\protect\citeauthoryear{Domann and Lommatzsch}{Domann and
  Lommatzsch}{2017}]%
        {domann2017highly}
\bibfield{author}{\bibinfo{person}{Jaschar Domann} {and}
  \bibinfo{person}{Andreas Lommatzsch}.} \bibinfo{year}{2017}\natexlab{}.
\newblock \showarticletitle{A highly available real-time news recommender based
  on Apache Spark}. In \bibinfo{booktitle}{\emph{Int. Conf. of the Cross-Lang.
  Eval. Forum for Eur. Lang.}} Springer, \bibinfo{pages}{161--172}.
\newblock


\bibitem[\protect\citeauthoryear{Doychev, Lawlor, Rafter, and Smyth}{Doychev
  et~al\mbox{.}}{2014}]%
        {doychev2014analysis}
\bibfield{author}{\bibinfo{person}{Doychin Doychev}, \bibinfo{person}{Aonghus
  Lawlor}, \bibinfo{person}{Rachael Rafter}, {and} \bibinfo{person}{Barry
  Smyth}.} \bibinfo{year}{2014}\natexlab{}.
\newblock \showarticletitle{An analysis of recommender algorithms for online
  news}. In \bibinfo{booktitle}{\emph{CLEF}}. \bibinfo{pages}{177--184}.
\newblock


\bibitem[\protect\citeauthoryear{Durairaj and Kumar}{Durairaj and
  Kumar}{2014}]%
        {durairaj2014news}
\bibfield{author}{\bibinfo{person}{M Durairaj} {and} \bibinfo{person}{K~Muthu
  Kumar}.} \bibinfo{year}{2014}\natexlab{}.
\newblock \showarticletitle{News recommendation systems using web mining: a
  study}. In \bibinfo{booktitle}{\emph{International Journal of Engineering
  Trends and Technology}}, Vol.~\bibinfo{volume}{126}.
  \bibinfo{pages}{293--299}.
\newblock


\bibitem[\protect\citeauthoryear{Dwivedi and Arya}{Dwivedi and Arya}{2016}]%
        {dwivedi2016survey}
\bibfield{author}{\bibinfo{person}{Sanjay~K Dwivedi} {and}
  \bibinfo{person}{Chandrakala Arya}.} \bibinfo{year}{2016}\natexlab{}.
\newblock \showarticletitle{A survey of news recommendation approaches}. In
  \bibinfo{booktitle}{\emph{ICTBIG}}. IEEE, \bibinfo{pages}{1--6}.
\newblock


\bibitem[\protect\citeauthoryear{Epure, Kille, Ingvaldsen, Deneckere, Salinesi,
  and Albayrak}{Epure et~al\mbox{.}}{2017}]%
        {epure2017recommending}
\bibfield{author}{\bibinfo{person}{Elena~Viorica Epure},
  \bibinfo{person}{Benjamin Kille}, \bibinfo{person}{Jon~Espen Ingvaldsen},
  \bibinfo{person}{Rebecca Deneckere}, \bibinfo{person}{Camille Salinesi},
  {and} \bibinfo{person}{Sahin Albayrak}.} \bibinfo{year}{2017}\natexlab{}.
\newblock \showarticletitle{Recommending personalized news in short user
  sessions}. In \bibinfo{booktitle}{\emph{Recsys}}. \bibinfo{pages}{121--129}.
\newblock


\bibitem[\protect\citeauthoryear{Feng, Khan, Rahman, and Ahmad}{Feng
  et~al\mbox{.}}{2020}]%
        {feng2020news}
\bibfield{author}{\bibinfo{person}{Chong Feng}, \bibinfo{person}{Muzammil
  Khan}, \bibinfo{person}{Arif~Ur Rahman}, {and} \bibinfo{person}{Arshad
  Ahmad}.} \bibinfo{year}{2020}\natexlab{}.
\newblock \showarticletitle{News Recommendation Systems-Accomplishments,
  Challenges \& Future Directions}.
\newblock \bibinfo{journal}{\emph{IEEE Access}}  \bibinfo{volume}{8}
  (\bibinfo{year}{2020}), \bibinfo{pages}{16702--16725}.
\newblock


\bibitem[\protect\citeauthoryear{Fortuna, Fortuna, and Mladeni{\'c}}{Fortuna
  et~al\mbox{.}}{2010}]%
        {fortuna2010real}
\bibfield{author}{\bibinfo{person}{Bla{\v{z}} Fortuna},
  \bibinfo{person}{Carolina Fortuna}, {and} \bibinfo{person}{Dunja
  Mladeni{\'c}}.} \bibinfo{year}{2010}\natexlab{}.
\newblock \showarticletitle{Real-time news recommender system}. In
  \bibinfo{booktitle}{\emph{ECML PKDD}}. Springer, \bibinfo{pages}{583--586}.
\newblock


\bibitem[\protect\citeauthoryear{Frasincar, Borsje, and Levering}{Frasincar
  et~al\mbox{.}}{2009}]%
        {frasincar2009semantic}
\bibfield{author}{\bibinfo{person}{Flavius Frasincar}, \bibinfo{person}{Jethro
  Borsje}, {and} \bibinfo{person}{Leonard Levering}.}
  \bibinfo{year}{2009}\natexlab{}.
\newblock \showarticletitle{A semantic web-based approach for building
  personalized news services}.
\newblock \bibinfo{journal}{\emph{Int. Journal of E-Business Res.}}
  \bibinfo{volume}{5}, \bibinfo{number}{3} (\bibinfo{year}{2009}),
  \bibinfo{pages}{35--53}.
\newblock


\bibitem[\protect\citeauthoryear{Fu, Li, Zhang, Wu, Ma, Huang, and Jiang}{Fu
  et~al\mbox{.}}{2020}]%
        {fu2020recurrent}
\bibfield{author}{\bibinfo{person}{Jinlan Fu}, \bibinfo{person}{Yi Li},
  \bibinfo{person}{Qi Zhang}, \bibinfo{person}{Qinzhuo Wu},
  \bibinfo{person}{Renfeng Ma}, \bibinfo{person}{Xuanjing Huang}, {and}
  \bibinfo{person}{Yu-Gang Jiang}.} \bibinfo{year}{2020}\natexlab{}.
\newblock \showarticletitle{Recurrent memory reasoning network for expert
  finding in community question answering}. In
  \bibinfo{booktitle}{\emph{WSDM}}. \bibinfo{pages}{187--195}.
\newblock


\bibitem[\protect\citeauthoryear{Gabriel De~Souza, Jannach, and
  Da~Cunha}{Gabriel De~Souza et~al\mbox{.}}{2019}]%
        {gabriel2019contextual}
\bibfield{author}{\bibinfo{person}{P~Moreira Gabriel De~Souza},
  \bibinfo{person}{Dietmar Jannach}, {and} \bibinfo{person}{Adilson~Marques
  Da~Cunha}.} \bibinfo{year}{2019}\natexlab{}.
\newblock \showarticletitle{Contextual Hybrid Session-Based News Recommendation
  With Recurrent Neural Networks}.
\newblock \bibinfo{journal}{\emph{IEEE Access}}  \bibinfo{volume}{7}
  (\bibinfo{year}{2019}), \bibinfo{pages}{169185--169203}.
\newblock


\bibitem[\protect\citeauthoryear{Gabrilovich, Dumais, and Horvitz}{Gabrilovich
  et~al\mbox{.}}{2004}]%
        {gabrilovich2004newsjunkie}
\bibfield{author}{\bibinfo{person}{Evgeniy Gabrilovich}, \bibinfo{person}{Susan
  Dumais}, {and} \bibinfo{person}{Eric Horvitz}.}
  \bibinfo{year}{2004}\natexlab{}.
\newblock \showarticletitle{Newsjunkie: providing personalized newsfeeds via
  analysis of information novelty}. In \bibinfo{booktitle}{\emph{WWW}}.
  \bibinfo{pages}{482--490}.
\newblock


\bibitem[\protect\citeauthoryear{Gao, Li, Han, and Ma}{Gao
  et~al\mbox{.}}{2009}]%
        {gao2009infoslim}
\bibfield{author}{\bibinfo{person}{Feng Gao}, \bibinfo{person}{Yuhong Li},
  \bibinfo{person}{Li Han}, {and} \bibinfo{person}{Jian Ma}.}
  \bibinfo{year}{2009}\natexlab{}.
\newblock \showarticletitle{InfoSlim: an ontology-content based personalized
  mobile news recommendation system}. In \bibinfo{booktitle}{\emph{WiCOM}}.
  IEEE, \bibinfo{pages}{1--4}.
\newblock


\bibitem[\protect\citeauthoryear{Gao, Xin, Liu, Wang, Lu, Li, Fan, and Guo}{Gao
  et~al\mbox{.}}{2018}]%
        {gao2018fine}
\bibfield{author}{\bibinfo{person}{Jie Gao}, \bibinfo{person}{Xin Xin},
  \bibinfo{person}{Junshuai Liu}, \bibinfo{person}{Rui Wang},
  \bibinfo{person}{Jing Lu}, \bibinfo{person}{Biao Li}, \bibinfo{person}{Xin
  Fan}, {and} \bibinfo{person}{Ping Guo}.} \bibinfo{year}{2018}\natexlab{}.
\newblock \showarticletitle{Fine-Grained Deep Knowledge-Aware Network for News
  Recommendation with Self-Attention}. In \bibinfo{booktitle}{\emph{WI}}. IEEE,
  \bibinfo{pages}{81--88}.
\newblock


\bibitem[\protect\citeauthoryear{Garcin, Dimitrakakis, and Faltings}{Garcin
  et~al\mbox{.}}{2013}]%
        {garcin2013personalized}
\bibfield{author}{\bibinfo{person}{Florent Garcin}, \bibinfo{person}{Christos
  Dimitrakakis}, {and} \bibinfo{person}{Boi Faltings}.}
  \bibinfo{year}{2013}\natexlab{}.
\newblock \showarticletitle{Personalized news recommendation with context
  trees}. In \bibinfo{booktitle}{\emph{Recsys}}. \bibinfo{pages}{105--112}.
\newblock


\bibitem[\protect\citeauthoryear{Garcin and Faltings}{Garcin and
  Faltings}{2013}]%
        {garcin2013pen}
\bibfield{author}{\bibinfo{person}{Florent Garcin} {and} \bibinfo{person}{Boi
  Faltings}.} \bibinfo{year}{2013}\natexlab{}.
\newblock \showarticletitle{Pen recsys: A personalized news recommender systems
  framework}. In \bibinfo{booktitle}{\emph{Proceedings of the international
  news recommender systems workshop and challenge}}. \bibinfo{pages}{3--9}.
\newblock


\bibitem[\protect\citeauthoryear{Garcin, Zhou, Faltings, and Schickel}{Garcin
  et~al\mbox{.}}{2012}]%
        {garcin2012personalized}
\bibfield{author}{\bibinfo{person}{Florent Garcin}, \bibinfo{person}{Kai Zhou},
  \bibinfo{person}{Boi Faltings}, {and} \bibinfo{person}{Vincent Schickel}.}
  \bibinfo{year}{2012}\natexlab{}.
\newblock \showarticletitle{Personalized news recommendation based on
  collaborative filtering}. In \bibinfo{booktitle}{\emph{WI-IAT}},
  Vol.~\bibinfo{volume}{1}. IEEE, \bibinfo{pages}{437--441}.
\newblock


\bibitem[\protect\citeauthoryear{Ge, Wu, Wu, Qi, and Huang}{Ge
  et~al\mbox{.}}{2020}]%
        {ge2020graph}
\bibfield{author}{\bibinfo{person}{Suyu Ge}, \bibinfo{person}{Chuhan Wu},
  \bibinfo{person}{Fangzhao Wu}, \bibinfo{person}{Tao Qi}, {and}
  \bibinfo{person}{Yongfeng Huang}.} \bibinfo{year}{2020}\natexlab{}.
\newblock \showarticletitle{Graph Enhanced Representation Learning for News
  Recommendation}. In \bibinfo{booktitle}{\emph{WWW}}.
  \bibinfo{pages}{2863--2869}.
\newblock


\bibitem[\protect\citeauthoryear{Gers, Schmidhuber, and Cummins}{Gers
  et~al\mbox{.}}{2000}]%
        {gers2000learning}
\bibfield{author}{\bibinfo{person}{Felix~A Gers}, \bibinfo{person}{J{\"u}rgen
  Schmidhuber}, {and} \bibinfo{person}{Fred Cummins}.}
  \bibinfo{year}{2000}\natexlab{}.
\newblock \showarticletitle{Learning to forget: Continual prediction with
  LSTM}.
\newblock \bibinfo{journal}{\emph{Neural computation}} \bibinfo{volume}{12},
  \bibinfo{number}{10} (\bibinfo{year}{2000}), \bibinfo{pages}{2451--2471}.
\newblock


\bibitem[\protect\citeauthoryear{Gershman, Wolfe, Fink, and Carbonell}{Gershman
  et~al\mbox{.}}{2011}]%
        {gershman2011news}
\bibfield{author}{\bibinfo{person}{Anatole Gershman}, \bibinfo{person}{Travis
  Wolfe}, \bibinfo{person}{Eugene Fink}, {and} \bibinfo{person}{Jaime~G
  Carbonell}.} \bibinfo{year}{2011}\natexlab{}.
\newblock \showarticletitle{News personalization using support vector
  machines}.
\newblock  (\bibinfo{year}{2011}).
\newblock


\bibitem[\protect\citeauthoryear{Gharahighehi, Vens, and Pliakos}{Gharahighehi
  et~al\mbox{.}}{2020}]%
        {gharahighehi2020multi}
\bibfield{author}{\bibinfo{person}{Alireza Gharahighehi},
  \bibinfo{person}{Celine Vens}, {and} \bibinfo{person}{Konstantinos Pliakos}.}
  \bibinfo{year}{2020}\natexlab{}.
\newblock \showarticletitle{Multi-stakeholder news recommendation using
  hypergraph learning}. In \bibinfo{booktitle}{\emph{INRA@ECML PKDD}}.
\newblock


\bibitem[\protect\citeauthoryear{Gharahighehi, Vens, and Pliakos}{Gharahighehi
  et~al\mbox{.}}{2021}]%
        {gharahighehi2021fair}
\bibfield{author}{\bibinfo{person}{Alireza Gharahighehi},
  \bibinfo{person}{Celine Vens}, {and} \bibinfo{person}{Konstantinos Pliakos}.}
  \bibinfo{year}{2021}\natexlab{}.
\newblock \showarticletitle{Fair multi-stakeholder news recommender system with
  hypergraph ranking}.
\newblock \bibinfo{journal}{\emph{Information Processing \& Management}}
  \bibinfo{volume}{58}, \bibinfo{number}{5} (\bibinfo{year}{2021}),
  \bibinfo{pages}{102663}.
\newblock


\bibitem[\protect\citeauthoryear{Goossen, IJntema, Frasincar, Hogenboom, and
  Kaymak}{Goossen et~al\mbox{.}}{2011}]%
        {goossen2011news}
\bibfield{author}{\bibinfo{person}{Frank Goossen}, \bibinfo{person}{Wouter
  IJntema}, \bibinfo{person}{Flavius Frasincar}, \bibinfo{person}{Frederik
  Hogenboom}, {and} \bibinfo{person}{Uzay Kaymak}.}
  \bibinfo{year}{2011}\natexlab{}.
\newblock \showarticletitle{News personalization using the CF-IDF semantic
  recommender}. In \bibinfo{booktitle}{\emph{WIMS}}. \bibinfo{pages}{1--12}.
\newblock


\bibitem[\protect\citeauthoryear{Gu, Dong, Zeng, and He}{Gu
  et~al\mbox{.}}{2014}]%
        {gu2014effective}
\bibfield{author}{\bibinfo{person}{Wanrong Gu}, \bibinfo{person}{Shoubin Dong},
  \bibinfo{person}{Zhizhao Zeng}, {and} \bibinfo{person}{Jinchao He}.}
  \bibinfo{year}{2014}\natexlab{}.
\newblock \showarticletitle{An effective news recommendation method for
  microblog user}.
\newblock \bibinfo{journal}{\emph{The Scientific World Journal}}
  \bibinfo{volume}{2014} (\bibinfo{year}{2014}).
\newblock


\bibitem[\protect\citeauthoryear{Gulcehre, Denil, Malinowski, Razavi, Pascanu,
  Hermann, Battaglia, Bapst, Raposo, Santoro, et~al\mbox{.}}{Gulcehre
  et~al\mbox{.}}{2018}]%
        {gulcehre2018hyperbolic}
\bibfield{author}{\bibinfo{person}{Caglar Gulcehre}, \bibinfo{person}{Misha
  Denil}, \bibinfo{person}{Mateusz Malinowski}, \bibinfo{person}{Ali Razavi},
  \bibinfo{person}{Razvan Pascanu}, \bibinfo{person}{Karl~Moritz Hermann},
  \bibinfo{person}{Peter Battaglia}, \bibinfo{person}{Victor Bapst},
  \bibinfo{person}{David Raposo}, \bibinfo{person}{Adam Santoro},
  {et~al\mbox{.}}} \bibinfo{year}{2018}\natexlab{}.
\newblock \showarticletitle{Hyperbolic attention networks}.
\newblock \bibinfo{journal}{\emph{arXiv preprint arXiv:1805.09786}}
  (\bibinfo{year}{2018}).
\newblock


\bibitem[\protect\citeauthoryear{Gulla, Zhang, Liu, {\"O}zg{\"o}bek, and
  Su}{Gulla et~al\mbox{.}}{2017}]%
        {gulla2017adressa}
\bibfield{author}{\bibinfo{person}{Jon~Atle Gulla}, \bibinfo{person}{Lemei
  Zhang}, \bibinfo{person}{Peng Liu}, \bibinfo{person}{{\"O}zlem
  {\"O}zg{\"o}bek}, {and} \bibinfo{person}{Xiaomeng Su}.}
  \bibinfo{year}{2017}\natexlab{}.
\newblock \showarticletitle{The Adressa dataset for news recommendation}. In
  \bibinfo{booktitle}{\emph{WI}}. \bibinfo{pages}{1042--1048}.
\newblock


\bibitem[\protect\citeauthoryear{Han}{Han}{2020}]%
        {han2020personalized}
\bibfield{author}{\bibinfo{person}{Kunni Han}.}
  \bibinfo{year}{2020}\natexlab{}.
\newblock \showarticletitle{Personalized News Recommendation and Simulation
  Based on Improved Collaborative Filtering Algorithm}.
\newblock \bibinfo{journal}{\emph{Complexity}}  \bibinfo{volume}{2020}
  (\bibinfo{year}{2020}).
\newblock


\bibitem[\protect\citeauthoryear{Han, Huang, and Liu}{Han
  et~al\mbox{.}}{2021}]%
        {han2021neural}
\bibfield{author}{\bibinfo{person}{Songqiao Han}, \bibinfo{person}{Hailiang
  Huang}, {and} \bibinfo{person}{Jiangwei Liu}.}
  \bibinfo{year}{2021}\natexlab{}.
\newblock \showarticletitle{Neural news recommendation with event extraction}.
\newblock \bibinfo{journal}{\emph{arXiv preprint arXiv:2111.05068}}
  (\bibinfo{year}{2021}).
\newblock


\bibitem[\protect\citeauthoryear{Harandi and Gulla}{Harandi and Gulla}{2015}]%
        {harandi2015survey}
\bibfield{author}{\bibinfo{person}{Mahboobeh Harandi} {and}
  \bibinfo{person}{Jon~Atle Gulla}.} \bibinfo{year}{2015}\natexlab{}.
\newblock \showarticletitle{Survey of User Profiling in News Recommender
  Systems}. In \bibinfo{booktitle}{\emph{INRA@ECML PKDD}}.
  \bibinfo{pages}{20--26}.
\newblock


\bibitem[\protect\citeauthoryear{Hendrickx, Smets, and Ballon}{Hendrickx
  et~al\mbox{.}}{2021}]%
        {hendrickx2021news}
\bibfield{author}{\bibinfo{person}{Jonathan Hendrickx},
  \bibinfo{person}{Annelien Smets}, {and} \bibinfo{person}{Pieter Ballon}.}
  \bibinfo{year}{2021}\natexlab{}.
\newblock \showarticletitle{News Recommender Systems and News Diversity, Two of
  a Kind? A Case Study from a Small Media Market}.
\newblock \bibinfo{journal}{\emph{Journalism and Media}} \bibinfo{volume}{2},
  \bibinfo{number}{3} (\bibinfo{year}{2021}), \bibinfo{pages}{515--528}.
\newblock


\bibitem[\protect\citeauthoryear{Hidasi, Karatzoglou, Baltrunas, and
  Tikk}{Hidasi et~al\mbox{.}}{2016}]%
        {hidasi2016session}
\bibfield{author}{\bibinfo{person}{Bal{\'a}zs Hidasi},
  \bibinfo{person}{Alexandros Karatzoglou}, \bibinfo{person}{Linas Baltrunas},
  {and} \bibinfo{person}{D Tikk}.} \bibinfo{year}{2016}\natexlab{}.
\newblock \showarticletitle{Session-based recommendations with recurrent neural
  networks}. In \bibinfo{booktitle}{\emph{ICLR}}.
\newblock


\bibitem[\protect\citeauthoryear{Hogenboom, Capelle, and Moerland}{Hogenboom
  et~al\mbox{.}}{2013}]%
        {hogenboom2013news}
\bibfield{author}{\bibinfo{person}{Frederik Hogenboom}, \bibinfo{person}{Michel
  Capelle}, {and} \bibinfo{person}{Marnix Moerland}.}
  \bibinfo{year}{2013}\natexlab{}.
\newblock \showarticletitle{News Recommendation using Semantics with the
  Bing-SF-IDF Approach}. In \bibinfo{booktitle}{\emph{ER}}. Springer,
  \bibinfo{pages}{160--169}.
\newblock


\bibitem[\protect\citeauthoryear{Hogenboom, Capelle, Moerland, and
  Frasincar}{Hogenboom et~al\mbox{.}}{2014}]%
        {hogenboom2014bing}
\bibfield{author}{\bibinfo{person}{Frederik Hogenboom}, \bibinfo{person}{Michel
  Capelle}, \bibinfo{person}{Marnix Moerland}, {and} \bibinfo{person}{Flavius
  Frasincar}.} \bibinfo{year}{2014}\natexlab{}.
\newblock \showarticletitle{Bing-SF-IDF+ semantics-driven news recommendation}.
  In \bibinfo{booktitle}{\emph{WWW}}. \bibinfo{pages}{291--292}.
\newblock


\bibitem[\protect\citeauthoryear{Hsieh, Yang, Wei, Naaman, and Estrin}{Hsieh
  et~al\mbox{.}}{2016}]%
        {hsieh2016immersive}
\bibfield{author}{\bibinfo{person}{Cheng-Kang Hsieh}, \bibinfo{person}{Longqi
  Yang}, \bibinfo{person}{Honghao Wei}, \bibinfo{person}{Mor Naaman}, {and}
  \bibinfo{person}{Deborah Estrin}.} \bibinfo{year}{2016}\natexlab{}.
\newblock \showarticletitle{Immersive recommendation: News and event
  recommendations using personal digital traces}. In
  \bibinfo{booktitle}{\emph{WWW}}. \bibinfo{pages}{51--62}.
\newblock


\bibitem[\protect\citeauthoryear{Hu, Li, Shi, Yang, and Shao}{Hu
  et~al\mbox{.}}{2020a}]%
        {hu2020graph}
\bibfield{author}{\bibinfo{person}{Linmei Hu}, \bibinfo{person}{Chen Li},
  \bibinfo{person}{Chuan Shi}, \bibinfo{person}{Cheng Yang}, {and}
  \bibinfo{person}{Chao Shao}.} \bibinfo{year}{2020}\natexlab{a}.
\newblock \showarticletitle{Graph neural news recommendation with long-term and
  short-term interest modeling}.
\newblock \bibinfo{journal}{\emph{Information Processing \& Management}}
  \bibinfo{volume}{57}, \bibinfo{number}{2} (\bibinfo{year}{2020}),
  \bibinfo{pages}{102142}.
\newblock


\bibitem[\protect\citeauthoryear{Hu, Xu, Li, Yang, Shi, Duan, Xie, and Zhou}{Hu
  et~al\mbox{.}}{2020b}]%
        {hu2020graph2}
\bibfield{author}{\bibinfo{person}{Linmei Hu}, \bibinfo{person}{Siyong Xu},
  \bibinfo{person}{Chen Li}, \bibinfo{person}{Cheng Yang},
  \bibinfo{person}{Chuan Shi}, \bibinfo{person}{Nan Duan},
  \bibinfo{person}{Xing Xie}, {and} \bibinfo{person}{Ming Zhou}.}
  \bibinfo{year}{2020}\natexlab{b}.
\newblock \showarticletitle{Graph neural news recommendation with unsupervised
  preference disentanglement}. In \bibinfo{booktitle}{\emph{ACL}}.
  \bibinfo{pages}{4255--4264}.
\newblock


\bibitem[\protect\citeauthoryear{Huang, He, Gao, Deng, Acero, and Heck}{Huang
  et~al\mbox{.}}{2013}]%
        {huang2013learning}
\bibfield{author}{\bibinfo{person}{Po-Sen Huang}, \bibinfo{person}{Xiaodong
  He}, \bibinfo{person}{Jianfeng Gao}, \bibinfo{person}{Li Deng},
  \bibinfo{person}{Alex Acero}, {and} \bibinfo{person}{Larry Heck}.}
  \bibinfo{year}{2013}\natexlab{}.
\newblock \showarticletitle{Learning deep structured semantic models for web
  search using clickthrough data}. In \bibinfo{booktitle}{\emph{CIKM}}.
  \bibinfo{pages}{2333--2338}.
\newblock


\bibitem[\protect\citeauthoryear{Hutto and Gilbert}{Hutto and Gilbert}{2014}]%
        {hutto2014vader}
\bibfield{author}{\bibinfo{person}{Clayton Hutto} {and} \bibinfo{person}{Eric
  Gilbert}.} \bibinfo{year}{2014}\natexlab{}.
\newblock \showarticletitle{Vader: A parsimonious rule-based model for
  sentiment analysis of social media text}. In
  \bibinfo{booktitle}{\emph{ICWSM}}, Vol.~\bibinfo{volume}{8}.
\newblock


\bibitem[\protect\citeauthoryear{Iana, Alam, and Paulheim}{Iana
  et~al\mbox{.}}{2021}]%
        {iana2021survey}
\bibfield{author}{\bibinfo{person}{Andreea Iana}, \bibinfo{person}{Mehwish
  Alam}, {and} \bibinfo{person}{Heiko Paulheim}.}
  \bibinfo{year}{2021}\natexlab{}.
\newblock \showarticletitle{A Survey On Knowledge-Aware News Recommender
  Systems}.
\newblock  (\bibinfo{year}{2021}).
\newblock


\bibitem[\protect\citeauthoryear{Ilievski and Roy}{Ilievski and Roy}{2013}]%
        {ilievski2013personalized}
\bibfield{author}{\bibinfo{person}{Ilija Ilievski} {and} \bibinfo{person}{Sujoy
  Roy}.} \bibinfo{year}{2013}\natexlab{}.
\newblock \showarticletitle{Personalized news recommendation based on implicit
  feedback}. In \bibinfo{booktitle}{\emph{Proceedings of the international news
  recommender systems workshop and challenge}}. \bibinfo{pages}{10--15}.
\newblock


\bibitem[\protect\citeauthoryear{Islambouli, Ingram, and Gillet}{Islambouli
  et~al\mbox{.}}{2021}]%
        {islambouli2021user}
\bibfield{author}{\bibinfo{person}{Rania Islambouli}, \bibinfo{person}{Sandy
  Ingram}, {and} \bibinfo{person}{Denis Gillet}.}
  \bibinfo{year}{2021}\natexlab{}.
\newblock \showarticletitle{A User Centered News Recommendation System}. In
  \bibinfo{booktitle}{\emph{Proceedings of the 4th Workshop on Human Factors in
  Hypertext}}. \bibinfo{pages}{15--16}.
\newblock


\bibitem[\protect\citeauthoryear{Ji, He, Xu, Liu, and Zhao}{Ji
  et~al\mbox{.}}{2015}]%
        {ji2015knowledge}
\bibfield{author}{\bibinfo{person}{Guoliang Ji}, \bibinfo{person}{Shizhu He},
  \bibinfo{person}{Liheng Xu}, \bibinfo{person}{Kang Liu}, {and}
  \bibinfo{person}{Jun Zhao}.} \bibinfo{year}{2015}\natexlab{}.
\newblock \showarticletitle{Knowledge graph embedding via dynamic mapping
  matrix}. In \bibinfo{booktitle}{\emph{ACL-IJCNLP}}.
  \bibinfo{pages}{687--696}.
\newblock


\bibitem[\protect\citeauthoryear{Ji, Hong, Shangguan, Wang, and Ma}{Ji
  et~al\mbox{.}}{2016}]%
        {ji2016regularized}
\bibfield{author}{\bibinfo{person}{Youchun Ji}, \bibinfo{person}{Wenxing Hong},
  \bibinfo{person}{Yali Shangguan}, \bibinfo{person}{Huan Wang}, {and}
  \bibinfo{person}{Jing Ma}.} \bibinfo{year}{2016}\natexlab{}.
\newblock \showarticletitle{Regularized singular value decomposition in news
  recommendation system}. In \bibinfo{booktitle}{\emph{ICCSE}}. IEEE,
  \bibinfo{pages}{621--626}.
\newblock


\bibitem[\protect\citeauthoryear{Ji, Wu, Liu, and {\'I}{\~n}igo}{Ji
  et~al\mbox{.}}{2021a}]%
        {ji2021attention}
\bibfield{author}{\bibinfo{person}{Zhenyan Ji}, \bibinfo{person}{Mengdan Wu},
  \bibinfo{person}{Jirui Liu}, {and} \bibinfo{person}{Jos{\'e}
  Enrique~Armend{\'a}riz {\'I}{\~n}igo}.} \bibinfo{year}{2021}\natexlab{a}.
\newblock \showarticletitle{Attention-Based Graph Neural Network for News
  Recommendation}. In \bibinfo{booktitle}{\emph{IJCNN}}. IEEE,
  \bibinfo{pages}{1--8}.
\newblock


\bibitem[\protect\citeauthoryear{Ji, Wu, Yang, and {\'I}{\~n}igo}{Ji
  et~al\mbox{.}}{2021b}]%
        {ji2021temporal}
\bibfield{author}{\bibinfo{person}{Zhenyan Ji}, \bibinfo{person}{Mengdan Wu},
  \bibinfo{person}{Hong Yang}, {and} \bibinfo{person}{Jos{\'e}
  Enrique~Armend{\'a}riz {\'I}{\~n}igo}.} \bibinfo{year}{2021}\natexlab{b}.
\newblock \showarticletitle{Temporal sensitive heterogeneous graph neural
  network for news recommendation}.
\newblock \bibinfo{journal}{\emph{Future Generation Computer Systems}}
  (\bibinfo{year}{2021}).
\newblock


\bibitem[\protect\citeauthoryear{Jia, Li, Zhang, He, and Zhu}{Jia
  et~al\mbox{.}}{2021}]%
        {jia2021rmbert}
\bibfield{author}{\bibinfo{person}{Qinglin Jia}, \bibinfo{person}{Jingjie Li},
  \bibinfo{person}{Qi Zhang}, \bibinfo{person}{Xiuqiang He}, {and}
  \bibinfo{person}{Jieming Zhu}.} \bibinfo{year}{2021}\natexlab{}.
\newblock \showarticletitle{RMBERT: News Recommendation via Recurrent Reasoning
  Memory Network over BERT}. In \bibinfo{booktitle}{\emph{SIGIR}}.
  \bibinfo{pages}{1773--1777}.
\newblock


\bibitem[\protect\citeauthoryear{Jonnalagedda and Gauch}{Jonnalagedda and
  Gauch}{2013}]%
        {jonnalagedda2013personalized}
\bibfield{author}{\bibinfo{person}{Nirmal Jonnalagedda} {and}
  \bibinfo{person}{Susan Gauch}.} \bibinfo{year}{2013}\natexlab{}.
\newblock \showarticletitle{Personalized news recommendation using twitter}. In
  \bibinfo{booktitle}{\emph{WI-IAT}}, Vol.~\bibinfo{volume}{3}. IEEE,
  \bibinfo{pages}{21--25}.
\newblock


\bibitem[\protect\citeauthoryear{Jonnalagedda, Gauch, Labille, and
  Alfarhood}{Jonnalagedda et~al\mbox{.}}{2016}]%
        {jonnalagedda2016incorporating}
\bibfield{author}{\bibinfo{person}{Nirmal Jonnalagedda}, \bibinfo{person}{Susan
  Gauch}, \bibinfo{person}{Kevin Labille}, {and} \bibinfo{person}{Sultan
  Alfarhood}.} \bibinfo{year}{2016}\natexlab{}.
\newblock \showarticletitle{Incorporating popularity in a personalized news
  recommender system}.
\newblock \bibinfo{journal}{\emph{PeerJ}}  \bibinfo{volume}{2}
  (\bibinfo{year}{2016}).
\newblock


\bibitem[\protect\citeauthoryear{Joseph and Jiang}{Joseph and Jiang}{2019}]%
        {joseph2019content}
\bibfield{author}{\bibinfo{person}{Kevin Joseph} {and} \bibinfo{person}{Hui
  Jiang}.} \bibinfo{year}{2019}\natexlab{}.
\newblock \showarticletitle{Content based News Recommendation via Shortest
  Entity Distance over Knowledge Graphs}. In
  \bibinfo{booktitle}{\emph{Companion Proceedings of WWW}}.
  \bibinfo{pages}{690--699}.
\newblock


\bibitem[\protect\citeauthoryear{Jugovac, Jannach, and Karimi}{Jugovac
  et~al\mbox{.}}{2018}]%
        {Streamingrec}
\bibfield{author}{\bibinfo{person}{Michael Jugovac}, \bibinfo{person}{Dietmar
  Jannach}, {and} \bibinfo{person}{Mozhgan Karimi}.}
  \bibinfo{year}{2018}\natexlab{}.
\newblock \showarticletitle{Streamingrec: A Framework for Benchmarking
  Stream-Based News Recommenders}. In \bibinfo{booktitle}{\emph{Recsys}}.
  \bibinfo{pages}{269–273}.
\newblock


\bibitem[\protect\citeauthoryear{Kang and McAuley}{Kang and McAuley}{2018}]%
        {kang2018self}
\bibfield{author}{\bibinfo{person}{Wang-Cheng Kang} {and}
  \bibinfo{person}{Julian McAuley}.} \bibinfo{year}{2018}\natexlab{}.
\newblock \showarticletitle{Self-attentive sequential recommendation}. In
  \bibinfo{booktitle}{\emph{ICDM}}. IEEE, \bibinfo{pages}{197--206}.
\newblock


\bibitem[\protect\citeauthoryear{Karimi, Jannach, and Jugovac}{Karimi
  et~al\mbox{.}}{2018}]%
        {karimi2018news}
\bibfield{author}{\bibinfo{person}{Mozhgan Karimi}, \bibinfo{person}{Dietmar
  Jannach}, {and} \bibinfo{person}{Michael Jugovac}.}
  \bibinfo{year}{2018}\natexlab{}.
\newblock \showarticletitle{News recommender systems--Survey and roads ahead}.
\newblock \bibinfo{journal}{\emph{Information Processing \& Management}}
  \bibinfo{volume}{54}, \bibinfo{number}{6} (\bibinfo{year}{2018}),
  \bibinfo{pages}{1203--1227}.
\newblock


\bibitem[\protect\citeauthoryear{Kazai, Yusof, and Clarke}{Kazai
  et~al\mbox{.}}{2016}]%
        {kazai2016personalised}
\bibfield{author}{\bibinfo{person}{Gabriella Kazai}, \bibinfo{person}{Iskander
  Yusof}, {and} \bibinfo{person}{Daoud Clarke}.}
  \bibinfo{year}{2016}\natexlab{}.
\newblock \showarticletitle{Personalised News and Blog Recommendations Based on
  User Location, Facebook and Twitter User Profiling}. In
  \bibinfo{booktitle}{\emph{SIGIR}}. \bibinfo{pages}{1129–1132}.
\newblock


\bibitem[\protect\citeauthoryear{Khattar, Kumar, and Varma}{Khattar
  et~al\mbox{.}}{2017}]%
        {Khattar2017leveraging}
\bibfield{author}{\bibinfo{person}{Dhruv Khattar}, \bibinfo{person}{Vaibhav
  Kumar}, {and} \bibinfo{person}{Vasudeva Varma}.}
  \bibinfo{year}{2017}\natexlab{}.
\newblock \showarticletitle{Leveraging Moderate User Data for News
  Recommendation}. In \bibinfo{booktitle}{\emph{SERecSys@ICDM}},
  \bibfield{editor}{\bibinfo{person}{Raju Gottumukkala}, \bibinfo{person}{Xia
  Ning}, \bibinfo{person}{Guozhu Dong}, \bibinfo{person}{Vijay Raghavan},
  \bibinfo{person}{Srinivas Aluru}, \bibinfo{person}{George Karypis},
  \bibinfo{person}{Lucio Miele}, {and} \bibinfo{person}{Xindong Wu}} (Eds.).
  \bibinfo{publisher}{{IEEE} Computer Society}, \bibinfo{pages}{757--760}.
\newblock


\bibitem[\protect\citeauthoryear{Khattar, Kumar, Varma, and Gupta}{Khattar
  et~al\mbox{.}}{2018}]%
        {khattar2018weave}
\bibfield{author}{\bibinfo{person}{Dhruv Khattar}, \bibinfo{person}{Vaibhav
  Kumar}, \bibinfo{person}{Vasudeva Varma}, {and} \bibinfo{person}{Manish
  Gupta}.} \bibinfo{year}{2018}\natexlab{}.
\newblock \showarticletitle{Weave\&rec: A word embedding based 3-d
  convolutional network for news recommendation}. In
  \bibinfo{booktitle}{\emph{CIKM}}. \bibinfo{pages}{1855--1858}.
\newblock


\bibitem[\protect\citeauthoryear{Kille, Hopfgartner, Brodt, and Heintz}{Kille
  et~al\mbox{.}}{2013}]%
        {Plista}
\bibfield{author}{\bibinfo{person}{Benjamin Kille}, \bibinfo{person}{Frank
  Hopfgartner}, \bibinfo{person}{Torben Brodt}, {and} \bibinfo{person}{Tobias
  Heintz}.} \bibinfo{year}{2013}\natexlab{}.
\newblock \showarticletitle{The Plista Dataset}. In
  \bibinfo{booktitle}{\emph{Proceedings of the international news recommender
  systems workshop and challenge}}. \bibinfo{pages}{16–23}.
\newblock


\bibitem[\protect\citeauthoryear{Kille, Lommatzsch, Hopfgartner, Larson, and
  Brodt}{Kille et~al\mbox{.}}{2017}]%
        {kille2017clef}
\bibfield{author}{\bibinfo{person}{Benjamin Kille}, \bibinfo{person}{Andreas
  Lommatzsch}, \bibinfo{person}{Frank Hopfgartner}, \bibinfo{person}{Martha
  Larson}, {and} \bibinfo{person}{Torben Brodt}.}
  \bibinfo{year}{2017}\natexlab{}.
\newblock \showarticletitle{CLEF 2017 newsreel overview: offline and online
  evaluation of stream-based news recommender systems}. In
  \bibinfo{booktitle}{\emph{CLEF}}. \bibinfo{publisher}{CEUR}.
\newblock


\bibitem[\protect\citeauthoryear{Kim}{Kim}{2014}]%
        {kim2014convolutional}
\bibfield{author}{\bibinfo{person}{Yoon Kim}.} \bibinfo{year}{2014}\natexlab{}.
\newblock \showarticletitle{Convolutional Neural Networks for Sentence
  Classification}. In \bibinfo{booktitle}{\emph{EMNLP}}.
  \bibinfo{pages}{1746--1751}.
\newblock


\bibitem[\protect\citeauthoryear{Kirshenbaum, Forman, and Dugan}{Kirshenbaum
  et~al\mbox{.}}{2012}]%
        {kirshenbaum2012live}
\bibfield{author}{\bibinfo{person}{Evan Kirshenbaum}, \bibinfo{person}{George
  Forman}, {and} \bibinfo{person}{Michael Dugan}.}
  \bibinfo{year}{2012}\natexlab{}.
\newblock \showarticletitle{A Live Comparison of Methods for Personalized
  Article Recommendation at Forbes.com}. In \bibinfo{booktitle}{\emph{ECML
  PKDD}}, \bibfield{editor}{\bibinfo{person}{Peter~A. Flach},
  \bibinfo{person}{Tijl De~Bie}, {and} \bibinfo{person}{Nello Cristianini}}
  (Eds.). \bibinfo{pages}{51--66}.
\newblock


\bibitem[\protect\citeauthoryear{Kompan and Bielikov{\'a}}{Kompan and
  Bielikov{\'a}}{2010}]%
        {kompan2010content}
\bibfield{author}{\bibinfo{person}{Michal Kompan} {and}
  \bibinfo{person}{M{\'a}ria Bielikov{\'a}}.} \bibinfo{year}{2010}\natexlab{}.
\newblock \showarticletitle{Content-Based News Recommendation}. In
  \bibinfo{booktitle}{\emph{EC-Web}},
  \bibfield{editor}{\bibinfo{person}{Francesco Buccafurri} {and}
  \bibinfo{person}{Giovanni Semeraro}} (Eds.). \bibinfo{pages}{61--72}.
\newblock


\bibitem[\protect\citeauthoryear{Kumar, Khattar, Gupta, Gupta, and Varma}{Kumar
  et~al\mbox{.}}{2017b}]%
        {kumar2017deep}
\bibfield{author}{\bibinfo{person}{Vaibhav Kumar}, \bibinfo{person}{Dhruv
  Khattar}, \bibinfo{person}{Shashank Gupta}, \bibinfo{person}{Manish Gupta},
  {and} \bibinfo{person}{Vasudeva Varma}.} \bibinfo{year}{2017}\natexlab{b}.
\newblock \showarticletitle{Deep Neural Architecture for News Recommendation.}.
  In \bibinfo{booktitle}{\emph{CLEF}}.
\newblock


\bibitem[\protect\citeauthoryear{Kumar, Khattar, Gupta, Gupta, and Varma}{Kumar
  et~al\mbox{.}}{2017c}]%
        {Khattar2017user}
\bibfield{author}{\bibinfo{person}{Vaibhav Kumar}, \bibinfo{person}{Dhruv
  Khattar}, \bibinfo{person}{Shashank Gupta}, \bibinfo{person}{Manish Gupta},
  {and} \bibinfo{person}{Vasudeva Varma}.} \bibinfo{year}{2017}\natexlab{c}.
\newblock \showarticletitle{User Profiling Based Deep Neural Network for
  Temporal News Recommendation}. In \bibinfo{booktitle}{\emph{SERecSys@ICDM}}.
  \bibinfo{publisher}{{IEEE} Computer Society}, \bibinfo{pages}{765--772}.
\newblock


\bibitem[\protect\citeauthoryear{Kumar, Khattar, Gupta, and Varma}{Kumar
  et~al\mbox{.}}{2017a}]%
        {kumar2017word}
\bibfield{author}{\bibinfo{person}{Vaibhav Kumar}, \bibinfo{person}{Dhruv
  Khattar}, \bibinfo{person}{Shashank Gupta}, {and} \bibinfo{person}{Vasudeva
  Varma}.} \bibinfo{year}{2017}\natexlab{a}.
\newblock \showarticletitle{Word semantics based 3-d convolutional neural
  networks for news recommendation}. In
  \bibinfo{booktitle}{\emph{SERecSys@ICDM}}. IEEE, \bibinfo{pages}{761--764}.
\newblock


\bibitem[\protect\citeauthoryear{Lang}{Lang}{1995}]%
        {lang1995newsweeder}
\bibfield{author}{\bibinfo{person}{Ken Lang}.} \bibinfo{year}{1995}\natexlab{}.
\newblock \showarticletitle{Newsweeder: Learning to filter netnews}.
\newblock In \bibinfo{booktitle}{\emph{Machine Learning}}.
  \bibinfo{publisher}{Elsevier}, \bibinfo{pages}{331--339}.
\newblock


\bibitem[\protect\citeauthoryear{Lavrenko, Schmill, Lawrie, Ogilvie, Jensen,
  and Allan}{Lavrenko et~al\mbox{.}}{2000}]%
        {lavrenko2000language}
\bibfield{author}{\bibinfo{person}{Victor Lavrenko}, \bibinfo{person}{Matt
  Schmill}, \bibinfo{person}{Dawn Lawrie}, \bibinfo{person}{Paul Ogilvie},
  \bibinfo{person}{David Jensen}, {and} \bibinfo{person}{James Allan}.}
  \bibinfo{year}{2000}\natexlab{}.
\newblock \showarticletitle{Language models for financial news recommendation}.
  In \bibinfo{booktitle}{\emph{CIKM}}. \bibinfo{pages}{389--396}.
\newblock


\bibitem[\protect\citeauthoryear{Lazer, Baum, Benkler, Berinsky, Greenhill,
  Menczer, Metzger, Nyhan, Pennycook, Rothschild, et~al\mbox{.}}{Lazer
  et~al\mbox{.}}{2018}]%
        {lazer2018science}
\bibfield{author}{\bibinfo{person}{David~MJ Lazer}, \bibinfo{person}{Matthew~A
  Baum}, \bibinfo{person}{Yochai Benkler}, \bibinfo{person}{Adam~J Berinsky},
  \bibinfo{person}{Kelly~M Greenhill}, \bibinfo{person}{Filippo Menczer},
  \bibinfo{person}{Miriam~J Metzger}, \bibinfo{person}{Brendan Nyhan},
  \bibinfo{person}{Gordon Pennycook}, \bibinfo{person}{David Rothschild},
  {et~al\mbox{.}}} \bibinfo{year}{2018}\natexlab{}.
\newblock \showarticletitle{The science of fake news}.
\newblock \bibinfo{journal}{\emph{Science}} \bibinfo{volume}{359},
  \bibinfo{number}{6380} (\bibinfo{year}{2018}), \bibinfo{pages}{1094--1096}.
\newblock


\bibitem[\protect\citeauthoryear{Le and Mikolov}{Le and Mikolov}{2014}]%
        {le2014distributed}
\bibfield{author}{\bibinfo{person}{Quoc Le} {and} \bibinfo{person}{Tomas
  Mikolov}.} \bibinfo{year}{2014}\natexlab{}.
\newblock \showarticletitle{Distributed representations of sentences and
  documents}. In \bibinfo{booktitle}{\emph{ICML}}. \bibinfo{pages}{1188--1196}.
\newblock


\bibitem[\protect\citeauthoryear{Lee, Oh, Seo, and Lee}{Lee
  et~al\mbox{.}}{2020}]%
        {lee2020news}
\bibfield{author}{\bibinfo{person}{Dongho Lee}, \bibinfo{person}{Byungkook Oh},
  \bibinfo{person}{Seungmin Seo}, {and} \bibinfo{person}{Kyong-Ho Lee}.}
  \bibinfo{year}{2020}\natexlab{}.
\newblock \showarticletitle{News Recommendation with Topic-Enriched Knowledge
  Graphs}. In \bibinfo{booktitle}{\emph{CIKM}}. \bibinfo{pages}{695--704}.
\newblock


\bibitem[\protect\citeauthoryear{Lee and Park}{Lee and Park}{2007}]%
        {lee2007moners}
\bibfield{author}{\bibinfo{person}{Hong~Joo Lee} {and}
  \bibinfo{person}{Sung~Joo Park}.} \bibinfo{year}{2007}\natexlab{}.
\newblock \showarticletitle{MONERS: A news recommender for the mobile web}.
\newblock \bibinfo{journal}{\emph{Expert Systems With Applications}}
  \bibinfo{volume}{32}, \bibinfo{number}{1} (\bibinfo{year}{2007}),
  \bibinfo{pages}{143--150}.
\newblock


\bibitem[\protect\citeauthoryear{Li, Tao, Wu, Feng, Zhao, and Yan}{Li
  et~al\mbox{.}}{2019}]%
        {li2019sampling}
\bibfield{author}{\bibinfo{person}{Jia Li}, \bibinfo{person}{Chongyang Tao},
  \bibinfo{person}{Wei Wu}, \bibinfo{person}{Yansong Feng},
  \bibinfo{person}{Dongyan Zhao}, {and} \bibinfo{person}{Rui Yan}.}
  \bibinfo{year}{2019}\natexlab{}.
\newblock \showarticletitle{Sampling matters! An empirical study of negative
  sampling strategies for learning of matching models in retrieval-based
  dialogue systems}. In \bibinfo{booktitle}{\emph{EMNLP-IJCNLP}}.
  \bibinfo{pages}{1291--1296}.
\newblock


\bibitem[\protect\citeauthoryear{Li, Chu, Langford, and Schapire}{Li
  et~al\mbox{.}}{2010a}]%
        {li2010contextual}
\bibfield{author}{\bibinfo{person}{Lihong Li}, \bibinfo{person}{Wei Chu},
  \bibinfo{person}{John Langford}, {and} \bibinfo{person}{Robert~E Schapire}.}
  \bibinfo{year}{2010}\natexlab{a}.
\newblock \showarticletitle{A contextual-bandit approach to personalized news
  article recommendation}. In \bibinfo{booktitle}{\emph{WWW}}.
  \bibinfo{pages}{661--670}.
\newblock


\bibitem[\protect\citeauthoryear{Li, Chu, Langford, and Wang}{Li
  et~al\mbox{.}}{2011a}]%
        {li2011unbiased}
\bibfield{author}{\bibinfo{person}{Lihong Li}, \bibinfo{person}{Wei Chu},
  \bibinfo{person}{John Langford}, {and} \bibinfo{person}{Xuanhui Wang}.}
  \bibinfo{year}{2011}\natexlab{a}.
\newblock \showarticletitle{Unbiased offline evaluation of
  contextual-bandit-based news article recommendation algorithms}. In
  \bibinfo{booktitle}{\emph{WSDM}}. \bibinfo{pages}{297--306}.
\newblock


\bibitem[\protect\citeauthoryear{Li and Li}{Li and Li}{2013}]%
        {li2013news}
\bibfield{author}{\bibinfo{person}{Lei Li} {and} \bibinfo{person}{Tao Li}.}
  \bibinfo{year}{2013}\natexlab{}.
\newblock \showarticletitle{News recommendation via hypergraph learning:
  encapsulation of user behavior and news content}. In
  \bibinfo{booktitle}{\emph{WSDM}}. \bibinfo{pages}{305--314}.
\newblock


\bibitem[\protect\citeauthoryear{Li, Wang, Li, Knox, and Padmanabhan}{Li
  et~al\mbox{.}}{2011b}]%
        {li2011scene}
\bibfield{author}{\bibinfo{person}{Lei Li}, \bibinfo{person}{Dingding Wang},
  \bibinfo{person}{Tao Li}, \bibinfo{person}{Daniel Knox}, {and}
  \bibinfo{person}{Balaji Padmanabhan}.} \bibinfo{year}{2011}\natexlab{b}.
\newblock \showarticletitle{SCENE: a scalable two-stage personalized news
  recommendation system}. In \bibinfo{booktitle}{\emph{SIGIR}}.
  \bibinfo{pages}{125--134}.
\newblock


\bibitem[\protect\citeauthoryear{Li, Wang, Zhu, and Li}{Li
  et~al\mbox{.}}{2011c}]%
        {li2011personalized}
\bibfield{author}{\bibinfo{person}{Lei Li}, \bibinfo{person}{Ding-Ding Wang},
  \bibinfo{person}{Shun-Zhi Zhu}, {and} \bibinfo{person}{Tao Li}.}
  \bibinfo{year}{2011}\natexlab{c}.
\newblock \showarticletitle{Personalized news recommendation: a review and an
  experimental investigation}.
\newblock \bibinfo{journal}{\emph{Journal of Computer Science and Technology}}
  \bibinfo{volume}{26}, \bibinfo{number}{5} (\bibinfo{year}{2011}),
  \bibinfo{pages}{754}.
\newblock


\bibitem[\protect\citeauthoryear{Li, Zheng, and Li}{Li et~al\mbox{.}}{2011d}]%
        {li2011logo}
\bibfield{author}{\bibinfo{person}{Lei Li}, \bibinfo{person}{Li Zheng}, {and}
  \bibinfo{person}{Tao Li}.} \bibinfo{year}{2011}\natexlab{d}.
\newblock \showarticletitle{Logo: a long-short user interest integration in
  personalized news recommendation}. In \bibinfo{booktitle}{\emph{Recsys}}.
  \bibinfo{pages}{317--320}.
\newblock


\bibitem[\protect\citeauthoryear{Li, Zheng, Yang, and Li}{Li
  et~al\mbox{.}}{2014}]%
        {li2014modeling}
\bibfield{author}{\bibinfo{person}{Lei Li}, \bibinfo{person}{Li Zheng},
  \bibinfo{person}{Fan Yang}, {and} \bibinfo{person}{Tao Li}.}
  \bibinfo{year}{2014}\natexlab{}.
\newblock \showarticletitle{Modeling and broadening temporal user interest in
  personalized news recommendation}.
\newblock \bibinfo{journal}{\emph{Expert Systems With Applications}}
  \bibinfo{volume}{41}, \bibinfo{number}{7} (\bibinfo{year}{2014}),
  \bibinfo{pages}{3168--3177}.
\newblock


\bibitem[\protect\citeauthoryear{Li and Wang}{Li and Wang}{2019}]%
        {li2019survey}
\bibfield{author}{\bibinfo{person}{Miaomiao Li} {and} \bibinfo{person}{Licheng
  Wang}.} \bibinfo{year}{2019}\natexlab{}.
\newblock \showarticletitle{A Survey on Personalized News Recommendation
  Technology}.
\newblock \bibinfo{journal}{\emph{IEEE Access}}  \bibinfo{volume}{7}
  (\bibinfo{year}{2019}), \bibinfo{pages}{145861--145879}.
\newblock


\bibitem[\protect\citeauthoryear{Li, Wang, Chen, and Lin}{Li
  et~al\mbox{.}}{2010b}]%
        {li2010user}
\bibfield{author}{\bibinfo{person}{Qing Li}, \bibinfo{person}{Jia Wang},
  \bibinfo{person}{Yuanzhu~Peter Chen}, {and} \bibinfo{person}{Zhangxi Lin}.}
  \bibinfo{year}{2010}\natexlab{b}.
\newblock \showarticletitle{User Comments for News Recommendation in
  Forum-Based Social Media}.
\newblock \bibinfo{journal}{\emph{Information Science}} \bibinfo{volume}{180},
  \bibinfo{number}{24} (\bibinfo{date}{Dec.} \bibinfo{year}{2010}),
  \bibinfo{pages}{4929–4939}.
\newblock


\bibitem[\protect\citeauthoryear{Li, Sahu, Talwalkar, and Smith}{Li
  et~al\mbox{.}}{2020}]%
        {li2020federated}
\bibfield{author}{\bibinfo{person}{Tian Li}, \bibinfo{person}{Anit~Kumar Sahu},
  \bibinfo{person}{Ameet Talwalkar}, {and} \bibinfo{person}{Virginia Smith}.}
  \bibinfo{year}{2020}\natexlab{}.
\newblock \showarticletitle{Federated Learning: Challenges, Methods, and Future
  Directions}.
\newblock \bibinfo{journal}{\emph{IEEE Signal Processing Magazine}}
  \bibinfo{volume}{37}, \bibinfo{number}{3} (\bibinfo{year}{2020}),
  \bibinfo{pages}{50--60}.
\newblock


\bibitem[\protect\citeauthoryear{Lian, Zhang, Xie, and Sun}{Lian
  et~al\mbox{.}}{2018}]%
        {lian2018towards}
\bibfield{author}{\bibinfo{person}{Jianxun Lian}, \bibinfo{person}{Fuzheng
  Zhang}, \bibinfo{person}{Xing Xie}, {and} \bibinfo{person}{Guangzhong Sun}.}
  \bibinfo{year}{2018}\natexlab{}.
\newblock \showarticletitle{Towards Better Representation Learning for
  Personalized News Recommendation: a Multi-Channel Deep Fusion Approach}. In
  \bibinfo{booktitle}{\emph{IJCAI}}. \bibinfo{pages}{3805--3811}.
\newblock


\bibitem[\protect\citeauthoryear{Liang, Loni, and Larson}{Liang
  et~al\mbox{.}}{2017}]%
        {liang2017clef}
\bibfield{author}{\bibinfo{person}{Yu Liang}, \bibinfo{person}{Babak Loni},
  {and} \bibinfo{person}{Martha Larson}.} \bibinfo{year}{2017}\natexlab{}.
\newblock \showarticletitle{CLEF NewsREEL 2017: Contextual Bandit News
  Recommendation.}. In \bibinfo{booktitle}{\emph{CLEF}}.
\newblock


\bibitem[\protect\citeauthoryear{Lin, Xie, Guan, Li, and Li}{Lin
  et~al\mbox{.}}{2014}]%
        {lin2014personalized}
\bibfield{author}{\bibinfo{person}{Chen Lin}, \bibinfo{person}{Runquan Xie},
  \bibinfo{person}{Xinjun Guan}, \bibinfo{person}{Lei Li}, {and}
  \bibinfo{person}{Tao Li}.} \bibinfo{year}{2014}\natexlab{}.
\newblock \showarticletitle{Personalized news recommendation via implicit
  social experts}.
\newblock \bibinfo{journal}{\emph{Information Science}}  \bibinfo{volume}{254}
  (\bibinfo{year}{2014}), \bibinfo{pages}{1--18}.
\newblock


\bibitem[\protect\citeauthoryear{Liu, Bai, Lian, Zhao, Sun, Wen, and Xie}{Liu
  et~al\mbox{.}}{2019}]%
        {liu2019news}
\bibfield{author}{\bibinfo{person}{Danyang Liu}, \bibinfo{person}{Ting Bai},
  \bibinfo{person}{Jianxun Lian}, \bibinfo{person}{Xin Zhao},
  \bibinfo{person}{Guangzhong Sun}, \bibinfo{person}{Ji-Rong Wen}, {and}
  \bibinfo{person}{Xing Xie}.} \bibinfo{year}{2019}\natexlab{}.
\newblock \showarticletitle{News Graph: An Enhanced Knowledge Graph for News
  Recommendation.}. In \bibinfo{booktitle}{\emph{KaRS@CIKM}}.
  \bibinfo{pages}{1--7}.
\newblock


\bibitem[\protect\citeauthoryear{Liu, Lian, Liu, Wang, Sun, and Xie}{Liu
  et~al\mbox{.}}{2021a}]%
        {liu2021reinforced}
\bibfield{author}{\bibinfo{person}{Danyang Liu}, \bibinfo{person}{Jianxun
  Lian}, \bibinfo{person}{Zheng Liu}, \bibinfo{person}{Xiting Wang},
  \bibinfo{person}{Guangzhong Sun}, {and} \bibinfo{person}{Xing Xie}.}
  \bibinfo{year}{2021}\natexlab{a}.
\newblock \showarticletitle{Reinforced Anchor Knowledge Graph Generation for
  News Recommendation Reasoning}. In \bibinfo{booktitle}{\emph{KDD}}.
  \bibinfo{pages}{1055--1065}.
\newblock


\bibitem[\protect\citeauthoryear{Liu, Lian, Wang, Qiao, Chen, Sun, and Xie}{Liu
  et~al\mbox{.}}{2020}]%
        {liu2020kred}
\bibfield{author}{\bibinfo{person}{Danyang Liu}, \bibinfo{person}{Jianxun
  Lian}, \bibinfo{person}{Shiyin Wang}, \bibinfo{person}{Ying Qiao},
  \bibinfo{person}{Jiun-Hung Chen}, \bibinfo{person}{Guangzhong Sun}, {and}
  \bibinfo{person}{Xing Xie}.} \bibinfo{year}{2020}\natexlab{}.
\newblock \showarticletitle{KRED: Knowledge-Aware Document Representation for
  News Recommendations}. In \bibinfo{booktitle}{\emph{Recsys}}.
  \bibinfo{pages}{200--209}.
\newblock


\bibitem[\protect\citeauthoryear{Liu, Dolan, and Pedersen}{Liu
  et~al\mbox{.}}{2010}]%
        {liu2010personalized}
\bibfield{author}{\bibinfo{person}{Jiahui Liu}, \bibinfo{person}{Peter Dolan},
  {and} \bibinfo{person}{Elin~R{\o}nby Pedersen}.}
  \bibinfo{year}{2010}\natexlab{}.
\newblock \showarticletitle{Personalized news recommendation based on click
  behavior}. In \bibinfo{booktitle}{\emph{IUI}}. \bibinfo{pages}{31--40}.
\newblock


\bibitem[\protect\citeauthoryear{Liu, Song, Li, Zhu, and Deng}{Liu
  et~al\mbox{.}}{2021c}]%
        {liu2021hybrid}
\bibfield{author}{\bibinfo{person}{Jing Liu}, \bibinfo{person}{Jinbao Song},
  \bibinfo{person}{Chen Li}, \bibinfo{person}{Xiaoya Zhu}, {and}
  \bibinfo{person}{Ruyi Deng}.} \bibinfo{year}{2021}\natexlab{c}.
\newblock \showarticletitle{A Hybrid News Recommendation Algorithm Based On
  K-means Clustering and Collaborative Filtering}. In
  \bibinfo{booktitle}{\emph{Journal of Physics: Conference Series}},
  Vol.~\bibinfo{volume}{1881}. IOP Publishing, \bibinfo{pages}{032050}.
\newblock


\bibitem[\protect\citeauthoryear{Liu, Shivaram, Culotta, Shapiro, and
  Bilgic}{Liu et~al\mbox{.}}{2021b}]%
        {liu2021interaction}
\bibfield{author}{\bibinfo{person}{Ping Liu}, \bibinfo{person}{Karthik
  Shivaram}, \bibinfo{person}{Aron Culotta}, \bibinfo{person}{Matthew~A
  Shapiro}, {and} \bibinfo{person}{Mustafa Bilgic}.}
  \bibinfo{year}{2021}\natexlab{b}.
\newblock \showarticletitle{The Interaction between Political Typology and
  Filter Bubbles in News Recommendation Algorithms}. In
  \bibinfo{booktitle}{\emph{WWW}}. \bibinfo{pages}{3791--3801}.
\newblock


\bibitem[\protect\citeauthoryear{Lommatzsch}{Lommatzsch}{2014}]%
        {lommatzsch2014real}
\bibfield{author}{\bibinfo{person}{Andreas Lommatzsch}.}
  \bibinfo{year}{2014}\natexlab{}.
\newblock \showarticletitle{Real-time news recommendation using context-aware
  ensembles}. In \bibinfo{booktitle}{\emph{ECIR}}. Springer,
  \bibinfo{pages}{51--62}.
\newblock


\bibitem[\protect\citeauthoryear{Lommatzsch, Kille, Hopfgartner, Larson, Brodt,
  Seiler, and {\"O}zg{\"o}bek}{Lommatzsch et~al\mbox{.}}{2017}]%
        {lommatzsch2017clef}
\bibfield{author}{\bibinfo{person}{Andreas Lommatzsch},
  \bibinfo{person}{Benjamin Kille}, \bibinfo{person}{Frank Hopfgartner},
  \bibinfo{person}{Martha Larson}, \bibinfo{person}{Torben Brodt},
  \bibinfo{person}{Jonas Seiler}, {and} \bibinfo{person}{{\"O}zlem
  {\"O}zg{\"o}bek}.} \bibinfo{year}{2017}\natexlab{}.
\newblock \showarticletitle{CLEF 2017 NewsREEL overview: A stream-based
  recommender task for evaluation and education}. In
  \bibinfo{booktitle}{\emph{CLEF}}. Springer, \bibinfo{pages}{239--254}.
\newblock


\bibitem[\protect\citeauthoryear{Lommatzsch, Kille, Hopfgartner, and
  Ramming}{Lommatzsch et~al\mbox{.}}{2018}]%
        {lommatzsch2018newsreel}
\bibfield{author}{\bibinfo{person}{Andreas Lommatzsch},
  \bibinfo{person}{Benjamin Kille}, \bibinfo{person}{Frank Hopfgartner}, {and}
  \bibinfo{person}{Leif Ramming}.} \bibinfo{year}{2018}\natexlab{}.
\newblock \showarticletitle{NewsREEL Multimedia at MediaEval 2018: News
  Recommendation with Image and Text Content}. In
  \bibinfo{booktitle}{\emph{MediaEval}}. CEUR-WS.
\newblock


\bibitem[\protect\citeauthoryear{Lu, Batra, Parikh, and Lee}{Lu
  et~al\mbox{.}}{2019}]%
        {lu2019vilbert}
\bibfield{author}{\bibinfo{person}{Jiasen Lu}, \bibinfo{person}{Dhruv Batra},
  \bibinfo{person}{Devi Parikh}, {and} \bibinfo{person}{Stefan Lee}.}
  \bibinfo{year}{2019}\natexlab{}.
\newblock \showarticletitle{Vilbert: Pretraining task-agnostic visiolinguistic
  representations for vision-and-language tasks}.
\newblock  (\bibinfo{year}{2019}), \bibinfo{pages}{13--23}.
\newblock


\bibitem[\protect\citeauthoryear{Lu and Liu}{Lu and Liu}{2016}]%
        {lu2016hier}
\bibfield{author}{\bibinfo{person}{Meilian Lu} {and} \bibinfo{person}{Jinliang
  Liu}.} \bibinfo{year}{2016}\natexlab{}.
\newblock \showarticletitle{Hier-UIM: A hierarchy user interest model for
  personalized news recommender}. In \bibinfo{booktitle}{\emph{CCIS}}. IEEE,
  \bibinfo{pages}{249--254}.
\newblock


\bibitem[\protect\citeauthoryear{Ludmann}{Ludmann}{2017}]%
        {ludmann2017recommending}
\bibfield{author}{\bibinfo{person}{Cornelius~A Ludmann}.}
  \bibinfo{year}{2017}\natexlab{}.
\newblock \showarticletitle{Recommending News Articles in the CLEF News
  Recommendation Evaluation Lab with the Data Stream Management System
  Odysseus.}. In \bibinfo{booktitle}{\emph{CLEF}}.
\newblock


\bibitem[\protect\citeauthoryear{Luo, Fan, and Keim}{Luo et~al\mbox{.}}{2008}]%
        {videorec}
\bibfield{author}{\bibinfo{person}{Hangzai Luo}, \bibinfo{person}{Jianping
  Fan}, {and} \bibinfo{person}{Daniel~A. Keim}.}
  \bibinfo{year}{2008}\natexlab{}.
\newblock \showarticletitle{Personalized News Video Recommendation}. In
  \bibinfo{booktitle}{\emph{ACM MM}}. \bibinfo{pages}{1001--1002}.
\newblock


\bibitem[\protect\citeauthoryear{Luostarinen and Kohonen}{Luostarinen and
  Kohonen}{2013}]%
        {luostarinen2013using}
\bibfield{author}{\bibinfo{person}{Tapio Luostarinen} {and}
  \bibinfo{person}{Oskar Kohonen}.} \bibinfo{year}{2013}\natexlab{}.
\newblock \showarticletitle{Using topic models in content-based news
  recommender systems}. In \bibinfo{booktitle}{\emph{NODALIDA}}.
  \bibinfo{pages}{239--251}.
\newblock


\bibitem[\protect\citeauthoryear{Ma, Na, Wang, Chen, and Xu}{Ma
  et~al\mbox{.}}{2021}]%
        {ma2021graph}
\bibfield{author}{\bibinfo{person}{Mingyuan Ma}, \bibinfo{person}{Sen Na},
  \bibinfo{person}{Hongyu Wang}, \bibinfo{person}{Congzhou Chen}, {and}
  \bibinfo{person}{Jin Xu}.} \bibinfo{year}{2021}\natexlab{}.
\newblock \showarticletitle{The graph-based behavior-aware recommendation for
  interactive news}.
\newblock \bibinfo{journal}{\emph{Applied Intelligence}}
  (\bibinfo{year}{2021}), \bibinfo{pages}{1--17}.
\newblock


\bibitem[\protect\citeauthoryear{Makhortykh and Wijermars}{Makhortykh and
  Wijermars}{2021}]%
        {makhortykh2021can}
\bibfield{author}{\bibinfo{person}{Mykola Makhortykh} {and}
  \bibinfo{person}{Mari{\"e}lle Wijermars}.} \bibinfo{year}{2021}\natexlab{}.
\newblock \showarticletitle{Can Filter Bubbles Protect Information Freedom?
  Discussions of Algorithmic News Recommenders in Eastern Europe}.
\newblock \bibinfo{journal}{\emph{Digital Journalism}} (\bibinfo{year}{2021}),
  \bibinfo{pages}{1--25}.
\newblock


\bibitem[\protect\citeauthoryear{Manoharan and Senthilkumar}{Manoharan and
  Senthilkumar}{2020}]%
        {manoharan2020intelligent}
\bibfield{author}{\bibinfo{person}{Saravanapriya Manoharan} {and}
  \bibinfo{person}{Radha Senthilkumar}.} \bibinfo{year}{2020}\natexlab{}.
\newblock \showarticletitle{An intelligent fuzzy rule-based personalized news
  recommendation using social media mining}.
\newblock \bibinfo{journal}{\emph{Computational intelligence and neuroscience}}
   \bibinfo{volume}{2020} (\bibinfo{year}{2020}).
\newblock


\bibitem[\protect\citeauthoryear{Mao, Zeng, and Wong}{Mao
  et~al\mbox{.}}{2021}]%
        {mao2021neuralnews}
\bibfield{author}{\bibinfo{person}{Zhiming Mao}, \bibinfo{person}{Xingshan
  Zeng}, {and} \bibinfo{person}{Kam-Fai Wong}.}
  \bibinfo{year}{2021}\natexlab{}.
\newblock \showarticletitle{Neural News Recommendation with Collaborative News
  Encoding and Structural User Encoding}. In \bibinfo{booktitle}{\emph{EMNLP:
  Findings}}. \bibinfo{pages}{46--55}.
\newblock


\bibitem[\protect\citeauthoryear{McMahan, Moore, Ramage, Hampson, and
  y~Arcas}{McMahan et~al\mbox{.}}{2017}]%
        {mcmahan2017communication}
\bibfield{author}{\bibinfo{person}{Brendan McMahan}, \bibinfo{person}{Eider
  Moore}, \bibinfo{person}{Daniel Ramage}, \bibinfo{person}{Seth Hampson},
  {and} \bibinfo{person}{Blaise~Aguera y Arcas}.}
  \bibinfo{year}{2017}\natexlab{}.
\newblock \showarticletitle{Communication-Efficient Learning of Deep Networks
  from Decentralized Data}. In \bibinfo{booktitle}{\emph{AISTATS}}.
  \bibinfo{pages}{1273--1282}.
\newblock


\bibitem[\protect\citeauthoryear{Meng, Shi, Hao, and Su}{Meng
  et~al\mbox{.}}{2021}]%
        {meng2021dcan}
\bibfield{author}{\bibinfo{person}{Lingkang Meng}, \bibinfo{person}{Chongyang
  Shi}, \bibinfo{person}{Shufeng Hao}, {and} \bibinfo{person}{Xiangrui Su}.}
  \bibinfo{year}{2021}\natexlab{}.
\newblock \showarticletitle{DCAN: Deep Co-Attention Network by Modeling User
  Preference and News Lifecycle for News Recommendation}. In
  \bibinfo{booktitle}{\emph{DASFAA}}. Springer, \bibinfo{pages}{100--114}.
\newblock


\bibitem[\protect\citeauthoryear{Mikolov, Sutskever, Chen, Corrado, and
  Dean}{Mikolov et~al\mbox{.}}{2013}]%
        {mikolov2013distributed}
\bibfield{author}{\bibinfo{person}{Tomas Mikolov}, \bibinfo{person}{Ilya
  Sutskever}, \bibinfo{person}{Kai Chen}, \bibinfo{person}{Greg~S Corrado},
  {and} \bibinfo{person}{Jeff Dean}.} \bibinfo{year}{2013}\natexlab{}.
\newblock \showarticletitle{Distributed representations of words and phrases
  and their compositionality}. In \bibinfo{booktitle}{\emph{NIPS}}.
  \bibinfo{pages}{3111--3119}.
\newblock


\bibitem[\protect\citeauthoryear{Moerland, Hogenboom, Capelle, and
  Frasincar}{Moerland et~al\mbox{.}}{2013}]%
        {moerland2013semantics}
\bibfield{author}{\bibinfo{person}{Marnix Moerland}, \bibinfo{person}{Frederik
  Hogenboom}, \bibinfo{person}{Michel Capelle}, {and} \bibinfo{person}{Flavius
  Frasincar}.} \bibinfo{year}{2013}\natexlab{}.
\newblock \showarticletitle{Semantics-based news recommendation with SF-IDF+}.
  In \bibinfo{booktitle}{\emph{WIMS}}. \bibinfo{pages}{1--8}.
\newblock


\bibitem[\protect\citeauthoryear{Mookiah, Eberle, and Mondal}{Mookiah
  et~al\mbox{.}}{2018}]%
        {mookiah2018personalized}
\bibfield{author}{\bibinfo{person}{Lenin Mookiah}, \bibinfo{person}{William
  Eberle}, {and} \bibinfo{person}{Maitrayi Mondal}.}
  \bibinfo{year}{2018}\natexlab{}.
\newblock \showarticletitle{Personalized news recommendation using graph-based
  approach}.
\newblock \bibinfo{journal}{\emph{Intelligent Data Analysis}}
  \bibinfo{volume}{22}, \bibinfo{number}{4} (\bibinfo{year}{2018}),
  \bibinfo{pages}{881--909}.
\newblock


\bibitem[\protect\citeauthoryear{Nguyen, Arch-Int, and Arch-Int}{Nguyen
  et~al\mbox{.}}{2015}]%
        {nguyen2015semantically}
\bibfield{author}{\bibinfo{person}{Cuong Dinh~Hoa Nguyen},
  \bibinfo{person}{Ngamnij Arch-Int}, {and} \bibinfo{person}{Somjit Arch-Int}.}
  \bibinfo{year}{2015}\natexlab{}.
\newblock \showarticletitle{A semantically hybrid framework of personalizing
  news recommendations}.
\newblock \bibinfo{journal}{\emph{International Journal of Innovative
  Computing, Information and Control}} \bibinfo{volume}{11},
  \bibinfo{number}{6} (\bibinfo{year}{2015}), \bibinfo{pages}{1947--1963}.
\newblock


\bibitem[\protect\citeauthoryear{Noh, Oh, and Park}{Noh et~al\mbox{.}}{2014}]%
        {noh2014location}
\bibfield{author}{\bibinfo{person}{Yunseok Noh}, \bibinfo{person}{Yong-Hwan
  Oh}, {and} \bibinfo{person}{Seong-Bae Park}.}
  \bibinfo{year}{2014}\natexlab{}.
\newblock \showarticletitle{A location-based personalized news recommendation}.
  In \bibinfo{booktitle}{\emph{BigComp}}. IEEE, \bibinfo{pages}{99--104}.
\newblock


\bibitem[\protect\citeauthoryear{Okura, Tagami, Ono, and Tajima}{Okura
  et~al\mbox{.}}{2017}]%
        {okura2017embedding}
\bibfield{author}{\bibinfo{person}{Shumpei Okura}, \bibinfo{person}{Yukihiro
  Tagami}, \bibinfo{person}{Shingo Ono}, {and} \bibinfo{person}{Akira Tajima}.}
  \bibinfo{year}{2017}\natexlab{}.
\newblock \showarticletitle{Embedding-based news recommendation for millions of
  users}. In \bibinfo{booktitle}{\emph{KDD}}. \bibinfo{pages}{1933--1942}.
\newblock


\bibitem[\protect\citeauthoryear{Oord, Li, and Vinyals}{Oord
  et~al\mbox{.}}{2018}]%
        {oord2018representation}
\bibfield{author}{\bibinfo{person}{Aaron van~den Oord}, \bibinfo{person}{Yazhe
  Li}, {and} \bibinfo{person}{Oriol Vinyals}.} \bibinfo{year}{2018}\natexlab{}.
\newblock \showarticletitle{Representation learning with contrastive predictive
  coding}.
\newblock \bibinfo{journal}{\emph{arXiv preprint arXiv:1807.03748}}
  (\bibinfo{year}{2018}).
\newblock


\bibitem[\protect\citeauthoryear{{\"O}zg{\"o}bek, Gulla, and
  Erdur}{{\"O}zg{\"o}bek et~al\mbox{.}}{2014}]%
        {ozgobek2014survey}
\bibfield{author}{\bibinfo{person}{{\"O}zlem {\"O}zg{\"o}bek},
  \bibinfo{person}{Jon~Atle Gulla}, {and} \bibinfo{person}{Riza~Cenk Erdur}.}
  \bibinfo{year}{2014}\natexlab{}.
\newblock \showarticletitle{A Survey on Challenges and Methods in News
  Recommendation}. In \bibinfo{booktitle}{\emph{WEBIST}}.
  \bibinfo{pages}{278--285}.
\newblock


\bibitem[\protect\citeauthoryear{Parizi and Kazemifard}{Parizi and
  Kazemifard}{2015}]%
        {parizi2015emotional}
\bibfield{author}{\bibinfo{person}{Ali~Hakimi Parizi} {and}
  \bibinfo{person}{Mohammad Kazemifard}.} \bibinfo{year}{2015}\natexlab{}.
\newblock \showarticletitle{Emotional news recommender system}. In
  \bibinfo{booktitle}{\emph{ICCS}}. IEEE, \bibinfo{pages}{37--41}.
\newblock


\bibitem[\protect\citeauthoryear{Parizi, Kazemifard, and Asghari}{Parizi
  et~al\mbox{.}}{2016}]%
        {parizi2016emonews}
\bibfield{author}{\bibinfo{person}{Ali~Hakimi Parizi},
  \bibinfo{person}{Mohammad Kazemifard}, {and} \bibinfo{person}{Mohsen
  Asghari}.} \bibinfo{year}{2016}\natexlab{}.
\newblock \showarticletitle{EmoNews: an Emotional News Recommender System.}
\newblock \bibinfo{journal}{\emph{Journal of Digital Information Management}}
  \bibinfo{volume}{14}, \bibinfo{number}{6} (\bibinfo{year}{2016}).
\newblock


\bibitem[\protect\citeauthoryear{Park, Lee, and Choi}{Park
  et~al\mbox{.}}{2017}]%
        {park2017deep}
\bibfield{author}{\bibinfo{person}{Keunchan Park}, \bibinfo{person}{Jisoo Lee},
  {and} \bibinfo{person}{Jaeho Choi}.} \bibinfo{year}{2017}\natexlab{}.
\newblock \showarticletitle{Deep neural networks for news recommendations}. In
  \bibinfo{booktitle}{\emph{CIKM}}. \bibinfo{pages}{2255--2258}.
\newblock


\bibitem[\protect\citeauthoryear{Patankar, Bose, and Khanna}{Patankar
  et~al\mbox{.}}{2019}]%
        {patankar2019bias}
\bibfield{author}{\bibinfo{person}{Anish Patankar}, \bibinfo{person}{Joy Bose},
  {and} \bibinfo{person}{Harshit Khanna}.} \bibinfo{year}{2019}\natexlab{}.
\newblock \showarticletitle{A bias aware news recommendation system}. In
  \bibinfo{booktitle}{\emph{ICSC}}. IEEE, \bibinfo{pages}{232--238}.
\newblock


\bibitem[\protect\citeauthoryear{Pennington, Socher, and Manning}{Pennington
  et~al\mbox{.}}{2014}]%
        {pennington2014glove}
\bibfield{author}{\bibinfo{person}{Jeffrey Pennington},
  \bibinfo{person}{Richard Socher}, {and} \bibinfo{person}{Christopher
  Manning}.} \bibinfo{year}{2014}\natexlab{}.
\newblock \showarticletitle{Glove: Global vectors for word representation}. In
  \bibinfo{booktitle}{\emph{EMNLP}}. \bibinfo{pages}{1532--1543}.
\newblock


\bibitem[\protect\citeauthoryear{Phelan, McCarthy, Bennett, and Smyth}{Phelan
  et~al\mbox{.}}{2011a}]%
        {phelan2011using}
\bibfield{author}{\bibinfo{person}{Owen Phelan}, \bibinfo{person}{Kevin
  McCarthy}, \bibinfo{person}{Mike Bennett}, {and} \bibinfo{person}{Barry
  Smyth}.} \bibinfo{year}{2011}\natexlab{a}.
\newblock \showarticletitle{On using the real-time web for news recommendation
  \& discovery}. In \bibinfo{booktitle}{\emph{WWW}}. \bibinfo{pages}{103--104}.
\newblock


\bibitem[\protect\citeauthoryear{Phelan, McCarthy, Bennett, and Smyth}{Phelan
  et~al\mbox{.}}{2011b}]%
        {phelan2011terms}
\bibfield{author}{\bibinfo{person}{Owen Phelan}, \bibinfo{person}{Kevin
  McCarthy}, \bibinfo{person}{Mike Bennett}, {and} \bibinfo{person}{Barry
  Smyth}.} \bibinfo{year}{2011}\natexlab{b}.
\newblock \showarticletitle{Terms of a feather: Content-based news
  recommendation and discovery using twitter}. In
  \bibinfo{booktitle}{\emph{ECIR}}. Springer, \bibinfo{pages}{448--459}.
\newblock


\bibitem[\protect\citeauthoryear{Phelan, McCarthy, and Smyth}{Phelan
  et~al\mbox{.}}{2009}]%
        {phelan2009using}
\bibfield{author}{\bibinfo{person}{Owen Phelan}, \bibinfo{person}{Kevin
  McCarthy}, {and} \bibinfo{person}{Barry Smyth}.}
  \bibinfo{year}{2009}\natexlab{}.
\newblock \showarticletitle{Using twitter to recommend real-time topical news}.
  In \bibinfo{booktitle}{\emph{Recsys}}. \bibinfo{pages}{385--388}.
\newblock


\bibitem[\protect\citeauthoryear{Prawesh and Padmanabhan}{Prawesh and
  Padmanabhan}{2021}]%
        {prawesh2021complex}
\bibfield{author}{\bibinfo{person}{Shankar Prawesh} {and}
  \bibinfo{person}{Balaji Padmanabhan}.} \bibinfo{year}{2021}\natexlab{}.
\newblock \showarticletitle{A complex systems perspective of news recommender
  systems: Guiding emergent outcomes with feedback models}.
\newblock \bibinfo{journal}{\emph{Plos one}} \bibinfo{volume}{16},
  \bibinfo{number}{1} (\bibinfo{year}{2021}), \bibinfo{pages}{e0245096}.
\newblock


\bibitem[\protect\citeauthoryear{Qi, Wu, Wu, and Huang}{Qi
  et~al\mbox{.}}{2021a}]%
        {qi2021kim}
\bibfield{author}{\bibinfo{person}{Tao Qi}, \bibinfo{person}{Fangzhao Wu},
  \bibinfo{person}{Chuhan Wu}, {and} \bibinfo{person}{Yongfeng Huang}.}
  \bibinfo{year}{2021}\natexlab{a}.
\newblock \showarticletitle{Personalized News Recommendation with
  Knowledge-aware Interactive Matching}. In \bibinfo{booktitle}{\emph{SIGIR}}.
  \bibinfo{pages}{61--70}.
\newblock


\bibitem[\protect\citeauthoryear{Qi, Wu, Wu, and Huang}{Qi
  et~al\mbox{.}}{2021b}]%
        {qi2021pprec}
\bibfield{author}{\bibinfo{person}{Tao Qi}, \bibinfo{person}{Fangzhao Wu},
  \bibinfo{person}{Chuhan Wu}, {and} \bibinfo{person}{Yongfeng Huang}.}
  \bibinfo{year}{2021}\natexlab{b}.
\newblock \showarticletitle{PP-Rec: News Recommendation with Personalized User
  Interest and Time-aware News Popularity}. In \bibinfo{booktitle}{\emph{ACL}}.
  \bibinfo{pages}{5457--5467}.
\newblock


\bibitem[\protect\citeauthoryear{Qi, Wu, Wu, Huang, and Xie}{Qi
  et~al\mbox{.}}{2020}]%
        {qi2020privacy}
\bibfield{author}{\bibinfo{person}{Tao Qi}, \bibinfo{person}{Fangzhao Wu},
  \bibinfo{person}{Chuhan Wu}, \bibinfo{person}{Yongfeng Huang}, {and}
  \bibinfo{person}{Xing Xie}.} \bibinfo{year}{2020}\natexlab{}.
\newblock \showarticletitle{Privacy-Preserving News Recommendation Model
  Learning}. In \bibinfo{booktitle}{\emph{EMNLP: Findings}}.
  \bibinfo{pages}{1423--1432}.
\newblock


\bibitem[\protect\citeauthoryear{Qi, Wu, Wu, Huang, and Xie}{Qi
  et~al\mbox{.}}{2021c}]%
        {qi2021uni}
\bibfield{author}{\bibinfo{person}{Tao Qi}, \bibinfo{person}{Fangzhao Wu},
  \bibinfo{person}{Chuhan Wu}, \bibinfo{person}{Yongfeng Huang}, {and}
  \bibinfo{person}{Xing Xie}.} \bibinfo{year}{2021}\natexlab{c}.
\newblock \showarticletitle{Uni-FedRec: A Unified Privacy-Preserving News
  Recommendation Framework for Model Training and Online Serving}. In
  \bibinfo{booktitle}{\emph{EMNLP: Findings}}. \bibinfo{pages}{1438--1448}.
\newblock


\bibitem[\protect\citeauthoryear{Qi, Wu, Wu, Yang, Yu, Xie, and Huang}{Qi
  et~al\mbox{.}}{2021d}]%
        {qi2021hierec}
\bibfield{author}{\bibinfo{person}{Tao Qi}, \bibinfo{person}{Fangzhao Wu},
  \bibinfo{person}{Chuhan Wu}, \bibinfo{person}{Peiru Yang},
  \bibinfo{person}{Yang Yu}, \bibinfo{person}{Xing Xie}, {and}
  \bibinfo{person}{Yongfeng Huang}.} \bibinfo{year}{2021}\natexlab{d}.
\newblock \showarticletitle{HieRec: Hierarchical User Interest Modeling for
  Personalized News Recommendation}. In \bibinfo{booktitle}{\emph{ACL}}.
\newblock


\bibitem[\protect\citeauthoryear{Qian, Zhao, Li, Fang, Zhao, Sheng, and
  Cui}{Qian et~al\mbox{.}}{2019}]%
        {qian2019interaction}
\bibfield{author}{\bibinfo{person}{Yongye Qian}, \bibinfo{person}{Pengpeng
  Zhao}, \bibinfo{person}{Zhixu Li}, \bibinfo{person}{Junhua Fang},
  \bibinfo{person}{Lei Zhao}, \bibinfo{person}{Victor~S Sheng}, {and}
  \bibinfo{person}{Zhiming Cui}.} \bibinfo{year}{2019}\natexlab{}.
\newblock \showarticletitle{Interaction Graph Neural Network for News
  Recommendation}. In \bibinfo{booktitle}{\emph{WISE}}. Springer,
  \bibinfo{pages}{599--614}.
\newblock


\bibitem[\protect\citeauthoryear{Qin}{Qin}{2020}]%
        {qin2020research}
\bibfield{author}{\bibinfo{person}{Jing Qin}.} \bibinfo{year}{2020}\natexlab{}.
\newblock \showarticletitle{Research Progress of News Recommendation Methods}.
\newblock \bibinfo{journal}{\emph{arXiv preprint arXiv:2012.02360}}
  (\bibinfo{year}{2020}).
\newblock


\bibitem[\protect\citeauthoryear{Rao, Jia, Feng, and Zhao}{Rao
  et~al\mbox{.}}{2013a}]%
        {rao2013personalized}
\bibfield{author}{\bibinfo{person}{Junyang Rao}, \bibinfo{person}{Aixia Jia},
  \bibinfo{person}{Yansong Feng}, {and} \bibinfo{person}{Dongyan Zhao}.}
  \bibinfo{year}{2013}\natexlab{a}.
\newblock \showarticletitle{Personalized news recommendation using ontologies
  harvested from the web}. In \bibinfo{booktitle}{\emph{WAIM}}. Springer,
  \bibinfo{pages}{781--787}.
\newblock


\bibitem[\protect\citeauthoryear{Rao, Jia, Feng, and Zhao}{Rao
  et~al\mbox{.}}{2013b}]%
        {rao2013taxonomy}
\bibfield{author}{\bibinfo{person}{Junyang Rao}, \bibinfo{person}{Aixia Jia},
  \bibinfo{person}{Yansong Feng}, {and} \bibinfo{person}{Dongyan Zhao}.}
  \bibinfo{year}{2013}\natexlab{b}.
\newblock \showarticletitle{Taxonomy based personalized news recommendation:
  Novelty and diversity}. In \bibinfo{booktitle}{\emph{WISE}}. Springer,
  \bibinfo{pages}{209--218}.
\newblock


\bibitem[\protect\citeauthoryear{Raza and Ding}{Raza and Ding}{2021a}]%
        {raza2021deep}
\bibfield{author}{\bibinfo{person}{Shaina Raza} {and} \bibinfo{person}{Chen
  Ding}.} \bibinfo{year}{2021}\natexlab{a}.
\newblock \showarticletitle{Deep Dynamic Neural Network to trade-off between
  Accuracy and Diversity in a News Recommender System}.
\newblock \bibinfo{journal}{\emph{arXiv preprint arXiv:2103.08458}}
  (\bibinfo{year}{2021}).
\newblock


\bibitem[\protect\citeauthoryear{Raza and Ding}{Raza and Ding}{2021b}]%
        {raza2021news}
\bibfield{author}{\bibinfo{person}{Shaina Raza} {and} \bibinfo{person}{Chen
  Ding}.} \bibinfo{year}{2021}\natexlab{b}.
\newblock \showarticletitle{News recommender system: a review of recent
  progress, challenges, and opportunities}.
\newblock \bibinfo{journal}{\emph{Artificial Intelligence Review}}
  (\bibinfo{year}{2021}), \bibinfo{pages}{1--52}.
\newblock


\bibitem[\protect\citeauthoryear{Ren, Long, and Xu}{Ren et~al\mbox{.}}{2019}]%
        {REN2019113115}
\bibfield{author}{\bibinfo{person}{Jiangtao Ren}, \bibinfo{person}{Jiawei
  Long}, {and} \bibinfo{person}{Zhikang Xu}.} \bibinfo{year}{2019}\natexlab{}.
\newblock \showarticletitle{Financial news recommendation based on graph
  embeddings}.
\newblock \bibinfo{journal}{\emph{Decision Support Systems}}
  \bibinfo{volume}{125} (\bibinfo{year}{2019}), \bibinfo{pages}{113115}.
\newblock


\bibitem[\protect\citeauthoryear{Resnick, Iacovou, Suchak, Bergstrom, and
  Riedl}{Resnick et~al\mbox{.}}{1994}]%
        {resnick1994grouplens}
\bibfield{author}{\bibinfo{person}{Paul Resnick}, \bibinfo{person}{Neophytos
  Iacovou}, \bibinfo{person}{Mitesh Suchak}, \bibinfo{person}{Peter Bergstrom},
  {and} \bibinfo{person}{John Riedl}.} \bibinfo{year}{1994}\natexlab{}.
\newblock \showarticletitle{GroupLens: an open architecture for collaborative
  filtering of netnews}. In \bibinfo{booktitle}{\emph{CSCW}}.
  \bibinfo{pages}{175--186}.
\newblock


\bibitem[\protect\citeauthoryear{Reuver, Fokkens, Verberne, Toivonen, and
  Boggia}{Reuver et~al\mbox{.}}{2021}]%
        {reuver2021no}
\bibfield{author}{\bibinfo{person}{Myrthe Reuver}, \bibinfo{person}{Antske
  Fokkens}, \bibinfo{person}{Suzan Verberne}, \bibinfo{person}{H Toivonen},
  {and} \bibinfo{person}{M Boggia}.} \bibinfo{year}{2021}\natexlab{}.
\newblock \showarticletitle{No NLP task should be an island:
  multi-disciplinarity for diversity in news recommender systems}.
\newblock \bibinfo{journal}{\emph{Proceedings of the EACL Hackashop on news
  media content analysis and automated report generation}}
  (\bibinfo{year}{2021}), \bibinfo{pages}{45--55}.
\newblock


\bibitem[\protect\citeauthoryear{Santosh, Saha, and Ganguly}{Santosh
  et~al\mbox{.}}{2020}]%
        {santosh2020mvl}
\bibfield{author}{\bibinfo{person}{TYSS Santosh}, \bibinfo{person}{Avirup
  Saha}, {and} \bibinfo{person}{Niloy Ganguly}.}
  \bibinfo{year}{2020}\natexlab{}.
\newblock \showarticletitle{MVL: Multi-View Learning for News Recommendation}.
  In \bibinfo{booktitle}{\emph{SIGIR}}. \bibinfo{pages}{1873--1876}.
\newblock


\bibitem[\protect\citeauthoryear{Saranya and Sadhasivam}{Saranya and
  Sadhasivam}{2012}]%
        {saranya2012personalized}
\bibfield{author}{\bibinfo{person}{KG Saranya} {and} \bibinfo{person}{G~Sudha
  Sadhasivam}.} \bibinfo{year}{2012}\natexlab{}.
\newblock \showarticletitle{A personalized online news recommendation system}.
\newblock \bibinfo{journal}{\emph{International Journal of Computer
  Applications}} \bibinfo{volume}{57}, \bibinfo{number}{18}
  (\bibinfo{year}{2012}).
\newblock


\bibitem[\protect\citeauthoryear{{Shan Liu}, {Yao Dong}, and {Jianping
  Chai}}{{Shan Liu} et~al\mbox{.}}{2016}]%
        {7924826}
\bibfield{author}{\bibinfo{person}{{Shan Liu}}, \bibinfo{person}{{Yao Dong}},
  {and} \bibinfo{person}{{Jianping Chai}}.} \bibinfo{year}{2016}\natexlab{}.
\newblock \showarticletitle{Research of personalized news recommendation system
  based on hybrid collaborative filtering algorithm}. In
  \bibinfo{booktitle}{\emph{ICCC}}. \bibinfo{pages}{865--869}.
\newblock


\bibitem[\protect\citeauthoryear{Shapira, Shoval, Tractinsky, and
  Meyer}{Shapira et~al\mbox{.}}{2009}]%
        {shapira2009epaper}
\bibfield{author}{\bibinfo{person}{Bracha Shapira}, \bibinfo{person}{Peretz
  Shoval}, \bibinfo{person}{Noam Tractinsky}, {and} \bibinfo{person}{Joachim
  Meyer}.} \bibinfo{year}{2009}\natexlab{}.
\newblock \showarticletitle{ePaper: A personalized mobile newspaper}.
\newblock \bibinfo{journal}{\emph{Journal of the American Society for
  Information Science and Technology}} \bibinfo{volume}{60},
  \bibinfo{number}{11} (\bibinfo{year}{2009}), \bibinfo{pages}{2333--2346}.
\newblock


\bibitem[\protect\citeauthoryear{Sheu, Chu, Qi, and Li}{Sheu
  et~al\mbox{.}}{2021}]%
        {sheu2021knowledge}
\bibfield{author}{\bibinfo{person}{Heng-Shiou Sheu}, \bibinfo{person}{Zhixuan
  Chu}, \bibinfo{person}{Daiqing Qi}, {and} \bibinfo{person}{Sheng Li}.}
  \bibinfo{year}{2021}\natexlab{}.
\newblock \showarticletitle{Knowledge-Guided Article Embedding Refinement for
  Session-Based News Recommendation}.
\newblock \bibinfo{journal}{\emph{TNNLS}} (\bibinfo{year}{2021}).
\newblock


\bibitem[\protect\citeauthoryear{Sheu and Li}{Sheu and Li}{2020}]%
        {sheu2020context}
\bibfield{author}{\bibinfo{person}{Heng-Shiou Sheu} {and}
  \bibinfo{person}{Sheng Li}.} \bibinfo{year}{2020}\natexlab{}.
\newblock \showarticletitle{Context-aware graph embedding for session-based
  news recommendation}. In \bibinfo{booktitle}{\emph{Recsys}}.
  \bibinfo{pages}{657--662}.
\newblock


\bibitem[\protect\citeauthoryear{Shi, Ma, Wang, Zhang, Fang, Xu, Liu, and
  Ma}{Shi et~al\mbox{.}}{2021}]%
        {shi2021wg4rec}
\bibfield{author}{\bibinfo{person}{Shaoyun Shi}, \bibinfo{person}{Weizhi Ma},
  \bibinfo{person}{Zhen Wang}, \bibinfo{person}{Min Zhang},
  \bibinfo{person}{Kun Fang}, \bibinfo{person}{Jingfang Xu},
  \bibinfo{person}{Yiqun Liu}, {and} \bibinfo{person}{Shaoping Ma}.}
  \bibinfo{year}{2021}\natexlab{}.
\newblock \showarticletitle{WG4Rec: Modeling Textual Content with Word Graph
  for News Recommendation}. In \bibinfo{booktitle}{\emph{CIKM}}.
  \bibinfo{pages}{1651--1660}.
\newblock


\bibitem[\protect\citeauthoryear{Shu, Sliva, Wang, Tang, and Liu}{Shu
  et~al\mbox{.}}{2017}]%
        {shu2017fake}
\bibfield{author}{\bibinfo{person}{Kai Shu}, \bibinfo{person}{Amy Sliva},
  \bibinfo{person}{Suhang Wang}, \bibinfo{person}{Jiliang Tang}, {and}
  \bibinfo{person}{Huan Liu}.} \bibinfo{year}{2017}\natexlab{}.
\newblock \showarticletitle{Fake news detection on social media: A data mining
  perspective}.
\newblock \bibinfo{journal}{\emph{KDD}} \bibinfo{volume}{19},
  \bibinfo{number}{1} (\bibinfo{year}{2017}), \bibinfo{pages}{22--36}.
\newblock


\bibitem[\protect\citeauthoryear{Son, Kim, and Park}{Son et~al\mbox{.}}{2013}]%
        {son2013location}
\bibfield{author}{\bibinfo{person}{Jeong-Woo Son}, \bibinfo{person}{A-Yeong
  Kim}, {and} \bibinfo{person}{Seong-Bae Park}.}
  \bibinfo{year}{2013}\natexlab{}.
\newblock \showarticletitle{A location-based news article recommendation with
  explicit localized semantic analysis}. In \bibinfo{booktitle}{\emph{SIGIR}}.
  \bibinfo{pages}{293--302}.
\newblock


\bibitem[\protect\citeauthoryear{Song, Zhang, Shi, Li, Ma, and Wu}{Song
  et~al\mbox{.}}{2021}]%
        {song2021dql}
\bibfield{author}{\bibinfo{person}{Zhanghan Song}, \bibinfo{person}{Dian
  Zhang}, \bibinfo{person}{Xiaochuan Shi}, \bibinfo{person}{Wei Li},
  \bibinfo{person}{Chao Ma}, {and} \bibinfo{person}{Libing Wu}.}
  \bibinfo{year}{2021}\natexlab{}.
\newblock \showarticletitle{DEN-DQL: Quick Convergent Deep Q-Learning with
  Double Exploration Networks for News Recommendation}. In
  \bibinfo{booktitle}{\emph{IJCNN}}. IEEE, \bibinfo{pages}{1--8}.
\newblock


\bibitem[\protect\citeauthoryear{Sood and Kaur}{Sood and Kaur}{2014a}]%
        {sood2014preference}
\bibfield{author}{\bibinfo{person}{Mansi Sood} {and} \bibinfo{person}{Harmeet
  Kaur}.} \bibinfo{year}{2014}\natexlab{a}.
\newblock \showarticletitle{Preference based personalized news recommender
  system}.
\newblock \bibinfo{journal}{\emph{International Journal of Advanced Computer
  Research}} \bibinfo{volume}{4}, \bibinfo{number}{2} (\bibinfo{year}{2014}),
  \bibinfo{pages}{575}.
\newblock


\bibitem[\protect\citeauthoryear{Sood and Kaur}{Sood and Kaur}{2014b}]%
        {sood2014survey}
\bibfield{author}{\bibinfo{person}{Mansi Sood} {and} \bibinfo{person}{Harmeet
  Kaur}.} \bibinfo{year}{2014}\natexlab{b}.
\newblock \showarticletitle{Survey on news recommendation}.
\newblock \bibinfo{journal}{\emph{International Journal of Advanced Research in
  Electrical, Electronics and Instrumentation Engineering}}
  \bibinfo{volume}{3}, \bibinfo{number}{6} (\bibinfo{year}{2014}),
  \bibinfo{pages}{9972--9977}.
\newblock


\bibitem[\protect\citeauthoryear{Sottocornola, Symeonidis, and
  Zanker}{Sottocornola et~al\mbox{.}}{2018}]%
        {sottocornola2018session}
\bibfield{author}{\bibinfo{person}{Gabriele Sottocornola},
  \bibinfo{person}{Panagiotis Symeonidis}, {and} \bibinfo{person}{Markus
  Zanker}.} \bibinfo{year}{2018}\natexlab{}.
\newblock \showarticletitle{Session-based news recommendations}. In
  \bibinfo{booktitle}{\emph{Companion Proceedings of WWW}}.
  \bibinfo{pages}{1395--1399}.
\newblock


\bibitem[\protect\citeauthoryear{Sun, Liu, Wu, Pei, Lin, Ou, and Jiang}{Sun
  et~al\mbox{.}}{2019}]%
        {sun2019bert4rec}
\bibfield{author}{\bibinfo{person}{Fei Sun}, \bibinfo{person}{Jun Liu},
  \bibinfo{person}{Jian Wu}, \bibinfo{person}{Changhua Pei},
  \bibinfo{person}{Xiao Lin}, \bibinfo{person}{Wenwu Ou}, {and}
  \bibinfo{person}{Peng Jiang}.} \bibinfo{year}{2019}\natexlab{}.
\newblock \showarticletitle{BERT4Rec: Sequential recommendation with
  bidirectional encoder representations from transformer}. In
  \bibinfo{booktitle}{\emph{CIKM}}. \bibinfo{pages}{1441--1450}.
\newblock


\bibitem[\protect\citeauthoryear{Sun, Cheng, Zuberi, P{\'e}rez, and
  Volkovs}{Sun et~al\mbox{.}}{2021a}]%
        {sun2021hgcf}
\bibfield{author}{\bibinfo{person}{Jianing Sun}, \bibinfo{person}{Zhaoyue
  Cheng}, \bibinfo{person}{Saba Zuberi}, \bibinfo{person}{Felipe P{\'e}rez},
  {and} \bibinfo{person}{Maksims Volkovs}.} \bibinfo{year}{2021}\natexlab{a}.
\newblock \showarticletitle{HGCF: Hyperbolic Graph Convolution Networks for
  Collaborative Filtering}. In \bibinfo{booktitle}{\emph{WWW}}.
  \bibinfo{pages}{593--601}.
\newblock


\bibitem[\protect\citeauthoryear{Sun, Yi, Zeng, Li, He, Qiao, and Zhou}{Sun
  et~al\mbox{.}}{2021b}]%
        {sun2021hybrid}
\bibfield{author}{\bibinfo{person}{Yumin Sun}, \bibinfo{person}{Fangzhou Yi},
  \bibinfo{person}{Cheng Zeng}, \bibinfo{person}{Bing Li},
  \bibinfo{person}{Peng He}, \bibinfo{person}{Jinxia Qiao}, {and}
  \bibinfo{person}{Yinghui Zhou}.} \bibinfo{year}{2021}\natexlab{b}.
\newblock \showarticletitle{A Hybrid Approach to News Recommendation Based on
  Knowledge Graph and Long Short-Term User Preferences}. In
  \bibinfo{booktitle}{\emph{SCC}}. IEEE, \bibinfo{pages}{165--173}.
\newblock


\bibitem[\protect\citeauthoryear{Symeonidis, Kirjackaja, and Zanker}{Symeonidis
  et~al\mbox{.}}{2021}]%
        {symeonidis2021session}
\bibfield{author}{\bibinfo{person}{Panagiotis Symeonidis},
  \bibinfo{person}{Lidija Kirjackaja}, {and} \bibinfo{person}{Markus Zanker}.}
  \bibinfo{year}{2021}\natexlab{}.
\newblock \showarticletitle{Session-based news recommendations using SimRank on
  multi-modal graphs}.
\newblock \bibinfo{journal}{\emph{Expert Systems with Applications}}
  \bibinfo{volume}{180} (\bibinfo{year}{2021}), \bibinfo{pages}{115028}.
\newblock


\bibitem[\protect\citeauthoryear{Tavakolifard, Gulla, Almeroth, Ingvaldesn,
  Nygreen, and Berg}{Tavakolifard et~al\mbox{.}}{2013}]%
        {tavakolifard2013tailored}
\bibfield{author}{\bibinfo{person}{Mozhgan Tavakolifard},
  \bibinfo{person}{Jon~Atle Gulla}, \bibinfo{person}{Kevin~C Almeroth},
  \bibinfo{person}{Jon~Espen Ingvaldesn}, \bibinfo{person}{Gaute Nygreen},
  {and} \bibinfo{person}{Erik Berg}.} \bibinfo{year}{2013}\natexlab{}.
\newblock \showarticletitle{Tailored news in the palm of your hand: a
  multi-perspective transparent approach to news recommendation}. In
  \bibinfo{booktitle}{\emph{WWW}}. \bibinfo{pages}{305--308}.
\newblock


\bibitem[\protect\citeauthoryear{Tian, Yang, Ren, Wang, Wu, Wang, and Li}{Tian
  et~al\mbox{.}}{2021}]%
        {tian2021joint}
\bibfield{author}{\bibinfo{person}{Yu Tian}, \bibinfo{person}{Yuhao Yang},
  \bibinfo{person}{Xudong Ren}, \bibinfo{person}{Pengfei Wang},
  \bibinfo{person}{Fangzhao Wu}, \bibinfo{person}{Qian Wang}, {and}
  \bibinfo{person}{Chenliang Li}.} \bibinfo{year}{2021}\natexlab{}.
\newblock \showarticletitle{Joint Knowledge Pruning and Recurrent Graph
  Convolution for News Recommendation}. In \bibinfo{booktitle}{\emph{SIGIR}}.
  \bibinfo{pages}{51--60}.
\newblock


\bibitem[\protect\citeauthoryear{Tiwari, Kumar, Jethwani, Kumar, and
  Dadhich}{Tiwari et~al\mbox{.}}{2022}]%
        {tiwari2022pntrs}
\bibfield{author}{\bibinfo{person}{Sunita Tiwari}, \bibinfo{person}{Sushil
  Kumar}, \bibinfo{person}{Vikas Jethwani}, \bibinfo{person}{Deepak Kumar},
  {and} \bibinfo{person}{Vyoma Dadhich}.} \bibinfo{year}{2022}\natexlab{}.
\newblock \showarticletitle{PNTRS: personalized news and tweet recommendation
  system}.
\newblock \bibinfo{journal}{\emph{JCIT}} \bibinfo{volume}{24},
  \bibinfo{number}{3} (\bibinfo{year}{2022}), \bibinfo{pages}{1--19}.
\newblock


\bibitem[\protect\citeauthoryear{Tran, Hamad, Zaib, Aljubairy, Sheng, Zhang,
  Tran, and Khoa}{Tran et~al\mbox{.}}{2021}]%
        {tran2021deep}
\bibfield{author}{\bibinfo{person}{Dai~Hoang Tran}, \bibinfo{person}{Salma
  Hamad}, \bibinfo{person}{Munazza Zaib}, \bibinfo{person}{Abdulwahab
  Aljubairy}, \bibinfo{person}{Quan~Z Sheng}, \bibinfo{person}{Wei~Emma Zhang},
  \bibinfo{person}{Nguyen~H Tran}, {and} \bibinfo{person}{Nguyen Lu~Dang
  Khoa}.} \bibinfo{year}{2021}\natexlab{}.
\newblock \showarticletitle{Deep News Recommendation with Contextual User
  Profiling and Multifaceted Article Representation}. In
  \bibinfo{booktitle}{\emph{WISE}}. Springer, \bibinfo{pages}{237--251}.
\newblock


\bibitem[\protect\citeauthoryear{Tran, Tran, and Uong}{Tran
  et~al\mbox{.}}{2010}]%
        {tran2010user}
\bibfield{author}{\bibinfo{person}{Mai-Vu Tran}, \bibinfo{person}{Xuan-Tu
  Tran}, {and} \bibinfo{person}{Huy-Long Uong}.}
  \bibinfo{year}{2010}\natexlab{}.
\newblock \showarticletitle{User interest analysis with hidden topic in news
  recommendation system}. In \bibinfo{booktitle}{\emph{IALP}}. IEEE,
  \bibinfo{pages}{211--214}.
\newblock


\bibitem[\protect\citeauthoryear{Trevisiol, Aiello, Schifanella, and
  Jaimes}{Trevisiol et~al\mbox{.}}{2014}]%
        {trevisiol2014cold}
\bibfield{author}{\bibinfo{person}{Michele Trevisiol},
  \bibinfo{person}{Luca~Maria Aiello}, \bibinfo{person}{Rossano Schifanella},
  {and} \bibinfo{person}{Alejandro Jaimes}.} \bibinfo{year}{2014}\natexlab{}.
\newblock \showarticletitle{Cold-start news recommendation with
  domain-dependent browse graph}. In \bibinfo{booktitle}{\emph{Recsys}}.
  \bibinfo{pages}{81--88}.
\newblock


\bibitem[\protect\citeauthoryear{Viana and Soares}{Viana and Soares}{2017}]%
        {viana2017hybrid}
\bibfield{author}{\bibinfo{person}{Paula Viana} {and}
  \bibinfo{person}{M{\'a}rcio Soares}.} \bibinfo{year}{2017}\natexlab{}.
\newblock \showarticletitle{A hybrid approach for personalized news
  recommendation in a mobility scenario using long-short user interest}.
\newblock \bibinfo{journal}{\emph{International Journal on Artificial
  Intelligence Tools}} \bibinfo{volume}{26}, \bibinfo{number}{02}
  (\bibinfo{year}{2017}), \bibinfo{pages}{1760012}.
\newblock


\bibitem[\protect\citeauthoryear{Vinh, Tay, Zhang, Cong, and Li}{Vinh
  et~al\mbox{.}}{2018}]%
        {vinh2018hyperbolic}
\bibfield{author}{\bibinfo{person}{Tran Dang~Quang Vinh}, \bibinfo{person}{Yi
  Tay}, \bibinfo{person}{Shuai Zhang}, \bibinfo{person}{Gao Cong}, {and}
  \bibinfo{person}{Xiao-Li Li}.} \bibinfo{year}{2018}\natexlab{}.
\newblock \showarticletitle{Hyperbolic recommender systems}.
\newblock \bibinfo{journal}{\emph{arXiv preprint arXiv:1809.01703}}
  (\bibinfo{year}{2018}).
\newblock


\bibitem[\protect\citeauthoryear{Wang, Kim, Bang, Singh, Kociuba, Pomerville,
  and Liu}{Wang et~al\mbox{.}}{2020a}]%
        {wang2020discovery}
\bibfield{author}{\bibinfo{person}{Chong Wang}, \bibinfo{person}{Lisa Kim},
  \bibinfo{person}{Grace Bang}, \bibinfo{person}{Himani Singh},
  \bibinfo{person}{Russell Kociuba}, \bibinfo{person}{Steven Pomerville}, {and}
  \bibinfo{person}{Xiaomo Liu}.} \bibinfo{year}{2020}\natexlab{a}.
\newblock \showarticletitle{Discovery News: A Generic Framework for Financial
  News Recommendation}. In \bibinfo{booktitle}{\emph{AAAI}},
  Vol.~\bibinfo{volume}{34}. \bibinfo{pages}{13390--13395}.
\newblock


\bibitem[\protect\citeauthoryear{Wang, Wu, Liu, and Xie}{Wang
  et~al\mbox{.}}{2020c}]%
        {wang2020fine}
\bibfield{author}{\bibinfo{person}{Heyuan Wang}, \bibinfo{person}{Fangzhao Wu},
  \bibinfo{person}{Zheng Liu}, {and} \bibinfo{person}{Xing Xie}.}
  \bibinfo{year}{2020}\natexlab{c}.
\newblock \showarticletitle{Fine-grained Interest Matching for Neural News
  Recommendation}. In \bibinfo{booktitle}{\emph{ACL}}.
  \bibinfo{pages}{836--845}.
\newblock


\bibitem[\protect\citeauthoryear{Wang, Zhang, Xie, and Guo}{Wang
  et~al\mbox{.}}{2018}]%
        {wang2018dkn}
\bibfield{author}{\bibinfo{person}{Hongwei Wang}, \bibinfo{person}{Fuzheng
  Zhang}, \bibinfo{person}{Xing Xie}, {and} \bibinfo{person}{Minyi Guo}.}
  \bibinfo{year}{2018}\natexlab{}.
\newblock \showarticletitle{DKN: Deep knowledge-aware network for news
  recommendation}. In \bibinfo{booktitle}{\emph{WWW}}.
  \bibinfo{pages}{1835--1844}.
\newblock


\bibitem[\protect\citeauthoryear{Wang, Wei, dos Santos, Wang, Nallapati,
  Arnold, Xiang, and Philip}{Wang et~al\mbox{.}}{2020b}]%
        {wang2020h2kgat}
\bibfield{author}{\bibinfo{person}{Shen Wang}, \bibinfo{person}{Xiaokai Wei},
  \bibinfo{person}{Cicero dos Santos}, \bibinfo{person}{Zhiguo Wang},
  \bibinfo{person}{Ramesh Nallapati}, \bibinfo{person}{Andrew Arnold},
  \bibinfo{person}{Bing Xiang}, {and} \bibinfo{person}{S~Yu Philip}.}
  \bibinfo{year}{2020}\natexlab{b}.
\newblock \showarticletitle{H2KGAT: Hierarchical Hyperbolic Knowledge Graph
  Attention Network}. In \bibinfo{booktitle}{\emph{EMNLP}}.
  \bibinfo{pages}{4952--4962}.
\newblock


\bibitem[\protect\citeauthoryear{Wang, Yu, Ren, Tao, Zhang, Yu, and Wang}{Wang
  et~al\mbox{.}}{2017}]%
        {wang2017dynamic}
\bibfield{author}{\bibinfo{person}{Xuejian Wang}, \bibinfo{person}{Lantao Yu},
  \bibinfo{person}{Kan Ren}, \bibinfo{person}{Guanyu Tao},
  \bibinfo{person}{Weinan Zhang}, \bibinfo{person}{Yong Yu}, {and}
  \bibinfo{person}{Jun Wang}.} \bibinfo{year}{2017}\natexlab{}.
\newblock \showarticletitle{Dynamic attention deep model for article
  recommendation by learning human editors' demonstration}. In
  \bibinfo{booktitle}{\emph{KDD}}. \bibinfo{pages}{2051--2059}.
\newblock


\bibitem[\protect\citeauthoryear{Wei, Zhou, Cimini, Wu, Liu, and Zhang}{Wei
  et~al\mbox{.}}{2011}]%
        {wei2011effective}
\bibfield{author}{\bibinfo{person}{Dong Wei}, \bibinfo{person}{Tao Zhou},
  \bibinfo{person}{Giulio Cimini}, \bibinfo{person}{Pei Wu},
  \bibinfo{person}{Weiping Liu}, {and} \bibinfo{person}{Yi-Cheng Zhang}.}
  \bibinfo{year}{2011}\natexlab{}.
\newblock \showarticletitle{Effective mechanism for social recommendation of
  news}.
\newblock \bibinfo{journal}{\emph{Physica A: Statistical Mechanics and its
  Applications}} \bibinfo{volume}{390}, \bibinfo{number}{11}
  (\bibinfo{year}{2011}), \bibinfo{pages}{2117--2126}.
\newblock


\bibitem[\protect\citeauthoryear{Wei, Wei, Lei, et~al\mbox{.}}{Wei
  et~al\mbox{.}}{2021}]%
        {wei2021news}
\bibfield{author}{\bibinfo{person}{Guiying Wei}, \bibinfo{person}{Yimeng Wei},
  \bibinfo{person}{Jincheng Lei}, {et~al\mbox{.}}}
  \bibinfo{year}{2021}\natexlab{}.
\newblock \showarticletitle{News Recommendation Based on Click-Through Rate
  Prediction Model}. In \bibinfo{booktitle}{\emph{LISS}}. Springer Nature,
  \bibinfo{pages}{373}.
\newblock


\bibitem[\protect\citeauthoryear{Wen, Fang, and Guan}{Wen
  et~al\mbox{.}}{2012}]%
        {wen2012hybrid}
\bibfield{author}{\bibinfo{person}{Hao Wen}, \bibinfo{person}{Liping Fang},
  {and} \bibinfo{person}{Ling Guan}.} \bibinfo{year}{2012}\natexlab{}.
\newblock \showarticletitle{A hybrid approach for personalized recommendation
  of news on the Web}.
\newblock \bibinfo{journal}{\emph{Expert Systems With Applications}}
  \bibinfo{volume}{39}, \bibinfo{number}{5} (\bibinfo{year}{2012}),
  \bibinfo{pages}{5806--5814}.
\newblock


\bibitem[\protect\citeauthoryear{Wu, Wu, An, Huang, Huang, and Xie}{Wu
  et~al\mbox{.}}{2019b}]%
        {wu2019neuralnaml}
\bibfield{author}{\bibinfo{person}{Chuhan Wu}, \bibinfo{person}{Fangzhao Wu},
  \bibinfo{person}{Mingxiao An}, \bibinfo{person}{Jianqiang Huang},
  \bibinfo{person}{Yongfeng Huang}, {and} \bibinfo{person}{Xing Xie}.}
  \bibinfo{year}{2019}\natexlab{b}.
\newblock \showarticletitle{Neural news recommendation with attentive
  multi-view learning}. In \bibinfo{booktitle}{\emph{IJCAI}}. AAAI Press,
  \bibinfo{pages}{3863--3869}.
\newblock


\bibitem[\protect\citeauthoryear{Wu, Wu, An, Huang, Huang, and Xie}{Wu
  et~al\mbox{.}}{2019c}]%
        {wu2019npa}
\bibfield{author}{\bibinfo{person}{Chuhan Wu}, \bibinfo{person}{Fangzhao Wu},
  \bibinfo{person}{Mingxiao An}, \bibinfo{person}{Jianqiang Huang},
  \bibinfo{person}{Yongfeng Huang}, {and} \bibinfo{person}{Xing Xie}.}
  \bibinfo{year}{2019}\natexlab{c}.
\newblock \showarticletitle{Npa: Neural news recommendation with personalized
  attention}. In \bibinfo{booktitle}{\emph{KDD}}. \bibinfo{pages}{2576--2584}.
\newblock


\bibitem[\protect\citeauthoryear{Wu, Wu, An, Huang, and Xie}{Wu
  et~al\mbox{.}}{2019a}]%
        {wu2019neural}
\bibfield{author}{\bibinfo{person}{Chuhan Wu}, \bibinfo{person}{Fangzhao Wu},
  \bibinfo{person}{Mingxiao An}, \bibinfo{person}{Yongfeng Huang}, {and}
  \bibinfo{person}{Xing Xie}.} \bibinfo{year}{2019}\natexlab{a}.
\newblock \showarticletitle{Neural News Recommendation with Topic-Aware News
  Representation}. In \bibinfo{booktitle}{\emph{ACL}}.
  \bibinfo{pages}{1154--1159}.
\newblock


\bibitem[\protect\citeauthoryear{Wu, Wu, An, Qi, Huang, Huang, and Xie}{Wu
  et~al\mbox{.}}{2019d}]%
        {wu2019neuralnrhub}
\bibfield{author}{\bibinfo{person}{Chuhan Wu}, \bibinfo{person}{Fangzhao Wu},
  \bibinfo{person}{Mingxiao An}, \bibinfo{person}{Tao Qi},
  \bibinfo{person}{Jianqiang Huang}, \bibinfo{person}{Yongfeng Huang}, {and}
  \bibinfo{person}{Xing Xie}.} \bibinfo{year}{2019}\natexlab{d}.
\newblock \showarticletitle{Neural News Recommendation with Heterogeneous User
  Behavior}. In \bibinfo{booktitle}{\emph{EMNLP-IJCNLP}}.
  \bibinfo{pages}{4876--4885}.
\newblock


\bibitem[\protect\citeauthoryear{Wu, Wu, Ge, Qi, Huang, and Xie}{Wu
  et~al\mbox{.}}{2019e}]%
        {wu2019neuralnrms}
\bibfield{author}{\bibinfo{person}{Chuhan Wu}, \bibinfo{person}{Fangzhao Wu},
  \bibinfo{person}{Suyu Ge}, \bibinfo{person}{Tao Qi},
  \bibinfo{person}{Yongfeng Huang}, {and} \bibinfo{person}{Xing Xie}.}
  \bibinfo{year}{2019}\natexlab{e}.
\newblock \showarticletitle{Neural News Recommendation with Multi-Head
  Self-Attention}. In \bibinfo{booktitle}{\emph{EMNLP-IJCNLP}}.
  \bibinfo{pages}{6390--6395}.
\newblock


\bibitem[\protect\citeauthoryear{Wu, Wu, and Huang}{Wu et~al\mbox{.}}{2021a}]%
        {wu2021rethinking}
\bibfield{author}{\bibinfo{person}{Chuhan Wu}, \bibinfo{person}{Fangzhao Wu},
  {and} \bibinfo{person}{Yongfeng Huang}.} \bibinfo{year}{2021}\natexlab{a}.
\newblock \showarticletitle{Rethinking InfoNCE: How Many Negative Samples Do
  You Need?}
\newblock \bibinfo{journal}{\emph{arXiv preprint arXiv:2105.13003}}
  (\bibinfo{year}{2021}).
\newblock


\bibitem[\protect\citeauthoryear{Wu, Wu, Huang, and Xie}{Wu
  et~al\mbox{.}}{2020b}]%
        {wu2020ccf}
\bibfield{author}{\bibinfo{person}{Chuhan Wu}, \bibinfo{person}{Fangzhao Wu},
  \bibinfo{person}{Yongfeng Huang}, {and} \bibinfo{person}{Xing Xie}.}
  \bibinfo{year}{2020}\natexlab{b}.
\newblock \showarticletitle{Neural news recommendation with negative feedback}.
\newblock \bibinfo{journal}{\emph{CCF TPCI}} (\bibinfo{year}{2020}),
  \bibinfo{pages}{178--188}.
\newblock


\bibitem[\protect\citeauthoryear{Wu, Wu, Huang, and Xie}{Wu
  et~al\mbox{.}}{2021b}]%
        {wu2021uag}
\bibfield{author}{\bibinfo{person}{Chuhan Wu}, \bibinfo{person}{Fangzhao Wu},
  \bibinfo{person}{Yongfeng Huang}, {and} \bibinfo{person}{Xing Xie}.}
  \bibinfo{year}{2021}\natexlab{b}.
\newblock \showarticletitle{User-as-Graph: User Modeling with Heterogeneous
  Graph Pooling for News Recommendation}. In \bibinfo{booktitle}{\emph{IJCAI}}.
\newblock


\bibitem[\protect\citeauthoryear{Wu, Wu, Qi, and Huang}{Wu
  et~al\mbox{.}}{2020c}]%
        {wu2020clickbait}
\bibfield{author}{\bibinfo{person}{Chuhan Wu}, \bibinfo{person}{Fangzhao Wu},
  \bibinfo{person}{Tao Qi}, {and} \bibinfo{person}{Yongfeng Huang}.}
  \bibinfo{year}{2020}\natexlab{c}.
\newblock \showarticletitle{Clickbait Detection with Style-Aware Title Modeling
  and Co-attention}. In \bibinfo{booktitle}{\emph{CCL}}. Springer,
  \bibinfo{pages}{430--443}.
\newblock


\bibitem[\protect\citeauthoryear{Wu, Wu, Qi, and Huang}{Wu
  et~al\mbox{.}}{2020d}]%
        {wu2020sentirec}
\bibfield{author}{\bibinfo{person}{Chuhan Wu}, \bibinfo{person}{Fangzhao Wu},
  \bibinfo{person}{Tao Qi}, {and} \bibinfo{person}{Yongfeng Huang}.}
  \bibinfo{year}{2020}\natexlab{d}.
\newblock \showarticletitle{SentiRec: Sentiment Diversity-aware Neural News
  Recommendation}. In \bibinfo{booktitle}{\emph{AACL}}.
  \bibinfo{pages}{44--53}.
\newblock


\bibitem[\protect\citeauthoryear{Wu, Wu, Qi, and Huang}{Wu
  et~al\mbox{.}}{2020e}]%
        {wu2020cprs}
\bibfield{author}{\bibinfo{person}{Chuhan Wu}, \bibinfo{person}{Fangzhao Wu},
  \bibinfo{person}{Tao Qi}, {and} \bibinfo{person}{Yongfeng Huang}.}
  \bibinfo{year}{2020}\natexlab{e}.
\newblock \showarticletitle{User modeling with click preference and reading
  satisfaction for news recommendation}. In \bibinfo{booktitle}{\emph{IJCAI}}.
  \bibinfo{pages}{3023--3029}.
\newblock


\bibitem[\protect\citeauthoryear{Wu, Wu, Qi, and Huang}{Wu
  et~al\mbox{.}}{2021c}]%
        {wu2021empowering}
\bibfield{author}{\bibinfo{person}{Chuhan Wu}, \bibinfo{person}{Fangzhao Wu},
  \bibinfo{person}{Tao Qi}, {and} \bibinfo{person}{Yongfeng Huang}.}
  \bibinfo{year}{2021}\natexlab{c}.
\newblock \showarticletitle{Empowering News Recommendation with Pre-trained
  Language Models}. In \bibinfo{booktitle}{\emph{SIGIR}}.
  \bibinfo{pages}{1652--1656}.
\newblock


\bibitem[\protect\citeauthoryear{Wu, Wu, Qi, and Huang}{Wu
  et~al\mbox{.}}{2021d}]%
        {wu2021feedrec}
\bibfield{author}{\bibinfo{person}{Chuhan Wu}, \bibinfo{person}{Fangzhao Wu},
  \bibinfo{person}{Tao Qi}, {and} \bibinfo{person}{Yongfeng Huang}.}
  \bibinfo{year}{2021}\natexlab{d}.
\newblock \showarticletitle{FeedRec: News Feed Recommendation with Various User
  Feedbacks}.
\newblock \bibinfo{journal}{\emph{arXiv preprint arXiv:2102.04903}}
  (\bibinfo{year}{2021}).
\newblock


\bibitem[\protect\citeauthoryear{Wu, Wu, Qi, and Huang}{Wu
  et~al\mbox{.}}{2021e}]%
        {wu2021can}
\bibfield{author}{\bibinfo{person}{Chuhan Wu}, \bibinfo{person}{Fangzhao Wu},
  \bibinfo{person}{Tao Qi}, {and} \bibinfo{person}{Yongfeng Huang}.}
  \bibinfo{year}{2021}\natexlab{e}.
\newblock \showarticletitle{Is News Recommendation a Sequential Recommendation
  Task?}
\newblock \bibinfo{journal}{\emph{arXiv e-prints}} (\bibinfo{year}{2021}),
  \bibinfo{pages}{arXiv--2108}.
\newblock


\bibitem[\protect\citeauthoryear{Wu, Wu, Qi, and Huang}{Wu
  et~al\mbox{.}}{2021f}]%
        {wu2021mm}
\bibfield{author}{\bibinfo{person}{Chuhan Wu}, \bibinfo{person}{Fangzhao Wu},
  \bibinfo{person}{Tao Qi}, {and} \bibinfo{person}{Yongfeng Huang}.}
  \bibinfo{year}{2021}\natexlab{f}.
\newblock \showarticletitle{MM-Rec: Multimodal News Recommendation}.
\newblock \bibinfo{journal}{\emph{arXiv preprint arXiv:2104.07407}}
  (\bibinfo{year}{2021}).
\newblock


\bibitem[\protect\citeauthoryear{Wu, Wu, Qi, and Huang}{Wu
  et~al\mbox{.}}{2021g}]%
        {wu2021two}
\bibfield{author}{\bibinfo{person}{Chuhan Wu}, \bibinfo{person}{Fangzhao Wu},
  \bibinfo{person}{Tao Qi}, {and} \bibinfo{person}{Yongfeng Huang}.}
  \bibinfo{year}{2021}\natexlab{g}.
\newblock \showarticletitle{Two Birds with One Stone: Unified Model Learning
  for Both Recall and Ranking in News Recommendation}.
\newblock \bibinfo{journal}{\emph{arXiv preprint arXiv:2104.07404}}
  (\bibinfo{year}{2021}).
\newblock


\bibitem[\protect\citeauthoryear{Wu, Wu, Qi, Huang, and Xie}{Wu
  et~al\mbox{.}}{2021h}]%
        {wu2021fastformer}
\bibfield{author}{\bibinfo{person}{Chuhan Wu}, \bibinfo{person}{Fangzhao Wu},
  \bibinfo{person}{Tao Qi}, \bibinfo{person}{Yongfeng Huang}, {and}
  \bibinfo{person}{Xing Xie}.} \bibinfo{year}{2021}\natexlab{h}.
\newblock \showarticletitle{Fastformer: Additive attention can be all you
  need}.
\newblock \bibinfo{journal}{\emph{arXiv preprint arXiv:2108.09084}}
  (\bibinfo{year}{2021}).
\newblock


\bibitem[\protect\citeauthoryear{Wu, Wu, Qi, Lian, Huang, and Xie}{Wu
  et~al\mbox{.}}{2020f}]%
        {wu2020ptum}
\bibfield{author}{\bibinfo{person}{Chuhan Wu}, \bibinfo{person}{Fangzhao Wu},
  \bibinfo{person}{Tao Qi}, \bibinfo{person}{Jianxun Lian},
  \bibinfo{person}{Yongfeng Huang}, {and} \bibinfo{person}{Xing Xie}.}
  \bibinfo{year}{2020}\natexlab{f}.
\newblock \showarticletitle{PTUM: Pre-training User Model from Unlabeled User
  Behaviors via Self-supervision}. In \bibinfo{booktitle}{\emph{EMNLP:
  Findings}}. \bibinfo{pages}{1939--1944}.
\newblock


\bibitem[\protect\citeauthoryear{Wu, Wu, Wang, Huang, and Xie}{Wu
  et~al\mbox{.}}{2021i}]%
        {wu2021fairness}
\bibfield{author}{\bibinfo{person}{Chuhan Wu}, \bibinfo{person}{Fangzhao Wu},
  \bibinfo{person}{Xiting Wang}, \bibinfo{person}{Yongfeng Huang}, {and}
  \bibinfo{person}{Xing Xie}.} \bibinfo{year}{2021}\natexlab{i}.
\newblock \showarticletitle{FairRec:Fairness-aware News Recommendation with
  Decomposed Adversarial Learning}. In \bibinfo{booktitle}{\emph{AAAI}}.
  \bibinfo{pages}{4462--4469}.
\newblock


\bibitem[\protect\citeauthoryear{Wu, Wu, Yu, Qi, Huang, and Liu}{Wu
  et~al\mbox{.}}{2021j}]%
        {wu2021newsbert}
\bibfield{author}{\bibinfo{person}{Chuhan Wu}, \bibinfo{person}{Fangzhao Wu},
  \bibinfo{person}{Yang Yu}, \bibinfo{person}{Tao Qi},
  \bibinfo{person}{Yongfeng Huang}, {and} \bibinfo{person}{Qi Liu}.}
  \bibinfo{year}{2021}\natexlab{j}.
\newblock \showarticletitle{NewsBERT: Distilling Pre-trained Language Model for
  Intelligent News Application}. In \bibinfo{booktitle}{\emph{EMNLP:
  Findings}}. \bibinfo{pages}{3285--3295}.
\newblock


\bibitem[\protect\citeauthoryear{Wu, Qiao, Chen, Wu, Qi, Lian, Liu, Xie, Gao,
  Wu, and Zhou}{Wu et~al\mbox{.}}{2020a}]%
        {wu2020mind}
\bibfield{author}{\bibinfo{person}{Fangzhao Wu}, \bibinfo{person}{Ying Qiao},
  \bibinfo{person}{Jiun-Hung Chen}, \bibinfo{person}{Chuhan Wu},
  \bibinfo{person}{Tao Qi}, \bibinfo{person}{Jianxun Lian},
  \bibinfo{person}{Danyang Liu}, \bibinfo{person}{Xing Xie},
  \bibinfo{person}{Jianfeng Gao}, \bibinfo{person}{Winnie Wu}, {and}
  \bibinfo{person}{Ming Zhou}.} \bibinfo{year}{2020}\natexlab{a}.
\newblock \showarticletitle{MIND: A Large-scale Dataset for News
  Recommendation}. In \bibinfo{booktitle}{\emph{ACL}}.
\newblock


\bibitem[\protect\citeauthoryear{Xiao, Liu, Shao, Di, and Xie}{Xiao
  et~al\mbox{.}}{2021}]%
        {xiao2021training}
\bibfield{author}{\bibinfo{person}{Shitao Xiao}, \bibinfo{person}{Zheng Liu},
  \bibinfo{person}{Yingxia Shao}, \bibinfo{person}{Tao Di}, {and}
  \bibinfo{person}{Xing Xie}.} \bibinfo{year}{2021}\natexlab{}.
\newblock \showarticletitle{Training Microsoft News Recommenders with
  Pretrained Language Models in the Loop}.
\newblock \bibinfo{journal}{\emph{arXiv e-prints}} (\bibinfo{year}{2021}),
  \bibinfo{pages}{arXiv--2102}.
\newblock


\bibitem[\protect\citeauthoryear{Xiao, Ai, Hsu, Wang, and Jiao}{Xiao
  et~al\mbox{.}}{2015}]%
        {xiao2015time}
\bibfield{author}{\bibinfo{person}{Yingyuan Xiao}, \bibinfo{person}{Pengqiang
  Ai}, \bibinfo{person}{Ching-hsien Hsu}, \bibinfo{person}{Hongya Wang}, {and}
  \bibinfo{person}{Xu Jiao}.} \bibinfo{year}{2015}\natexlab{}.
\newblock \showarticletitle{Time-ordered collaborative filtering for news
  recommendation}.
\newblock \bibinfo{journal}{\emph{China Communcations}} \bibinfo{volume}{12},
  \bibinfo{number}{12} (\bibinfo{year}{2015}), \bibinfo{pages}{53--62}.
\newblock


\bibitem[\protect\citeauthoryear{Xun, Zhang, Zhao, Zhu, Zhang, Li, He, He,
  Chua, and Wu}{Xun et~al\mbox{.}}{2021}]%
        {xun2021we}
\bibfield{author}{\bibinfo{person}{Jiahao Xun}, \bibinfo{person}{Shengyu
  Zhang}, \bibinfo{person}{Zhou Zhao}, \bibinfo{person}{Jieming Zhu},
  \bibinfo{person}{Qi Zhang}, \bibinfo{person}{Jingjie Li},
  \bibinfo{person}{Xiuqiang He}, \bibinfo{person}{Xiaofei He},
  \bibinfo{person}{Tat-Seng Chua}, {and} \bibinfo{person}{Fei Wu}.}
  \bibinfo{year}{2021}\natexlab{}.
\newblock \showarticletitle{Why Do We Click: Visual Impression-aware News
  Recommendation}. In \bibinfo{booktitle}{\emph{MM}}.
  \bibinfo{pages}{3881--3890}.
\newblock


\bibitem[\protect\citeauthoryear{Yang}{Yang}{2016}]%
        {yang2016effects}
\bibfield{author}{\bibinfo{person}{JungAe Yang}.}
  \bibinfo{year}{2016}\natexlab{}.
\newblock \showarticletitle{Effects of popularity-based news recommendations
  (“most-viewed”) on users' exposure to online news}.
\newblock \bibinfo{journal}{\emph{Media Psychology}} \bibinfo{volume}{19},
  \bibinfo{number}{2} (\bibinfo{year}{2016}), \bibinfo{pages}{243--271}.
\newblock


\bibitem[\protect\citeauthoryear{Yang, Wu, Wu, and Wang}{Yang
  et~al\mbox{.}}{2020}]%
        {yang2020double}
\bibfield{author}{\bibinfo{person}{Zhihong Yang}, \bibinfo{person}{Yuewei Wu},
  \bibinfo{person}{Muqing Wu}, {and} \bibinfo{person}{Yulong Wang}.}
  \bibinfo{year}{2020}\natexlab{}.
\newblock \showarticletitle{Double Cross \& Deep Network for News
  Recommendation}. In \bibinfo{booktitle}{\emph{ICEEIM}}. Atlantis Press,
  \bibinfo{pages}{101--106}.
\newblock


\bibitem[\protect\citeauthoryear{Yeung and Yang}{Yeung and Yang}{2010}]%
        {yeung2010proactive}
\bibfield{author}{\bibinfo{person}{Kam~Fung Yeung} {and}
  \bibinfo{person}{Yanyan Yang}.} \bibinfo{year}{2010}\natexlab{}.
\newblock \showarticletitle{A proactive personalized mobile news recommendation
  system}. In \bibinfo{booktitle}{\emph{Developments in E-systems
  Engineering}}. IEEE, \bibinfo{pages}{207--212}.
\newblock


\bibitem[\protect\citeauthoryear{Yi, Wu, Wu, Li, Sun, and Xie}{Yi
  et~al\mbox{.}}{2021a}]%
        {yi2021debiasedrec}
\bibfield{author}{\bibinfo{person}{Jingwei Yi}, \bibinfo{person}{Fangzhao Wu},
  \bibinfo{person}{Chuhan Wu}, \bibinfo{person}{Qifei Li},
  \bibinfo{person}{Guangzhong Sun}, {and} \bibinfo{person}{Xing Xie}.}
  \bibinfo{year}{2021}\natexlab{a}.
\newblock \showarticletitle{DebiasedRec: Bias-aware User Modeling and Click
  Prediction for Personalized News Recommendation}.
\newblock \bibinfo{journal}{\emph{arXiv preprint arXiv:2104.07360}}
  (\bibinfo{year}{2021}).
\newblock


\bibitem[\protect\citeauthoryear{Yi, Wu, Wu, Liu, Sun, and Xie}{Yi
  et~al\mbox{.}}{2021b}]%
        {yi2021efficient}
\bibfield{author}{\bibinfo{person}{Jingwei Yi}, \bibinfo{person}{Fangzhao Wu},
  \bibinfo{person}{Chuhan Wu}, \bibinfo{person}{Ruixuan Liu},
  \bibinfo{person}{Guangzhong Sun}, {and} \bibinfo{person}{Xing Xie}.}
  \bibinfo{year}{2021}\natexlab{b}.
\newblock \showarticletitle{Efficient-FedRec: Efficient Federated Learning
  Framework for Privacy-Preserving News Recommendation}. In
  \bibinfo{booktitle}{\emph{EMNLP}}. \bibinfo{pages}{2814--2824}.
\newblock


\bibitem[\protect\citeauthoryear{Yi, Hong, Zhong, Liu, and Rajan}{Yi
  et~al\mbox{.}}{2014}]%
        {yi2014beyond}
\bibfield{author}{\bibinfo{person}{Xing Yi}, \bibinfo{person}{Liangjie Hong},
  \bibinfo{person}{Erheng Zhong}, \bibinfo{person}{Nanthan~Nan Liu}, {and}
  \bibinfo{person}{Suju Rajan}.} \bibinfo{year}{2014}\natexlab{}.
\newblock \showarticletitle{Beyond clicks: dwell time for personalization}. In
  \bibinfo{booktitle}{\emph{Recsys}}. \bibinfo{pages}{113--120}.
\newblock


\bibitem[\protect\citeauthoryear{Yue and Joachims}{Yue and Joachims}{2009}]%
        {yue2009interactively}
\bibfield{author}{\bibinfo{person}{Yisong Yue} {and} \bibinfo{person}{Thorsten
  Joachims}.} \bibinfo{year}{2009}\natexlab{}.
\newblock \showarticletitle{Interactively optimizing information retrieval
  systems as a dueling bandits problem}. In \bibinfo{booktitle}{\emph{ICML}}.
  \bibinfo{pages}{1201--1208}.
\newblock


\bibitem[\protect\citeauthoryear{Zelen{\'\i}k and Bielikov{\'a}}{Zelen{\'\i}k
  and Bielikov{\'a}}{2011}]%
        {zelenik2011news}
\bibfield{author}{\bibinfo{person}{Dusan Zelen{\'\i}k} {and}
  \bibinfo{person}{M{\'a}ria Bielikov{\'a}}.} \bibinfo{year}{2011}\natexlab{}.
\newblock \showarticletitle{News Recommending based on Text Similarity and user
  Behaviour}. In \bibinfo{booktitle}{\emph{WEBIST}}. \bibinfo{pages}{302--307}.
\newblock


\bibitem[\protect\citeauthoryear{Zhang, Chen, and Ma}{Zhang
  et~al\mbox{.}}{2019a}]%
        {zhang2019dynamic}
\bibfield{author}{\bibinfo{person}{Hui Zhang}, \bibinfo{person}{Xu Chen}, {and}
  \bibinfo{person}{Shuai Ma}.} \bibinfo{year}{2019}\natexlab{a}.
\newblock \showarticletitle{Dynamic News Recommendation with Hierarchical
  Attention Network}. In \bibinfo{booktitle}{\emph{ICDM}}. IEEE,
  \bibinfo{pages}{1456--1461}.
\newblock


\bibitem[\protect\citeauthoryear{Zhang, Xin, Luo, and Guot}{Zhang
  et~al\mbox{.}}{2017}]%
        {zhang2017fine}
\bibfield{author}{\bibinfo{person}{Kuai Zhang}, \bibinfo{person}{Xin Xin},
  \bibinfo{person}{Pei Luo}, {and} \bibinfo{person}{Ping Guot}.}
  \bibinfo{year}{2017}\natexlab{}.
\newblock \showarticletitle{Fine-grained news recommendation by fusing matrix
  factorization, topic analysis and knowledge graph representation}. In
  \bibinfo{booktitle}{\emph{SMC}}. IEEE, \bibinfo{pages}{918--923}.
\newblock


\bibitem[\protect\citeauthoryear{Zhang, Liu, and Gulla}{Zhang
  et~al\mbox{.}}{2018}]%
        {zhang2018deep}
\bibfield{author}{\bibinfo{person}{Lemei Zhang}, \bibinfo{person}{Peng Liu},
  {and} \bibinfo{person}{Jon~Atle Gulla}.} \bibinfo{year}{2018}\natexlab{}.
\newblock \showarticletitle{A deep joint network for session-based news
  recommendations with contextual augmentation}.
\newblock In \bibinfo{booktitle}{\emph{HT}}. \bibinfo{pages}{201--209}.
\newblock


\bibitem[\protect\citeauthoryear{Zhang, Liu, and Gulla}{Zhang
  et~al\mbox{.}}{2019b}]%
        {zhang2019dynamicattention}
\bibfield{author}{\bibinfo{person}{Lemei Zhang}, \bibinfo{person}{Peng Liu},
  {and} \bibinfo{person}{Jon~Atle Gulla}.} \bibinfo{year}{2019}\natexlab{b}.
\newblock \showarticletitle{Dynamic attention-integrated neural network for
  session-based news recommendation}.
\newblock \bibinfo{journal}{\emph{Machine Learning}} \bibinfo{volume}{108},
  \bibinfo{number}{10} (\bibinfo{year}{2019}), \bibinfo{pages}{1851--1875}.
\newblock


\bibitem[\protect\citeauthoryear{Zhang, Dou, and Yao}{Zhang
  et~al\mbox{.}}{2021a}]%
        {zhang2021learning}
\bibfield{author}{\bibinfo{person}{Peitian Zhang}, \bibinfo{person}{Zhicheng
  Dou}, {and} \bibinfo{person}{Jing Yao}.} \bibinfo{year}{2021}\natexlab{a}.
\newblock \showarticletitle{Learning to Select Historical News Articles for
  Interaction based Neural News Recommendation}.
\newblock \bibinfo{journal}{\emph{arXiv preprint arXiv:2110.06459}}
  (\bibinfo{year}{2021}).
\newblock


\bibitem[\protect\citeauthoryear{Zhang, Jia, Wang, Li, Wang, and He}{Zhang
  et~al\mbox{.}}{2021b}]%
        {zhang2021amm}
\bibfield{author}{\bibinfo{person}{Qi Zhang}, \bibinfo{person}{Qinglin Jia},
  \bibinfo{person}{Chuyuan Wang}, \bibinfo{person}{Jingjie Li},
  \bibinfo{person}{Zhaowei Wang}, {and} \bibinfo{person}{Xiuqiang He}.}
  \bibinfo{year}{2021}\natexlab{b}.
\newblock \showarticletitle{AMM: Attentive Multi-field Matching for News
  Recommendation}. In \bibinfo{booktitle}{\emph{SIGIR}}.
  \bibinfo{pages}{1588--1592}.
\newblock


\bibitem[\protect\citeauthoryear{Zhang, Li, Jia, Wang, Zhu, Wang, and He}{Zhang
  et~al\mbox{.}}{2021c}]%
        {zhang2021unbert}
\bibfield{author}{\bibinfo{person}{Qi Zhang}, \bibinfo{person}{Jingjie Li},
  \bibinfo{person}{Qinglin Jia}, \bibinfo{person}{Chuyuan Wang},
  \bibinfo{person}{Jieming Zhu}, \bibinfo{person}{Zhaowei Wang}, {and}
  \bibinfo{person}{Xiuqiang He}.} \bibinfo{year}{2021}\natexlab{c}.
\newblock \showarticletitle{{UNBERT:} User-News Matching {BERT} for News
  Recommendation}. In \bibinfo{booktitle}{\emph{IJCAI}},
  \bibfield{editor}{\bibinfo{person}{Zhi{-}Hua Zhou}} (Ed.).
  \bibinfo{pages}{3356--3362}.
\newblock


\bibitem[\protect\citeauthoryear{Zhang and Ling}{Zhang and Ling}{2021}]%
        {zhang2021research}
\bibfield{author}{\bibinfo{person}{Weijia Zhang} {and} \bibinfo{person}{Feng
  Ling}.} \bibinfo{year}{2021}\natexlab{}.
\newblock \showarticletitle{Research on News Recommendation System Based on
  Deep Network and Personalized Needs}.
\newblock \bibinfo{journal}{\emph{Wireless Communications and Mobile
  Computing}}  \bibinfo{volume}{2021} (\bibinfo{year}{2021}).
\newblock


\bibitem[\protect\citeauthoryear{Zhang, Yang, and Xu}{Zhang
  et~al\mbox{.}}{2021d}]%
        {zhang2021combining}
\bibfield{author}{\bibinfo{person}{Xuanyu Zhang}, \bibinfo{person}{Qing Yang},
  {and} \bibinfo{person}{Dongliang Xu}.} \bibinfo{year}{2021}\natexlab{d}.
\newblock \showarticletitle{Combining Explicit Entity Graph with Implicit Text
  Information for News Recommendation}. In \bibinfo{booktitle}{\emph{Companion
  Proceedings of WWW}}. \bibinfo{pages}{412--416}.
\newblock


\bibitem[\protect\citeauthoryear{Zheng, Zhang, Zheng, Xiang, Yuan, Xie, and
  Li}{Zheng et~al\mbox{.}}{2018}]%
        {zheng2018drn}
\bibfield{author}{\bibinfo{person}{Guanjie Zheng}, \bibinfo{person}{Fuzheng
  Zhang}, \bibinfo{person}{Zihan Zheng}, \bibinfo{person}{Yang Xiang},
  \bibinfo{person}{Nicholas~Jing Yuan}, \bibinfo{person}{Xing Xie}, {and}
  \bibinfo{person}{Zhenhui Li}.} \bibinfo{year}{2018}\natexlab{}.
\newblock \showarticletitle{DRN: A deep reinforcement learning framework for
  news recommendation}. In \bibinfo{booktitle}{\emph{WWW}}.
  \bibinfo{pages}{167--176}.
\newblock


\bibitem[\protect\citeauthoryear{Zhou, Mou, Fan, Pi, Bian, Zhou, Zhu, and
  Gai}{Zhou et~al\mbox{.}}{2019}]%
        {zhou2019deep}
\bibfield{author}{\bibinfo{person}{Guorui Zhou}, \bibinfo{person}{Na Mou},
  \bibinfo{person}{Ying Fan}, \bibinfo{person}{Qi Pi}, \bibinfo{person}{Weijie
  Bian}, \bibinfo{person}{Chang Zhou}, \bibinfo{person}{Xiaoqiang Zhu}, {and}
  \bibinfo{person}{Kun Gai}.} \bibinfo{year}{2019}\natexlab{}.
\newblock \showarticletitle{Deep interest evolution network for click-through
  rate prediction}. In \bibinfo{booktitle}{\emph{AAAI}},
  Vol.~\bibinfo{volume}{33}. \bibinfo{pages}{5941--5948}.
\newblock


\bibitem[\protect\citeauthoryear{Zhou, Zhu, Song, Fan, Zhu, Ma, Yan, Jin, Li,
  and Gai}{Zhou et~al\mbox{.}}{2018}]%
        {zhou2018deep}
\bibfield{author}{\bibinfo{person}{Guorui Zhou}, \bibinfo{person}{Xiaoqiang
  Zhu}, \bibinfo{person}{Chenru Song}, \bibinfo{person}{Ying Fan},
  \bibinfo{person}{Han Zhu}, \bibinfo{person}{Xiao Ma},
  \bibinfo{person}{Yanghui Yan}, \bibinfo{person}{Junqi Jin},
  \bibinfo{person}{Han Li}, {and} \bibinfo{person}{Kun Gai}.}
  \bibinfo{year}{2018}\natexlab{}.
\newblock \showarticletitle{Deep interest network for click-through rate
  prediction}. In \bibinfo{booktitle}{\emph{KDD}}. \bibinfo{pages}{1059--1068}.
\newblock


\bibitem[\protect\citeauthoryear{Zhou, Wang, Zhao, Zhu, Wang, Zhang, Wang, and
  Wen}{Zhou et~al\mbox{.}}{2020}]%
        {zhou2020s3}
\bibfield{author}{\bibinfo{person}{Kun Zhou}, \bibinfo{person}{Hui Wang},
  \bibinfo{person}{Wayne~Xin Zhao}, \bibinfo{person}{Yutao Zhu},
  \bibinfo{person}{Sirui Wang}, \bibinfo{person}{Fuzheng Zhang},
  \bibinfo{person}{Zhongyuan Wang}, {and} \bibinfo{person}{Ji-Rong Wen}.}
  \bibinfo{year}{2020}\natexlab{}.
\newblock \showarticletitle{S3-rec: Self-supervised learning for sequential
  recommendation with mutual information maximization}. In
  \bibinfo{booktitle}{\emph{CIKM}}. \bibinfo{pages}{1893--1902}.
\newblock


\bibitem[\protect\citeauthoryear{Zhu, Zhou, Song, Tan, and Guo}{Zhu
  et~al\mbox{.}}{2019}]%
        {zhu2019dan}
\bibfield{author}{\bibinfo{person}{Qiannan Zhu}, \bibinfo{person}{Xiaofei
  Zhou}, \bibinfo{person}{Zeliang Song}, \bibinfo{person}{Jianlong Tan}, {and}
  \bibinfo{person}{Li Guo}.} \bibinfo{year}{2019}\natexlab{}.
\newblock \showarticletitle{Dan: Deep attention neural network for news
  recommendation}. In \bibinfo{booktitle}{\emph{AAAI}},
  Vol.~\bibinfo{volume}{33}. \bibinfo{pages}{5973--5980}.
\newblock


\bibitem[\protect\citeauthoryear{Zihayat, Ayanso, Zhao, Davoudi, and
  An}{Zihayat et~al\mbox{.}}{2019}]%
        {zihayat2019utility}
\bibfield{author}{\bibinfo{person}{Morteza Zihayat}, \bibinfo{person}{Anteneh
  Ayanso}, \bibinfo{person}{Xing Zhao}, \bibinfo{person}{Heidar Davoudi}, {and}
  \bibinfo{person}{Aijun An}.} \bibinfo{year}{2019}\natexlab{}.
\newblock \showarticletitle{A utility-based news recommendation system}.
\newblock \bibinfo{journal}{\emph{Decision Support Systems}}
  \bibinfo{volume}{117} (\bibinfo{year}{2019}), \bibinfo{pages}{14--27}.
\newblock


\end{thebibliography}
 
\end{document}